\documentclass[11pt]{article}

\usepackage[margin=1in,footskip=0.25in]{geometry}
\usepackage[colorlinks=true, linkcolor=red, urlcolor=blue, citecolor=gray]{hyperref}
\usepackage[usenames,dvipsnames,svgnames,table]{xcolor}

\usepackage{algorithm}
\usepackage{algorithmic}
\usepackage{amsfonts,amssymb,amsthm,amsmath}
\usepackage{bbm}
\usepackage{booktabs}
\usepackage{caption}
\usepackage{color}
\usepackage{enumitem}
\usepackage{float}
\usepackage{graphicx}
\usepackage{mathtools}
\usepackage{nicefrac}       
\usepackage{subcaption}
\usepackage{xspace}
\usepackage{url}
\usepackage{breakcites}
\usepackage[capitalize, nameinlink]{cleveref}
\usepackage{wrapfig}

\usepackage{contour}
\usepackage[normalem]{ulem}

\contourlength{0.8pt}

\usepackage{todonotes}

\makeatletter
\def\hlinewd#1{%
	\noalign{\ifnum0=`}\fi\hrule \@height #1 \futurelet
	\reserved@a\@xhline}
\makeatother

\newtheorem{theorem}{Theorem}

\newtheorem{corollary}[theorem]{Corollary}
\newtheorem{lemma}[theorem]{Lemma}

\newtheorem{definition}{Definition}

\newtheorem{importedtheorem}[theorem]{Imported Theorem}
\newtheorem{importedlemma}[theorem]{Imported Lemma}

\makeatletter
\newtheorem*{rep@theorem}{\rep@title}
\newcommand{\newreptheorem}[2]{%
	\newenvironment{rep#1}[1]{%
		\def\rep@title{\hyperref[##1]{#2~\ref*{##1}} Restated}%
		\begin{rep@theorem}}%
		{\end{rep@theorem}}}

\makeatother
\newreptheorem{theorem}{Theorem}
\newreptheorem{lemma}{Lemma}
\newreptheorem{corollary}{Corollary}
\newreptheorem{definition}{Definition}
\newreptheorem{importedtheorem}{Imported Theorem}

\newcommand\numberthis{\addtocounter{equation}{1}\tag{\theequation}}

\newcommand{\defeq}[0]{\ensuremath{\;{\vcentcolon=}\;}\xspace}
\newcommand{\spaneq}{\ensuremath{\,\mathrel{\overset{\makebox[0pt]{\mbox{\normalfont\tiny\sffamily span}}}{=}}\,}}

\newcommand{\R}{\mathbb{R}}

\newcommand{\bv}[1]{\mathbf{#1}}

\newcommand{\norm}[1]{\|#1\|}

\DeclareMathOperator{\diag}{\mathrm{diag}}
\DeclareMathOperator{\poly}{poly}

\DeclareMathOperator{\colspan}{span}

\DeclareMathOperator{\rank}{rank}

\colorlet{todo_background_normal}{white}
\definecolor{todo_background_dark}{RGB}{39,40,34}

\definecolor{advice_text}{RGB}{78, 12, 123}
\colorlet{advice_background}{todo_background_normal}

\definecolor{incomplete_text}{RGB}{204, 64, 84}
\colorlet{incomplete_background}{todo_background_normal}

\newcommand{\eps}[0]{\ensuremath{\varepsilon}}
\let\epsilon\eps

\newcommand{\tsfrac}[2]{{\textstyle\frac{#1}{#2}}}

\newcommand{\pij}[2]{p_{_{#1\hspace{-0.05em},\hspace{-0.05em}#2}}\hspace{-0.2em}}

\newcommand{\abs}[1]{\ensuremath{\vert{#1}\vert}\xspace} 
\newcommand{\mat}[1]{\mathbf{#1}} 
\renewcommand{\vec}[1]{\boldsymbol{\mathrm{#1}}} 
\newcommand{\vecalt}[1]{\boldsymbol{#1}} 
\newcommand{\normof}[1]{\|#1\|} 

\newcommand{\bmat}[1]{\begin{bmatrix} #1 \end{bmatrix}} 
\newcommand{\sbmat}[1]{\left[\begin{smallmatrix} #1 \end{smallmatrix}\right]} 

\newcommand{\mA}{\ensuremath{\mat{A}}\xspace}
\newcommand{\mB}{\ensuremath{\mat{B}}\xspace}
\newcommand{\mC}{\ensuremath{\mat{C}}\xspace}
\newcommand{\mD}{\ensuremath{\mat{D}}\xspace}
\newcommand{\mE}{\ensuremath{\mat{E}}\xspace}

\newcommand{\mG}{\ensuremath{\mat{G}}\xspace}

\newcommand{\mI}{\ensuremath{\mat{I}}\xspace}

\newcommand{\mK}{\ensuremath{\mat{K}}\xspace}

\newcommand{\mM}{\ensuremath{\mat{M}}\xspace}

\newcommand{\mP}{\ensuremath{\mat{P}}\xspace}
\newcommand{\mQ}{\ensuremath{\mat{Q}}\xspace}

\newcommand{\mS}{\ensuremath{\mat{S}}\xspace}

\newcommand{\mU}{\ensuremath{\mat{U}}\xspace}
\newcommand{\mV}{\ensuremath{\mat{V}}\xspace}
\newcommand{\mW}{\ensuremath{\mat{W}}\xspace}
\newcommand{\mX}{\ensuremath{\mat{X}}\xspace}
\newcommand{\mY}{\ensuremath{\mat{Y}}\xspace}
\newcommand{\mZ}{\ensuremath{\mat{Z}}\xspace}
\newcommand{\mLambda}{\ensuremath{\mat{\Lambda}}\xspace}
\newcommand{\mSigma}{\ensuremath{\mat{\Sigma}}\xspace}

\newcommand{\vc}{\ensuremath{\vec{c}}\xspace}

\newcommand{\ve}{\ensuremath{\vec{e}}\xspace}

\newcommand{\vg}{\ensuremath{\vec{g}}\xspace}

\newcommand{\vk}{\ensuremath{\vec{k}}\xspace}

\newcommand{\vm}{\ensuremath{\vec{m}}\xspace}

\newcommand{\vq}{\ensuremath{\vec{q}}\xspace}

\newcommand{\vx}{\ensuremath{\vec{x}}\xspace}

\newcommand{\vz}{\ensuremath{\vec{z}}\xspace}

\newcommand{\vsigma}{\ensuremath{\vecalt{\sigma}}\xspace}

\newcommand{\cI}{\ensuremath{{\mathcal I}}\xspace}

\newcommand{\cN}{\ensuremath{{\mathcal N}}\xspace}

\newcommand{\bbN}{\ensuremath{{\mathbb N}}\xspace}

\newcommand{\bbR}{\ensuremath{{\mathbb R}}\xspace}

  \usepackage{nth}
  \usepackage{intcalc}

\title{On the Unreasonable Effectiveness of Single Vector Krylov Methods for Low-Rank Approximation}

\author{
	Raphael A. Meyer \\ New York University\\ \texttt{ram900@nyu.edu}
	\and 
	Cameron Musco\\ University of Massachusetts Amherst\\ \texttt{cmusco@cs.umass.edu}
	\and
	Christopher Musco\\ New York University\\ \texttt{cmusco@nyu.edu}
}

\date{}

\begin{document}
\maketitle


\begin{abstract}
Krylov subspace methods are a ubiquitous tool for computing near-optimal rank $k$ approximations of large matrices.
While ``large block'' Krylov methods with block size \emph{at least} $k$ give the best known theoretical guarantees, block size one (a single vector) or a small constant is often preferred in practice.
Despite their popularity, we lack theoretical bounds on the performance of such ``small block'' Krylov methods for low-rank approximation. 

We address this gap between theory and practice by proving that small block Krylov methods essentially match all known low-rank approximation guarantees for large block methods.
Via a black-box reduction we show, for example, that the standard single vector Krylov method run for $t$ iterations obtains the same spectral norm and Frobenius norm error bounds as a Krylov method with block size $\ell \geq k$ run for $O(t/\ell)$ iterations, up to a {logarithmic dependence} on the smallest gap between sequential singular values.  
That is, for a given number of matrix-vector products, single vector methods are essentially as effective as  \emph{any choice of large block size}.

By combining our result with tail-bounds on eigenvalue gaps in random matrices, we prove that the dependence on the smallest singular value gap can be eliminated if the input matrix is perturbed by a small random matrix.
Further, we show that single vector methods match the more complex algorithm of [Bakshi et al. `22], which combines the results of multiple block sizes to achieve an improved algorithm for Schatten $p$-norm low-rank approximation. 
\end{abstract}


\section{Introduction}
\label{sec:intro}
Krylov subspace methods have been studied since the 1950s and remain our most reliable algorithms for approximating eigenvectors and singular vectors of large matrices. Krylov methods access a matrix $\bv{A}$ via repeated matrix multiplications (each considered an iteration of the method) either with a single vector or a block of vectors. There has been significant interest in analyzing how many iterations are required to obtain accurate eigenvector or singular vector approximations. Classic work studies both single vector \cite{Kaniel:1966,Paige:1971} and block methods \cite{CullumDonath:1974,KahanParlett:1976,Saad:1980,Saad:2011}. 

More recently, there has been interest in analyzing  Krylov subspace methods specifically for the downstream task of \emph{low-rank approximation}. Since the top $k$ singular vectors can be used to obtain an optimal rank-$k$ approximation, the goal is to understand how many iterations are required to compute approximate singular vectors that yield a near-optimal rank-$k$ approximation \cite{RokhlinSzlamTygert:2009,HalkoMartinssonTropp:2011,Woodruff:2014}. This problem differs from classical work because convergence to the actual top singular vectors is sufficient but \emph{not necessary} for obtaining an accurate low-rank approximation \cite{DrineasIpsen:2019}.

Prototypical single vector and  block Krylov methods for low-rank approximation are shown in \cref{alg:single-vec-krylov} and \cref{alg:block-krylov}.\footnote{
	 \Cref{alg:single-vec-krylov,alg:block-krylov} are examples of the simplest possible implementations of Krylov methods for low-rank approximation. In practice, various optimizations like the Lanczos recurrence are often applied, and additional care is necessary to ensure that the orthogonal basis for the Krylov subspace $\mK$ in computed in a numerically stable way \cite{Saad:2011}.
	While an important topic, this paper is not focused on the numerical stability of Lanczos methods.
	All derivations assume computation in the Real RAM model of arithmetic.
	}
For an $n \times d$ input $\mA$, both methods returns an $n \times k$ orthogonal $\mQ$ so that the rank-$k$ matrix $\mQ\mQ^\intercal\mA$ is a good approximation to $\mA$. Ideally, it is nearly as good as $\mA$'s optimal rank $k$ approximation, $\mA_k$, which is given via projection onto $\mA$'s top $k$ singular vectors.\footnote{
	For simplicity, we focus on computing an approximate left singular vector subspace spanned by $\mQ \in \R^{n \times k}$.
	If we instead care about computing right singular vectors, \Cref{alg:single-vec-krylov,alg:block-krylov} can be applied to \(\mA^\intercal\) instead.
	} 

\begin{algorithm}[t]
	\caption[Single Vector Krylov Method for Low-Rank Approximation]{Single Vector Krylov Method for Low-Rank Approximation}
	\label{alg:single-vec-krylov}
	{\bfseries input}: Matrix \(\mA\in\bbR^{n \times d}\). Target rank \(k\). Starting vector \(\vx\in\bbR^{n}\). Number of iterations \(t\). \\
	{\bfseries output}: Orthogonal matrix \(\mQ\in\bbR^{n \times k}\).\\
	\vspace{-1em}
	\begin{algorithmic}[1]
		\STATE Compute an orthonormal basis \mZ for \(\mK = [\, \vx,~ (\mA\mA^\intercal)\vx,~  (\mA\mA^\intercal)^2\vx,~\ldots,~ (\mA\mA^\intercal)^t \vx \,]\).
		\STATE Compute \(\mU_k\), the \(k\) top eigenvectors of \(\mM = \mZ^\intercal\mA\mA^\intercal\mZ\)
		\STATE {\bfseries return} \(\mQ = \mZ\mU_k\).
	\end{algorithmic}
\end{algorithm}

\begin{algorithm}[t]
	\caption{Block Krylov Method for Low-Rank Approximation}
	\label{alg:block-krylov}
	{\bfseries input}: Matrix \(\mA\in\bbR^{n \times d}\). Target rank \(k\). Starting block \(\mB\in\bbR^{n \times \ell}\). Number of iterations \(t\). \\
	{\bfseries output}: Orthogonal matrix \(\mQ\in\bbR^{n \times k}\).\\
	\vspace{-1em}
	\begin{algorithmic}[1]
		\STATE Compute an orthonormal basis \mZ for \(\mK = [\, \mB,~ (\mA\mA^\intercal)\mB,~ (\mA\mA^\intercal)^2\mB~ \ldots,~ (\mA\mA^\intercal)^t \mB \,]\).
		\STATE Compute \(\mU_k\), the \(k\) top eigenvectors of \(\mM = \mZ^\intercal\mA\mA^\intercal\mZ\)
		\STATE {\bfseries return} \(\mQ = \mZ\mU_k\).
	\end{algorithmic}
\end{algorithm}

\subsection{Large block methods and gap-free bounds}
\label{sec:why_large_block}
Most recent work on Krylov methods for low-rank approximation focuses on ``large block'' methods, where $\ell$ in \cref{alg:block-krylov} is chosen to be $\geq k$  \cite{RokhlinSzlamTygert:2009,HalkoMartinssonShkolnisky:2011,Gu:2015,MuscoMusco:2015,tropp2018analysis,Yuan:2018ub,Drineas:2018vu,tropp2018analysis}.
In this regime,  block methods are known to quickly converge to a near-optimal low-rank approximation.
For example, in just $O\left (\frac{\log(n/\eps)}{\sqrt{(\sigma_k-\sigma_{\ell+1})/\sigma_{k}}}\right)$ iterations, \Cref{alg:block-krylov} initialized with an i.i.d. random Gaussian matrix \(\mB\in\bbR^{n \times \ell}\) with block size \(\ell \geq k\), achieves with high probability the bound
\begin{align}
	\label{eq:guar_1}
	\normof{\mA - \mQ\mQ^\intercal\mA}_\xi &\leq (1+\eps) \normof{\mA - \mA_k}_\xi
\end{align}
 for any $\eps > 0$ and $\|\cdot \|_\xi$ being either the Frobenius or spectral norm \cite{MuscoMusco:2015}.
That is, convergence is linear with a rate depending on the square root of the relative gap from the $k^\text{th}$ singular value, $\sigma_k$, to the $(\ell+1)^\text{st}$ singular value, $\sigma_{\ell+1}$.
Even for $\ell$ mildly larger than $k$, this gap is often quite large.
For example, \cite{HalkoMartinssonTropp:2011} recommends setting $\ell = k+5$ or $k+10$.

Beyond such spectrum dependent guarantees, another advantage of large block Krylov methods is that they enjoy \emph{gap-independent} bounds, which do not involve any terms depending on $\mA$'s spectrum.
For example, a now standard result is that \Cref{alg:block-krylov} achieves \Cref{eq:guar_1} in just
$O(\frac{1}{\sqrt\eps}\log(\frac n\eps))$ iterations \cite{MuscoMusco:2015}.\footnote{Randomly initialized block power method with block size $k$ gives a similar bound, but with a suboptimal $1/\eps$ rather than $1/\sqrt{\eps}$ dependence on the error \cite{RokhlinSzlamTygert:2009,HalkoMartinssonTropp:2011}.} Further, this bound is essentially optimal among all methods that access $\bv A$ only through matrix-vector products \cite{simchowitz2018tight,Bakshi:2023uu}.
Bounds where the iteration complexity does not depend on properties of \mA, are called ``universal'' guarantees \cite{Urschel:2021}.
Universal bounds are useful in applications where properties like large spectral gaps cannot be ensured, but where worst-case accuracy guarantees are still desired \cite{Hegde:2016wp,LiLindermanSzlam:2017,Soltani:2018uj}.

In contrast to large block sizes, it is \emph{impossible} to prove gap-independent guarantees for single vector or small block Krylov iteration.
To see why, consider $\mA \in \R^{k \times d}$ that is all zeros, except that $\mA_{ii} = 1$ for $i = 1, \ldots, k$. I.e., 
$
\bv{A} = \begin{bmatrix}
		\mI_k & \bv{0}\\
	\end{bmatrix}.
$
where $\bv{I}_k$ denotes the $k\times k$ identity matrix.
If we run \Cref{alg:block-krylov} on this matrix with block size $\ell < k$, then it can be checked that the Krylov subspace $\bv{K}$ will have rank $\ell < k$, and thus any low-rank approximation obtained from the subspace cannot be near-optimal.
In general, bounds for small block methods must depend \emph{inversely} on the gaps between sequential singular values. In the above example, these gaps are equal to $0$.

\begin{figure}[tb]
	\centering
	\includegraphics[width=\textwidth,trim={0 0 0 .6cm},clip]{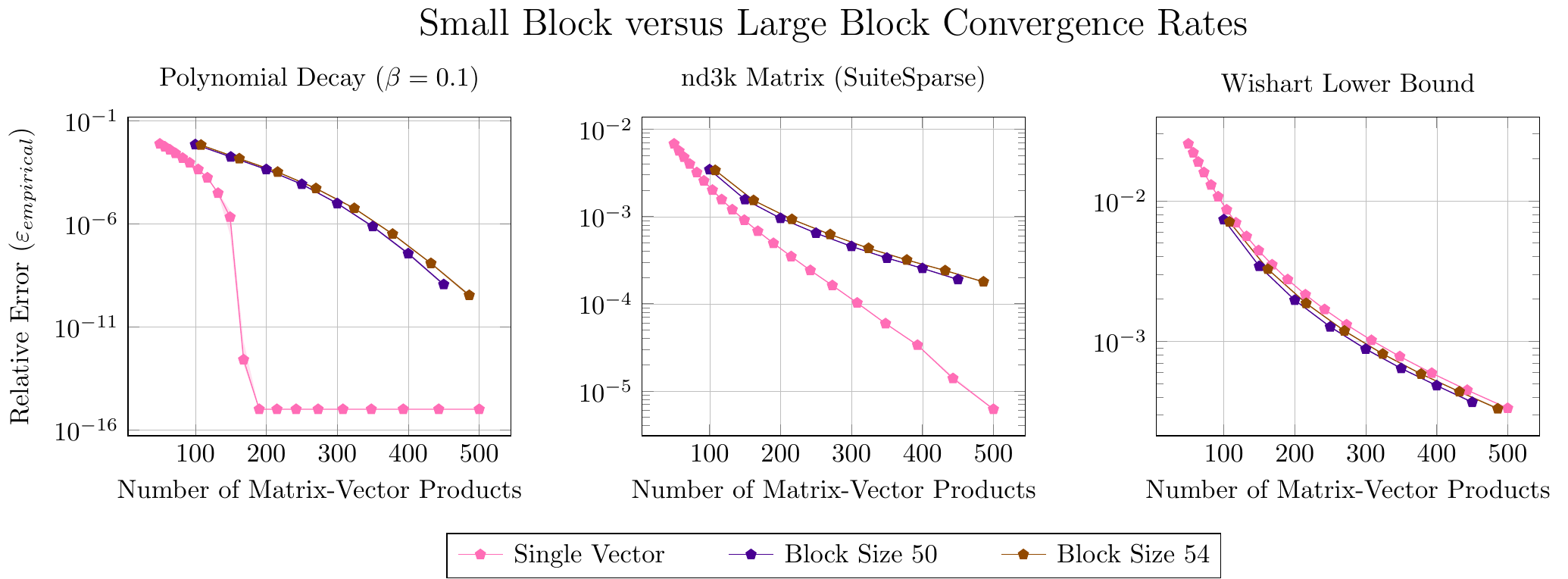}
	\caption{
		Comparison of the number of matrix-vector products needed for Krylov iteration to converge to an accurate rank $50$ approximation under different block sizes.
		The left figure uses a matrix with singular values decaying polynomially;  the middle figure uses a matrix from the SuiteSparse library; the right figure uses a worst-case matrix from the literature. See \Cref{sec:experiment-grid} for more details.
		In the first two plots, single vector Krylov outperforms large block methods. Even in the adversarially chosen hard instant on the right, it does not perform much worse. 
	}
	\label{fig:intro-plots}
\end{figure}

\subsection{Main contribution: the virtue of small block Krylov methods}
\label{sec:our-approach}
The inability of single vector and small block Krylov methods to offer gap-independent bounds has been a point of concern for the use of these methods in computing low-rank approximations \cite{LiLindermanSzlam:2017,MuscoMusco:2015,blockGaps}.
At the same time, in practice, low-rank approximation is frequently solved using iterative eigensolvers based on single vector or small block methods. Such methods are the standard in MATLAB, Julia, Python, and essentially all languages used for matrix computations \cite{Lehoucq:1998uq,Mathworks:2023va,SciPy-community:2023vp}.
These methods often perform very well, converging quickly to good low-rank approximations.
In fact, in our experience, they typically outperform large block  methods in terms of the number of matrix-vector products required to achieve a desired level of accuracy -- see \Cref{fig:intro-plots}.\footnote{The number of matrix-vector products used by an algorithm does not necessarily translate directly into the computational cost of the algorithm. For example, in many computing systems, it is faster to multiply a matrix $\mA$ by a block of $k$ vectors all at once, than to multiply by $k$ vectors chosen in sequence. Nevertheless, matrix-vector products are still a valuable measure of complexity for many problems where they dominate other runtime costs.}

The main goal of this paper is to explain this phenomenon. We ask:

\begin{quote}
	\textit{For low-rank approximation, when and why do small block Krylov methods require the same or fewer matrix-vector multiplications than large block Krylov methods?}
\end{quote}

We answer this question in a strong way by proving that small block methods nearly match or even improve on all known theoretical guarantees on the convergence of large-block methods for low-rank approximation.
In particular, up to a \emph{logarithmic dependence} on the smallest gap between singular values, the trade-off between accuracy and number of matrix-vector products achieved by small block methods matches the trade-off achieved by large block methods.
Since there are a variety of guarantees known for large block methods, this claim is broken down as a number of results throughout our paper. We state one such result as a concrete example:
\begin{theorem}
	\label{thm:single-vec-guarantee}
	For \(\mA \in \bbR^{n \times d}\), let \(g_{min} = \min_{i \in \{1,\ldots,k-1\}} \frac{\sigma_{i} - \sigma_{i+1}}{\sigma_i}\) be the smallest relative gap among the top \(k\) singular values.
	For any \(\eps, \delta \in (0,1)\), \Cref{alg:single-vec-krylov} initialized with an i.i.d. mean zero Gaussian vector $\vx$ and run for  \(t = O(\tsfrac{k}{\sqrt \eps} \log(\tsfrac1{g_{min}}) + \tsfrac1{\sqrt\eps} \log(\tsfrac{n}{\eps\delta}))\) iterations returns an orthogonal $\mQ \in \R^{n \times k}$ such that, with probability at least \(1-\delta\), letting $\norm{\cdot}_\xi$ be the spectral or Frobenius norm,
	\[
	\normof{\mA-\mQ\mQ^\intercal\mA}_\xi \leq (1+\eps)\normof{\mA-\mA_k}_\xi.
	\]
\end{theorem}
As discussed,  \cite{MuscoMusco:2015} prove that \cref{alg:block-krylov} with block size $k$ achieves an identical error bound in $O \left (\frac{\log(n/\eps \delta)}{\sqrt{\eps}}\right )$ iterations.
This translates to $O\left (\frac{k \log(n/\eps\delta)}{\sqrt{\eps}}\right )$ matrix-vector products, which  \Cref{thm:single-vec-guarantee} matches, except for the dependence on $\log({1}/{g_{min}})$.
At the same time, \Cref{thm:single-vec-guarantee} improves on the large block bound by separating the $\log (n/\eps \delta)$ and $k$ terms.

\medskip

\noindent\textbf{Remark.} 
Since it is a logarithmic instead of polynomial dependence, we consider the \(\log({1}/{g_{min}})\) term to be mild for typical problems. In experiments, it appears to have little impact on the observed convergence of the single vector Krylov method (see \cref{sec:experiments}). Indeed, except in adversarial cases, such as the identity matrix, where $g_{min}$ truly equals $0$, in finite precision, we cannot expect to resolve singular value gaps to accuracy better than machine precision. So, it is reasonable to think that in practice, this term should be at most a moderate constant. We make this intuition formal in \cref{sec:perturb}, showing that the dependence on $g_{min}$ can be eliminated in a smoothed analysis setting (i.e., when the input is perturbed by a small random matrix).

\medskip

\begin{figure}[b]
	\begin{minipage}[c]{0.4\textwidth}
		\includegraphics[width=.9\textwidth]{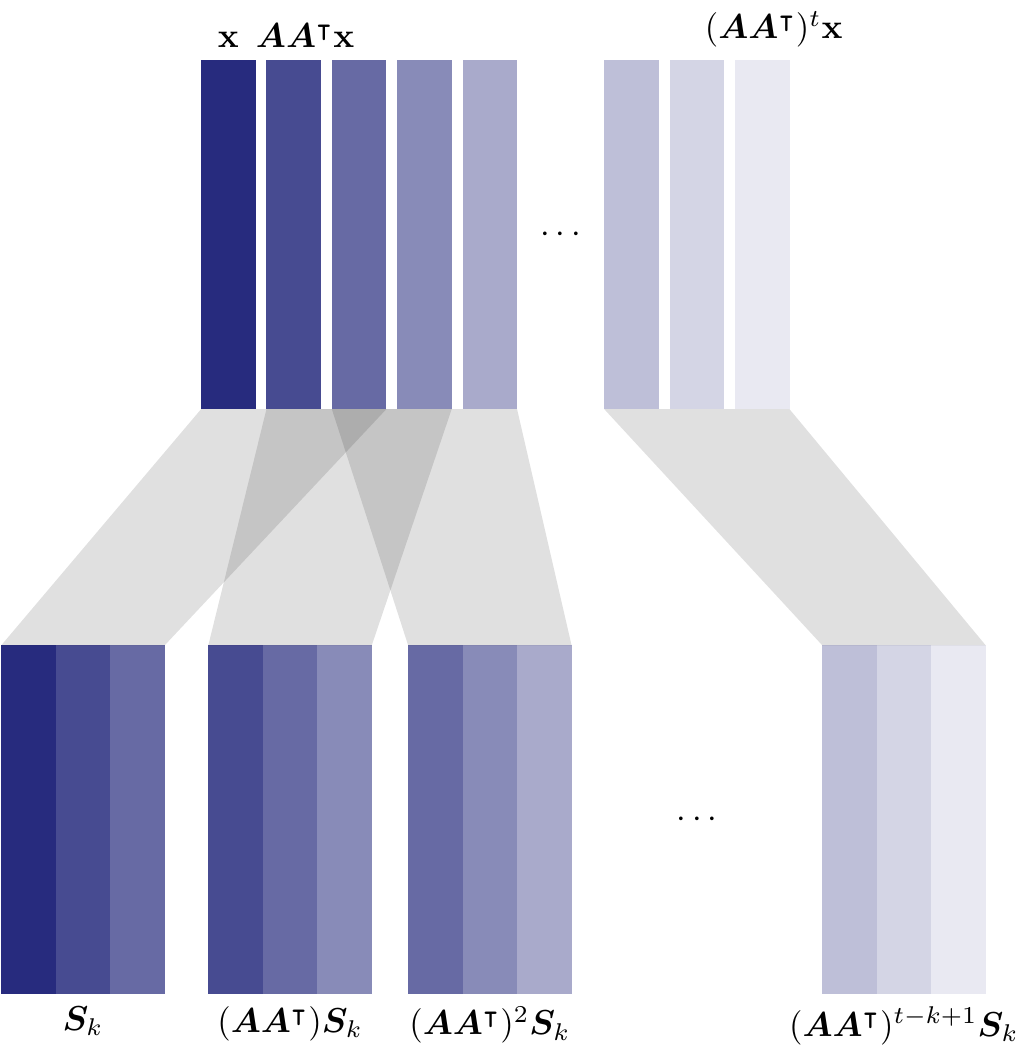}
	\end{minipage}\hfill
	\begin{minipage}[c]{0.55\textwidth}
		\caption{
			Our main analysis is based on the simple observation that the span of the Krylov subspace generated by a single vector Krylov method after $t$ iterations is exactly equivalent to that generated by a block Krylov method run for $t-k + 1$ iterations with starting block $\mS_k = [\, \vx,~ (\mA\mA^\intercal)\vx,~ \ldots,~ (\mA\mA^\intercal)^k \vx \,]$. This observation allows us to take advantage of existing results on block methods in a black-box way.
		} \label{fig:main-intuition}
	\end{minipage}
\end{figure}

Our proof for \cref{thm:single-vec-guarantee} (with some additional results) is given in \cref{sec:gap-dep-analysis}. Our approach is via a black-box reduction to the existing analysis for large block methods. In particular, we view the single-vector method of \Cref{alg:single-vec-krylov} as a block Krylov method in disguise. We observe that the span of the single-vector Krylov subspace $\mK$ (Line 1 of \Cref{alg:single-vec-krylov}) is exactly equivalent to the span of a block Krylov subspace generated from a specific starting matrix.
Concretely, suppose \Cref{alg:single-vec-krylov} is run for $t \geq k$ iterations and let \(\mS_k\in\bbR^{d \times k}\) equal the first \(k\) columns of $\mK$.
I.e.,
\begin{align}
	\label{eq:intro_our_b}
	\mS_k \defeq \bmat{\vx & \mA\mA^\intercal\vx & (\mA\mA^\intercal)^2\vx & \ldots & (\mA\mA^\intercal)^{k-1}\vx}.
\end{align}
Then we can check that for \(q = t - k + 1\): 
\begin{align}
	\label{eq:intro_our_krylov_space}
	\colspan(\mK) = \colspan\left(\bmat{\mS_k & \mA\mA^\intercal\mS_k  & (\mA\mA^\intercal)^{2}\mS_k & \ldots & (\mA\mA^\intercal)^{q}\mS_k}\right).
\end{align}
This equivalence is visualized in \Cref{fig:main-intuition}.
Since both \Cref{alg:single-vec-krylov} and \Cref{alg:block-krylov} only depend on the \emph{span} of the Krylov subspace they generate (through $\mZ$), the single vector method thus matches the block Krylov method run for $k-1$ fewer iterations, with the specific starting block $\mS_k$.

With this perspective, a naive hope might be to directly appeal to prior results on block Krylov iteration to analyze the single vector method. Unfortunately, these  results rely on the fact that the starting matrix $\mB$ is chosen at random, typically with i.i.d. Gaussian or sub-Gaussian entries \cite{HalkoMartinssonTropp:2011,MuscoMusco:2015,bakshi2022low}.
In contrast, $\mS_k$ is far from a random Gaussian matrix.
Its columns are highly dependent on each other.
To understand just how far $\mS_k$ is from an ideal starting matrix, note that, generally, a block Krylov subspace with $q$ blocks will have rank $qk$ when $\mB$ is a random Gaussian matrix.
In contrast, the block Krylov subspace $\bmat{\mS_k & \mA\mA^\intercal\mS_k & \ldots & (\mA\mA^\intercal)^{q}\mS_k}$ only has rank $t = q + k - 1$.

Surprisingly, however, we are still able to show that $\mS_k$ provides a (barely) good enough starting matrix for the block Krylov method in \Cref{alg:block-krylov} to succeed. To do so, we consider a natural definition of what it means to be a  ``good'' starting matrix. 
At a high-level, we need $\mS_k$ to have non-negligible inner product with all top $k$ singular vectors of $\mA$. While $\mS_k$ is exponentially worse in terms of starting inner product than a random $\mB$, this is made up for by the fact that it is far cheaper (in terms of matrix-vector products) to build a block Krylov subspace with $\mS_k$; we can compute a degree $q$ subspace using just $q+k$ matrix-vector products with $\mA$. In contrast, computing a degree $q$ block Krylov subspace with a random starting block  $\mB\in \R^{n \times k}$ requires $qk$ matrix-vector products.  The detailed proof is presented in  \Cref{sec:gap-dep-analysis}.

\subsection{Results and Paper Organization}
\label{sec:additional-results} 
\Cref{thm:single-vec-guarantee} is our main result, and its proof is contained entirely in \Cref{sec:gap-dep-analysis}.
In addition to this result, whose proof shows the crux of our argument that single vector methods converge quickly, we include several other bounds for single vector and small block methods.
We summarize these additional results below.
\Cref{sec:experiments} contains experiments which demonstrate that our bounds are predictive of the performance of these methods in practice.

\medskip
\noindent\textbf{Spectrum adaptive bounds, \cref{sec:spectral-decay}}.
The black-box nature of  \cref{thm:single-vec-guarantee}'s proof allows us to similarly adapt other results on large block Krylov methods to the single vector setting.
For example, we show that \cref{alg:single-vec-krylov} matches known ``spectrum dependent'' bounds for large block methods.
As discussed in \cref{sec:why_large_block}, the convergence rate of these bounds depends on \(g_{k\rightarrow \ell} = \frac{\sigma_k - \sigma_{\ell+1}}{\sigma_{k}}\), the gap between the $k^\text{th}$ and $(\ell+1)^\text{st}$ singular values.
Since this gap increases with $\ell$, there is a natural tradeoff: a larger block means more matrix-vector products per iteration, but fewer iterations. Our proof shows that single vector methods match any block size $\ell \geq k$ (up to a dependence on $\log(1/g_{min})$).
I.e., they automatically match the complexity of the method with \emph{best choice of large block size}, without need for parameter tuning.

\medskip
\noindent\textbf{Schatten p-norm low-rank approximation, \cref{sec:fast-frob,sec:fast-schatten}}. Recently, a combination of spectrum dependent and spectrum independent bounds have been used to give faster convergence rates for Schatten p-norm low rank approximation.
In particular, for constant $p$, \cite{bakshi2022low} show how to find a low-rank approximation achieving  $\normof{\mA-\mQ\mQ^\intercal\mA}_p \leq (1+\eps)\normof{\mA-\mA_k}_p$ using just $\tilde{O}(k/\eps^{1/3})$ matrix-vector products with $\mA$.
For the Frobenius norm (i.e., \(p=2\)), this is an improvement on the $\tilde{O}(k/\sqrt{\eps})$ required by \cite{MuscoMusco:2015}.
Their method requires running \cref{alg:block-krylov} with multiple choices of block size, and optimizing over the best Krylov subspace.
We show that, again up to a logarithmic dependence on $1/g_{min}$, the exact same guarantees can be obtained by simply running a single vector Krylov method.
In concurrent work, \cite{Bakshi:2023uu} show a similar result without a dependence on \(\log(1/g_{min})\) or \(\log(n)\).

\medskip
\noindent\textbf{Beyond block size 1, \cref{sec:beyond-single-vec}}.
While the above results focus on the single vector Krylov method, our bounds naturally generalize to other small block sizes between $1$ and $k$ (e.g. $4$, $10$, or $k-1$). For general small block size $b$, we show that dependence on  \(g_{min}\) can be replaced with a dependence on  the smallest ``$b^\text{th}$ order'' gap  \(g_{min,b} \defeq \min_{i \in \{1,\ldots,k-b\}} \frac{\sigma_{i} - \sigma_{i+b}}{\sigma_i}\).

\medskip
\noindent\textbf{Removing the gap dependence, \cref{sec:perturb}}.
While a dependence on singular value gaps is unavoidable for small block methods in the worst-case, the parameter seems to rarely have an impact in practice. We take a step towards explaining this observation via a  \emph{smoothed-analysis} result \cite{Spielman:2004vw,sankar2006smoothed}. Specifically, we leverage work in  random matrix theory on \textit{eigenvalue repulsion}, which shows that small spectral gaps in a matrix are brittle: adding a tiny amount of random noise to any matrix ensures that its singular value gaps are at worst inverse polynomial in the problem parameters. Using this fact, we present bounds that replace the dependence on $\log(1/g_{min})$ in our prior results with a dependence on $\log(\frac{n\kappa_k}{\delta\eps})$ for randomly perturbed matrices, where \(\kappa_k = \sigma_1/\sigma_k\) measures the conditioning of the top \(k\) singular values.
From an algorithm design perspective, the $\log(1/g_{min})$ can be removed even in the worst case by explicitly adding a random diagonal perturbation to $\mA$.

\medskip
\noindent\textbf{Single Vector Simultaneous Iteration, \cref{sec:single-vector-simultaneous-iteration}}.
Lastly, we describe a single vector analogue for simultaneous iteration.
While it converges somewhat more slowly, this method has the advantage of using less space than the single vector Krylov method, which needs to store the span for the entire Krylov subspace.
For instance, it allows us to store just \(k\) vectors while converging in \(\tilde O(k/\eps)\) iterations, in contrast to the \(\tilde O(k/\sqrt\eps)\) iterations required by the standard single vector Krylov method.
Since single vector simultaneous iteration only uses a single starting vector, its convergence still depends on \(\log(1/g_{min})\).

\subsection{Related Work}
\label{sec:related_work}

We briefly discuss additional prior work on low-rank approximation and Krylov methods, though since the literature is rich, so we cannot cover all relevant prior work. As discussed, early analyses of Krylov methods for approximating eigenvectors consider both single vector and block methods \cite{Saad:1980,GolubUnderwood:1977,KuczynskiWozniakowski:1992}. However, this work does not directly provide strong bounds for low-rank approximation, since convergence to the top singular vectors is not required to accurately solve the problem \cite{DrineasIpsen:2019}.

For low-rank approximation, large block methods have been more popular. In addition to prior work already discussed, this includes work on randomized sketching methods, which can be viewed as large block Krylov methods run for one or two iterations \cite{MartinssonRokhlinTygert:2006,CohenElderMusco:2015,ClarksonWoodruff:2013,DrineasMahoney:2016,MartinssonTropp:2020}. Sketching methods have become a mainstay technique in randomized numerical linear algebra. 

Work on single vector or small block methods for low-rank approximation has been more sparse. 
\cite{WangZhangZhang:2015} experimentally study small block methods, and suggest that large blocks are only worthwhile when singular value gaps are very small, when low precision suffices, or when making many passes over a matrix is expensive. \cite{Yuan:2018ub} theoretically studies the related problem of singular value approximation for all block sizes, and as in our work, obtains linear convergence rates depending on \(g_{k\rightarrow \ell}\).
They also show superlinear rates when \mA has a sufficiently quickly decaying spectrum.
While it is difficult to directly compare their results to ours on low-rank approximation, it would be interesting to consider such spectra in our setting. Finally, we note that
\cite{Allen-ZhuLi:2016} proves a result similar to \cref{thm:single-vec-guarantee} using an algorithm that in some ways is a single vector Krylov method. However, because the method iteratively restarts \(k\) times with \(k\) randomly chosen starting vectors, it ultimately returns a solution from a block Krylov subspace. 

Related to our results in \cref{sec:perturb}, we note that adding small random perturbations to avoid small singular value gaps or other conditioning issues is a technique that has been employed in several recent works focused on worst-case runtime bounds for linear algebraic problems \cite{Boutsidis:2016tk,Peng:2021uz,Banks:2022tn}.


\section{Notation }
\label{sec:prelims}

We use capital bold letters to denote matrices, lowercase bold letters to denote vectors, and lowercase non-bold letters to denote scalars. For a matrix $\mQ$, we let \(\vq_i\) denote the \(i^\text{th}\) column, $\colspan(\mQ)$ denote its column span, and $\mQ^\intercal$ denote its transpose.
Typically, \(\mA\in\bbR^{n \times d}\) denotes our input matrix.
We let \(\mA = \mU\mSigma\mV^\intercal\) denote the SVD of \mA, with \(\mU\in\bbR^{n \times n}, \mSigma\in\bbR^{n \times n}, \mV\in\bbR^{d \times n}\).
We let \(\sigma_1 \geq \sigma_2 \geq \ldots \ge \sigma_n \ge 0\) denote the singular values of \mA (the diagonal entries of $\mSigma$).
We let \(\mU_k\in\bbR^{n \times k}\) and \(\mV_k \in \bbR^{d \times k}\) denote the first \(k\) columns of \mU and \mV, and let \(\mSigma_k \in \bbR^{k \times k}\) denote the top \(k \times k\) principal submatrix of \mSigma.
Then \(\mA_k = \mU_k\mSigma_k\mV_k^\intercal\) is the best rank-\(k\) approximation to \mA in any unitarily invariant norm.
When \(\mA\) is square, we let \(\mA=\mU\mLambda\mU^\intercal\) be the eigendecomposition of \mA, where \(\lambda_1 \geq \lambda_2 \geq \ldots \geq \lambda_n\) are the eigenvalues of \mA (the diagonal entries of \mLambda).
We often work with symmetric positive semi-definite (PSD) matrices, which have all non-negative eigenvalues.
In this case, the singular values equal the eigenvalues.
We also work with matrix polynomials.
If \(p(x) = \sum_{i=1}^q c_i x^i\) is a polynomial and if \(\mA\) is square, \(p(\mA) \defeq \sum_{i=1}^q c_i \mA^i\).

We let \(\normof{\vx}_2\) denote the vector \(\ell_2\) norm, \(\normof{\mA}_2\) the spectral norm, \(\normof{\mA}_F\) the Frobenius norm, and \(\normof{\mA}_p = \left(\sum_{i=1}^p \sigma_i^p\right)^{1/p}\)  the Schatten \(p\)-norm. 
Wherever \(\normof{\mA}_\xi\) is used, the equation holds for both the spectral and Frobenius norms.
We let \([n] = \{1,\ldots,n\}\) be the set of integers between 1 and \(n\).
Finally, we let \(\cN(\vec0,\mI)\) denote the distribution over vectors whose entries are i.i.d. mean zero unit variance Gaussians. The dimension will be clear from context.

\section{Proof of \texorpdfstring{\Cref{thm:single-vec-guarantee}}{Theorem \ref*{thm:single-vec-guarantee}}}
\label{sec:gap-dep-analysis}

In this section, we prove \Cref{thm:single-vec-guarantee} by showing that \(\mS_k\) as described in \Cref{sec:our-approach} is a good enough starting matrix for block Krylov iteration. This proof serves as a foundation for all additional results in the paper.
Throughout, we assume that \(\mA\in\bbR^{n \times n}\) is square and positive semidefinite.
We shown in \Cref{app:reduct_to_psd_case} that this is without loss of generality: running \Cref{alg:single-vec-krylov} or \Cref{alg:block-krylov} on a matrix $\mC \in \R^{n \times d}$ with SVD \(\mC=\mU\mSigma\mV^\intercal\) yields an identical output to running the method on the PSD matrix \(\mA = (\mC\mC^\intercal)^{\nicefrac12} = \mU\mSigma\mU^\intercal\).
Further, all low-rank approximation and singular value approximation results guaranteed for the returned matrix $\mQ$ directly carry over from $\mA$ to $\mC$.

\subsection{A naive approach that actually works.}
\label{sec:naive-approach}
As discussed in \Cref{sec:our-approach}, our main approach is to view the single-vector method of \Cref{alg:single-vec-krylov} as a block Krylov method in disguise.
Specifically, recall the matrix \(\mS_k\) and the Krylov subspace from \Cref{eq:intro_our_b,eq:intro_our_krylov_space}.
Since we now assume that \mA is PSD, we have \(\mA\mA^\intercal=\mA^2\), so we can write
\begin{align}
	\label{eq:our_b}
	\mS_k \defeq \bmat{\vx & \mA^2\vx & \mA^4\vx & \ldots & \mA^{2(k-1)}\vx}
\end{align}
and
\begin{align}
	\label{eq:our_krylov_space}
	\colspan(\mK) = \colspan\left(\bmat{\mS_k & \mA^2\mS_k  & \ldots & \mA^{2q}\mS_k}\right),
\end{align}
where \(q \defeq t - k + 1\).
This matches the subspace spanned by \mK in \Cref{alg:block-krylov} with starting block \(\mB=\mS_k\). Since the output of  \Cref{alg:single-vec-krylov} and \Cref{alg:block-krylov} depend only on the \emph{span} of the Krylov subspace they construct, we will use this equivalence to  appeal to prior results on block Krylov methods to analyze \Cref{alg:single-vec-krylov}. To do so, we need to show that, even though \(\mS_k\) is very much unlike the i.i.d. random starting matrices used in prior work, it still provides a good enough starting matrix for convergence to a near-optimal low-rank approximation.

Towards that end, we use a natural definition of what it means to be a ``good'' starting matrix that specifically will allow us to leverage results on block Krylov methods from \cite{MuscoMusco:2015} in a black-box way.
That same definition suffices for other results as well \cite{Woodruff:2014,Drineas:2018vu}.
Intuitively, we require that a starting matrix \mB has nontrivial inner product with all of the top \(k\) singular vectors of \mA:

\begin{definition}[$(k,L)$-good Starting Matrix]
	\label{def:lgood}
	Let $\mA \in \R^{n \times d}$ be a matrix with top \(k\) left singular vectors $\mU_k \in \R^{n \times k}$.
	A matrix $\mB\in \R^{n \times k}$ is a $(k,L)$-good starting matrix for $\mA$ if, letting \(\mQ \in \R^{n \times k}\) be an orthonormal basis for \(\colspan(\mB)\), $\bv{U}_k^\intercal \bv{Q}$ is invertible and \(\normof{(\mU_k^\intercal\mQ)^{-1}}_2^2 \leq L\).
\end{definition}
The condition above is equivalent to requiring that all singular values of \(\mU_k^\intercal\mQ\) are at least \(\frac{1}{\sqrt L}\), or that all principle angles between \(\colspan(\mU_k)\) and \(\colspan(\mB)\)  have \(\cos(\theta_i) \geq \frac{1}{\sqrt L}\) \cite{Drineas:2018vu}.

Using \Cref{def:lgood}, we can immediately obtain bounds on the low-rank approximation error of a subspace $\mQ$ returned by \Cref{alg:block-krylov} when run with any $(k,L)$-good starting matrix.
We will use such bounds to analyze the single vector Krylov method of \Cref{alg:single-vec-krylov}, after proving that $\mS_k$ is $(k,L)$-good.
Consider the following bound, which depends logarithmically on $L$:

\begin{importedtheorem}[Theorem 1 of \cite{MuscoMusco:2015}]
	\label{impthm:poly-eps-low-rank}
	Let \(\mB\in\bbR^{n \times k}\) be any \((k,L)\)-good starting matrix (\Cref{def:lgood}) matrix for \mA.
	If we run Block Krylov iteration (\Cref{alg:block-krylov}) for \(q = O(\frac{1}{\sqrt{\eps}} \log(\frac{nL}{\eps}))\) iterations with starting block \mB, then the output  \(\mQ\in\bbR^{n \times k}\) satisfies
	\begin{align}
	\label{eq:norm_gen}
	\normof{\mA-\mQ\mQ^\intercal\mA}_{\xi} \leq (1+\eps) \normof{\mA-\mA_k}_{\xi}
	\end{align}
	and, letting \(\vq_i\) be the \(i^{th}\) column of \mQ,
	\begin{align}
	\label{eq:per_vec_gen}
		\left| \vq_i^\intercal \mA\mA^\intercal \vq_i -\sigma_i(\mA)^2\right | \leq \eps\sigma_{k+1}(\mA)^2.
	\end{align}
\end{importedtheorem}

\Cref{impthm:poly-eps-low-rank} is implicit in \cite{MuscoMusco:2015}, although, as stated in that work, it is specialized to when $\mB$ is a matrix with i.i.d. Gaussian entries.
In \Cref{app:musco_explanation} we discuss how the more general result stated above follows from \cite{MuscoMusco:2015}.
\Cref{impthm:poly-eps-low-rank} gives two different guarantees.
The first bounds low-rank approximation error, in both the Frobenius and spectral norms.
The second shows that the columns of \mQ can be used to estimate the top singular values of \mA. This can be called a ``Ritz value guarantee'' or a ``singular value guarantee''.

A random Gaussian matrix $\mB$ can be shown to be an $(k,O(nk))$-good starting matrix with high probability (see Lemma 4 in \cite{MuscoMusco:2015} or \cite{RudelsonVershynin:2010}). This is intuitive, since $\mB$ will span a uniformly random subspace, which has non-negligible inner product with any other fixed subspace, including the one spanned by \(\mU_k\). For a Gaussian starting block, \Cref{impthm:poly-eps-low-rank} therefore gives a bound of $O(\frac{1}{\sqrt{\eps}}\log(n/\eps))$ iterations to achieve \Cref{eq:norm_gen,eq:per_vec_gen}.

The fact that \(\mS_k\) satisfies \Cref{def:lgood} is less clear.
Our main technical contribution is to prove that it does, albeit with a much larger value of $L$ than in the Gaussian case: in \Cref{sec:submain} we show that, with probability at least $1-\delta$, $\mS_k$  is a $(k,L)$-good starting matrix for $L = O(\poly({n}/{\delta g_{\min}^{k}}))$.
Here \(g_{min} = \min_{i \in \{1,\ldots,k-1\}} \frac{\sigma_{i} - \sigma_{i+1}}{\sigma_i}\)  is the minimum gap between $\mA$'s top $k$ singular values.
Since \Cref{impthm:poly-eps-low-rank} depends logarithmically on $L$, it follows that block Krylov with starting block $\mS_k$ needs  $O\left(\frac{k}{\sqrt{\eps}}\log(\frac{1}{g_{\min}}) + \frac{1}{\sqrt{\eps}} \log(\frac{n}{\eps\delta})\right)$ iterations to achieve \Cref{eq:norm_gen,eq:per_vec_gen} -- yielding \Cref{thm:single-vec-guarantee}.

In other words, we require $k$ times as many iterations when starting with  $\mS_k$ instead of a fully random $\mB$.
However, notice that running \Cref{alg:block-krylov} for $q$ iterations with starting block $\mS_k$ only requires $q + k$ iterations of the single vector Krylov method from \Cref{alg:single-vec-krylov}, and thus $q+k$ matrix-vector products. In contrast, running the method with a random $\mB$ requires $qk$ matrix-vector products.
Ultimately, this allows us to achieve in \Cref{thm:single-vec-guarantee} a total complexity (in terms of matrix-vector products) that matches, and in some cases improves, the block Krylov method initialized with a Gaussian starting matrix, up to the dependence on $\log(1/g_{min})$.

\subsection{Main Technical Analysis}
\label{sec:submain}

Formally, we prove the following $(k,L)$-good guarantee for $\mS_k$.
In combination with \Cref{impthm:poly-eps-low-rank} and \Cref{eq:our_krylov_space}, this bound yields \Cref{thm:single-vec-guarantee}.
It will  also serve as the basis for all of the other results discussed in \Cref{sec:additional-results}.
\begin{theorem}
\label{thm:low-rank-approx}
Fix any PSD matrix \(\mA\in\bbR^{n \times n}\) with singular values $\sigma_1 \geq \ldots \geq \sigma_n$, let $\vg$ be a vector with i.i.d. mean zero Gaussian entries, and let $\mS_k = \bmat{\vg & \mA^2\vg & \mA^4\vg & \ldots & \mA^{2(k-1)}\vg}$.
For any $\delta \in (0,1)$, with probability at least $1-\delta$, $\mS_k$ is a $(k,L)$-good starting matrix for $\mA$ for $L = \frac{cnk^3 \log(\nicefrac n\delta)}{\delta^2 g_{\min}^{4k}}$.
Here $c=2.5\pi$ is a fixed constant and \(g_{min} = \min_{i \in \{1,\ldots,k-1\}} \frac{\sigma_{i} - \sigma_{i+1}}{\sigma_{i+1}}\).
\end{theorem}

To prove \Cref{thm:low-rank-approx}, we will need a simple bound on the minimum of $k$ independent Gaussian random variables:
\begin{lemma}
	\label{lem:chi-square-min-concentration}
	Let \(g_1,\ldots,g_k\sim\cN(0,1)\) and independent.
	Then, with probability at least \(\ge 1-\delta\), \(\min_i g_i^2 \geq \frac{2\delta^2}{\pi k^2}\).
\end{lemma}
\begin{proof}
We have that:
	\[
	\Pr[\min_i g_i^2 \geq t]
	= \big(1 - \Pr[{g_1}^2 \leq t]\big)^k
	= \big(1 - 2\Pr[0 \leq g_1 \leq \sqrt t]\big)^k
	\geq (1-\sqrt{1-e^{-2t/\pi}})^k.
	\]
	The last line uses the bound \(\Pr[0 \leq g_1 \leq \sqrt t] \leq \frac12 \sqrt{1-e^{-2x^2/\pi}}\) from \cite{chu1955bounds}.
	Setting the right hand side equal to \(1-\delta\) and solving for \(t\), we get:
	\begin{align*}
		t
		&= \tsfrac\pi2 \ln \tsfrac{1}{1-(1-(1-\delta)^{1/k})^2} \geq \tsfrac\pi2 (1-(1-\delta)^{1/k})^2  \geq \tsfrac\pi2 (\tsfrac\delta k)^2 = \tsfrac{\pi \delta^2}{2 k^2}.
	\end{align*}
In the first inequality we used that \(\ln\left(\frac{1}{1-x}\right) \geq x\). In the second we used that \(1-(1-x)^{1/k} \geq \tsfrac xk\).
\end{proof}

\begin{proof}[Proof of \Cref{thm:low-rank-approx}]
We first argue that \(\mU_k^\intercal\mS_k\) is invertible.
Observe that for any $\bv{x} \in \R^{k}$, \(\mS_k\vx=\hat p(\mA^2)\vg\) for some degree $k-1$ polynomial $\hat{p}$ with coefficients determined by the entries in $\bv{x}$.
We let \(\mA=\mU\mSigma\mU^\intercal\) be the SVD of \mA.
Then
\[ 
	\mU_k^\intercal \bv{S}_k \bv{x} = \mU_k^\intercal \hat p(\mA^2) \vg = \mU_k^\intercal\mU \hat p(\mSigma^2) \mU^\intercal \vg = \hat p(\mSigma_k^2)\tilde\vg,
\]
where \(\mSigma_k\) contains the top left \(k\) elements of \mSigma and \(\tilde\vg \defeq \mU_k^\intercal\vg \sim \mathcal{N}(\bv{0},\bv{I})\) by rotational invariance of the Gaussian distribution.
Note by \Cref{lem:chi-square-min-concentration} that \(\min_{i\in\{1, \ldots, k\}} \tilde{g}_i^2 \geq \frac{2\delta^2}{\pi k^2}\) with probability at least $1-\delta$.
So altogether, we can write 
\begin{align}\label{eq:denom}
\normof{\bv{U}_k^\intercal \bv{S}_k \bv{x}}_2^2 = \normof{\mU_k^\intercal\hat p(\mA^2)\tilde \vg}_2^2 = \sum_{i=1}^k (\hat p(\sigma_i^2))^2 \tilde g_i^2 \geq \frac{2\delta^2}{ \pi k^2} \sum_{i=1}^k (\hat p(\sigma_i^2))^2.
\end{align}
Since $\hat p$ has degree $k-1$, if none of the top $k$ singular values are repeated (i.e., $g_{\min} > 0$), we have that the right hand side is nonzero for any nonzero $\bv{x}$. Thus, $\bv{U}_k^\intercal \bv{S}_k$ is invertible.

Now, let \(\mQ\in\bbR^{n \times k}\) be any orthonormal basis for \(\colspan(\mS_k)\), so that \(\mS_k=\mQ\mC\) for some invertible matrix \(\mC\in\bbR^{k \times k}\) (observe that $\mS_k$ must be full rank since $\bv{U}_k^\intercal \bv{S}_k$ is invertible).
Since \(\mU_k^\intercal\mS_k\) is invertible, we then also know that \(\mU_k^\intercal\mQ\) is invertible, as required by \Cref{def:lgood}.
We also have \(\mS_k(\mU_k^\intercal\mS_k)^{-1} = \mQ\mC(\mU_k^\intercal\mQ\mC)^{-1} = \mQ(\mU_k^\intercal\mQ)^{-1}\) and therefore \(\normof{\mS_k(\mU_k^\intercal\mS_k)^{-1}}_2^2 = \normof{\bv Q(\mU_k^\intercal\mQ)^{-1}}_2^2 = \normof{(\mU_k^\intercal\mQ)^{-1}}_2^2 \).
So, to prove the theorem, it suffices to bound
\begin{align}
	\label{eq:to_prove_rearrange}
	\normof{\mS_k(\mU_k^\intercal\mS_k)^{-1}}_2^2
	= \max_{\vx} \frac{\normof{\mS_k(\mU_k^\intercal\mS_k)^{-1}\vx}_2^2}{\normof\vx_2^2}
	= \max_{\vx} \frac{\normof{\mS_k\vx}_2^2}{\normof{\mU_k^\intercal\mS_k\vx}_2^2}
	= \max_{\deg(\hat p) \leq k-1} \frac{\normof{\hat p(\mA^2)\vg}_2^2}{\normof{\mU_k^\intercal \hat p(\mA^2)\vg}_2^2}.
\end{align}
We already bounded the denominator in \eqref{eq:denom}. Thus, we turn to the numerator. Since $\bv g$ has i.i.d. mean zero, unit variance Gaussian entries we have for each $i$, $g_i^2 \leq 1+ 4\log(1/\delta)$ with probability at least $1- \delta$ by standard concentration bounds for chi-squared random variables \cite{LaurentMassart:2000}. So, by a union bound, \(\max_i g_i^2 \leq 5\log(\nicefrac n\delta)\) for $n > 2$. We thus have:
\begin{align}
	\label{eq:num_bound}
	\normof{\hat p(\mA^2)\vg}_2^2 \leq 5\log(\nicefrac n\delta) \sum_{i=1}^n (\hat p(\sigma_i^2))^2 \leq 5n \log (\nicefrac n\delta) \cdot\max_{i\in\{1, \ldots, n\}}  (\hat p(\sigma_i^2))^2.
\end{align}
Combining \eqref{eq:num_bound} and \eqref{eq:denom}, we conclude that
\begin{align}
	\label{eq:second_to_last}
	\normof{\mS_k(\mU_k^\intercal\mS_k)^{-1}}_2^2
	\leq \frac{5 \pi n k^2 \log(\tsfrac n\delta)}{2\delta^2}\cdot \max_{\deg(\hat p)\leq k-1} \frac{\max_{i\in\{1, \ldots, n\}} (\hat p(\sigma_i^2))^2}{\sum_{i=1}^k (\hat p(\sigma_i^2))^2}.
\end{align}
We now focus on bounding the maximum in \eqref{eq:second_to_last}.
Observe that if there were no gap between two of the top \(k\) singular values, then some nonzero polynomial \(\hat p\) could make the denominator zero by equaling zero on the at most \(k-1\) unique values in $\sigma_1, \ldots, \sigma_k$.
So, any bound on \eqref{eq:second_to_last} must depend on the minimum gap between singular values.
Second, note that if the maximum in the numerator is achieved for \(i \leq k\), then the overall ratio is trivially at most 1.
So, without loss of generality, we only consider \(\max_{i\in\{k+1,\ldots,n\}} (\hat p(\sigma_i^2))^2\) in the numerator.

To bound this ratio, we follow a similar broad approach to \cite{Saad:1980}, who bounds a related term by expanding \(\hat p\) as an interpolating polynomial.
Formally, we write \(\hat p\) as a Lagrange interpolating polynomial over \(\sigma_1^2,\ldots,\sigma_k^2\):
\[
	\hat p(x) = \sum_{i=1}^k \phi_i \ell_i(x)
	\hspace{0.5cm}\text{where}\hspace{0.5cm}
	\phi_i \defeq \hat p(\sigma_i^2),~~~~~
	\ell_i(x) \defeq \prod_{\substack{j\in\{1, \ldots, k\}\\j\neq i}} \frac{x-\sigma_j^2}{\sigma_i^2-\sigma_j^2},~~~~~
	i\in\{1,\ldots,k\}.
\]
For any \(0 \leq x \leq \sigma_k^2\) we have
\[
	\abs{\ell_i(x)} \leq \prod_{\substack{j\in\{1, \ldots, k\}\\j\neq i}} \frac{\sigma_j^2}{\abs{\sigma_i^2-\sigma_j^2}}  \leq \prod_{\substack{j\in\{1, \ldots, k\}\\j\neq i}} \frac{\sigma_j^2}{\abs{\sigma_i-\sigma_j}^2} \leq \frac1{g_{min}^{2k}},
\]
where the second inequality uses that $|\sigma_i^2 - \sigma_j^2| \ge |\sigma_i - \sigma_j|^2$ for all $\sigma_i,\sigma_j \ge 0$.
Next, we write \(\vecalt{\phi} = [\phi_1~\ldots~\phi_k] = [\hat p(\sigma_1)^2~\ldots~\hat p(\sigma_k^2)] \) and obtain:
\begin{align*}
	\frac{\max_{i\in\{k+1,\ldots,n\}} \abs{\hat p(\sigma_i^2)}^2}{\sum_{i=1}^k (\hat p(\sigma_i^2))^2}
	\leq \frac{\max_{x\in[0,\sigma_k^2]} \abs{\hat p(x)}^2}{\sum_{i=1}^k (\hat p(\sigma_i^2))^2} 
	&= \frac{\max_{x\in[0,\sigma_k^2]} \abs{\sum_{i=1}^k \phi_i \ell_i(x)}^2}{\sum_{i=1}^k (\hat p(\sigma_i^2))^2} \\
	&\leq \frac{\max_{x\in[0,\sigma_k^2]} (\normof{\vecalt\phi}_1 \max_i \abs{\ell_i(x)})^2}{\normof{\vecalt\phi}_2^2} \\
	&= \frac{\normof{\vecalt\phi}_1^2}{\normof{\vecalt\phi}_2^2} ~\max_{x\in[0,\sigma_k^2], i\in\{1, \ldots, k\}} \abs{\ell_i(x)}^2 \\
	&\leq \frac{k}{g_{min}^{4k}}.
\end{align*}
Finally, we plug back into \eqref{eq:second_to_last} to conclude that $\normof{(\mU_k^\intercal\mQ)^{-1}}_2^2 = \normof{\mS_k(\mU_k^\intercal\mS_k)^{-1}}_2^2
	\leq \frac{5 \pi n k^3}{2 g_{min}^{4k} \delta^2}\log(\tsfrac n\delta)$, which completes the proof.
\end{proof}

\subsection{Proof of \texorpdfstring{\Cref{thm:single-vec-guarantee}}{Theorem \ref*{thm:single-vec-guarantee}}}

As mentioned, \Cref{thm:single-vec-guarantee} follows directly by combining \Cref{impthm:poly-eps-low-rank} and  \Cref{thm:low-rank-approx}.
In particular, since $\mS_k$ is an $(k,L)$-good starting matrix for $L \leq \left(\frac{n}{\delta g_{\min}^k}\right)^c$ for a constant $c$, if we run the block Krylov method initialized with $\mS_k$, then with probability at least $1-\delta$, we obtain $\mQ$ achieving the guarantees of \Cref{thm:single-vec-guarantee} after
\begin{align*}
	q
	= O\left(\frac{1}{\sqrt{\eps}} \log\left(\frac{nL}{\eps}\right)\right)
	= O\left(\frac{k}{\sqrt{\eps}}\log\left(\frac{1}{g_{\min}}\right) + \frac{1}{\sqrt{\eps}} \log\left(\frac{n}{\eps\delta}\right)\right) \text{ iterations.}
\end{align*}
Moreover, by \Cref{eq:our_krylov_space}, the $\mQ$ returned by \Cref{alg:block-krylov} with $\mS_k$ as the starting block, is exactly the same as the $\mQ$ returned by the single vector Krylov method (\Cref{alg:single-vec-krylov}) after $q + k$ iterations.
Overall, we get the following full version of \Cref{thm:single-vec-guarantee}, which includes both the low-rank approximation and singular value guarantees from \Cref{impthm:poly-eps-low-rank}:

\begin{reptheorem}{thm:single-vec-guarantee}
For \(\mA \in \bbR^{n \times d}\), let \(g_{min} = \min_{i \in \{1,\ldots,k-1\}} \frac{\sigma_{i} - \sigma_{i+1}}{\sigma_i}\).
For any \(\eps, \delta \in (0,1)\), \Cref{alg:single-vec-krylov} initialized with \(\vx\sim\cN(\vec0,\mI)\) and run for  \(t = O(\tsfrac{k}{\sqrt \eps} \log(\tsfrac1{g_{min}}) + \tsfrac1{\sqrt\eps} \log(\tsfrac{n}{\eps\delta}))\) iterations returns an orthogonal $\mQ \in \R^{n \times k}$ such that, with probability at least \(1-\delta\),
\begin{align*}
	\normof{\mA-\mQ\mQ^\intercal\mA}_{\xi} &\leq (1+\eps) \normof{\mA-\mA_k}_{\xi} & &\text{and} & \left| \vq_i^\intercal \mA\mA^\intercal \vq_i -\sigma_i(\mA)^2 \right| &\leq \eps\sigma_{k+1}(\mA)^2.
\end{align*}
\end{reptheorem}

\section{Additional Applications of Main Result}
\label{sec:applications}

In this section, we leverage our analysis of the starting block $\bv{S}_k$ to give several results beyond \Cref{thm:single-vec-guarantee}.
In \Cref{sec:spectral-decay}, we start by generalizing from using \(\mS_k\) to using \(\mS_\ell\), which has \(\ell \geq k\) columns, and lets us obtain faster rates of convergence when the spectrum of \mA decays between singular values \(k\) and \(\ell+1\).
Next, in \Cref{sec:fast-frob}, we use the results of \Cref{thm:single-vec-guarantee} and \Cref{sec:spectral-decay} to get a faster rate of convergence in the Frobenius norm, simplifying the algorithm of \cite{bakshi2022low}.
In \Cref{sec:fast-schatten}, we generalize from the Frobenius norm to Schatten $p$-norms, also simplifying \cite{bakshi2022low}.
Lastly, in \Cref{sec:beyond-single-vec}, we generalize from single vector Krylov with block size \(b=1\) to small-block Krylov with any block size \(b < k\).

\subsection{Faster Convergence with Spectral Decay}
\label{sec:spectral-decay}

As discussed in \Cref{sec:intro}, in addition to spectrum-independent bounds, we know that large block Krylov methods achieve very fast convergence when using block size \(\ell \geq k\) if there is a sufficiently large gap between \(\sigma_k\) and \(\sigma_{\ell+1}\).
Formally, \cite{MuscoMusco:2015} show the following:
\begin{importedtheorem}[Theorem 13 of \cite{MuscoMusco:2015}]
	\label{impthm:spectral-decay-low-rank}
	Let \(\mB\in\bbR^{n \times \ell}\) be any \((\ell,L)\)-good starting matrix (\Cref{def:lgood}) matrix for \mA, for some \(\ell \geq k\).
	If we run Block Krylov iteration (\Cref{alg:block-krylov}) for \(q = O(\frac{1}{\sqrt{g_{k \rightarrow \ell}}} \log(\frac{nL}{\eps}))\) iterations with starting block \mB, where \(g_{k \rightarrow \ell} = \frac{\sigma_{k}-\sigma_{\ell+1}}{\sigma_{k}}\), then the output  \(\mQ\in\bbR^{n \times k}\) satisfies
\begin{align*}
		\normof{\mA-\mQ\mQ^\intercal\mA}_{\xi} &\leq (1+\eps) \normof{\mA-\mA_k}_{\xi} & &\text{and} & \left| \vq_i^\intercal \mA\mA^\intercal \vq_i -\sigma_i(\mA)^2 \right| &\leq \eps\sigma_{k+1}(\mA)^2.
\end{align*}
\end{importedtheorem}

Our second main result shows that the convergence rate of \Cref{impthm:spectral-decay-low-rank} applies to single vector Krylov as well. In particular, we can simply apply the same idea as was used to prove \Cref{thm:single-vec-guarantee}, but with ``simulated block size'' \(\ell \geq k\). Letting
\[
	\mS_\ell \defeq \bmat{\vx & \mA^2\vx & \mA^4\vx & \ldots & \mA^{2(\ell-1)}\vx}
	\hspace{0.5cm}\text{and}\hspace{0.5cm}
	\colspan(\mK) = \colspan\left(\bmat{\mS_\ell & \mA^2\mS_\ell  & \ldots & \mA^{2q}\mS_\ell}\right),
\]
we observe that single vector Krylov run for $t$ iterations exactly computes $\colspan(\mK)$ for $q = t-\ell+1$.
This lets us show that single vector Krylov run for \(t\) iterations essentially matches the convergence rate of block Krylov run for \(\approx t/\ell\) iterations:
\begin{theorem}[Spectral Decay Convergence]
\label{thm:decay-dependence}
For \(\mA \in\bbR^{n \times d}\) and \(\ell \geq k\), let \(g_{min} = \min_{i\in\{1,\ldots,\ell-1\}} \frac{\sigma_i - \sigma_{i+1}}{\sigma_{i+1}}\) and \(g_{k \rightarrow \ell} = \frac{\sigma_k - \sigma_{\ell+1}}{\sigma_{k}}\).
For any \(\eps,\delta\in(0,1)\), \Cref{alg:single-vec-krylov} initialized with \(\vx\sim\cN(\vec0,\mI)\) and run for \(t = O(\frac{\ell}{\sqrt{g_{k\rightarrow\ell}}} \log(\frac1{g_{\min}}) + \frac{1}{\sqrt{g_{k\rightarrow\ell}}}\log(\frac{n}{\delta\eps}))\) iterations returns an orthogonal \(\mQ\in\bbR^{n \times k}\) such that, with probability at least \(1-\delta\),
\begin{align*}
	\normof{\mA-\mQ\mQ^\intercal\mA}_{\xi} &\leq (1+\eps) \normof{\mA-\mA_k}_{\xi} & &\text{and} & \left| \vq_i^\intercal \mA\mA^\intercal \vq_i -\sigma_i(\mA)^2 \right| &\leq \eps\sigma_{k+1}(\mA)^2.
\end{align*}
\end{theorem}
\begin{proof}
By \Cref{thm:low-rank-approx}, we know that \(\mS_\ell\) is an \((\ell,L)\)-good starting matrix for block Krylov iteration on \mA, where \(L \leq (\frac{n}{\delta g_{min}^\ell})^c\) for a constant \(c\).
So, by \Cref{impthm:spectral-decay-low-rank}, we find that block Krylov iteration with starting block \(\mS_\ell\) converges in 
\[
	q
	= O\left(\frac{1}{\sqrt{g_{k\rightarrow \ell}}} \log\left(\frac{nL}{\eps}\right)\right)
	= O\left(\frac{\ell}{\sqrt{g_{k\rightarrow \ell}}} \log\left(\frac1{g_{min}}\right) + \frac{1}{\sqrt{g_{k\rightarrow\ell}}} \log\left(\frac{n}{\delta\eps}\right)\right)
\]
iterations.
Moreover, by \Cref{eq:our_krylov_space}, the \mQ returned by \Cref{alg:block-krylov} with \(\mS_\ell\) as the starting block is exactly the same as the \mQ returned by the single vector Krylov method (\Cref{alg:single-vec-krylov}) after \(q + \ell\) iterations, which completes the proof.
\end{proof}

\noindent\textbf{Comparison to Prior Bounds.}
In terms of the number of matrix-vector products computed, \Cref{thm:decay-dependence} can significantly improve upon the prior work for large \(k\).
\cite{MuscoMusco:2015} require using a block size \(\ell \geq k\) for \(t = O(\frac{1}{\sqrt{g_{k\rightarrow \ell}}} \log(\frac{n}{\delta\eps}))\) iterations, or equivalently for \(O(\frac{1}{\sqrt{g_{k\rightarrow \ell}}} \cdot \ell \log(\frac{n}{\delta\eps}))\) matrix-vector products.
Assuming that \(g_{min}\) is a constant, our guarantee for single vector Krylov methods only requires \(O(\frac{1}{\sqrt{g_{k\rightarrow \ell}}}(\ell + \log(\frac{n}{\eps\delta})))\) matrix-vector products.
That is, we reduce the product between \(\ell\) and \(\log(\frac{n}{\delta\eps})\) into a sum, suggesting a nearly \(\ell\)-fold speedup when we want high precision results.
Further, we obtain the above guarantee for any \(\ell \geq k\), meaning that single vector Krylov automatically competes with the bound for the \emph{best possible choice} of block size, without knowing it in advance. We observe in \Cref{sec:experiment-grid} that the above  theoretical advantages translate into practice -- for matrices with decaying singular value spectra, we find that single vector Krylov methods substantially outperform large block methods.

\medskip

\noindent\textbf{Comparison to Lower Bounds.}
It is worth comparing \Cref{thm:decay-dependence} to the lower bound of \cite{simchowitz2018tight}, who show that finding an orthogonal matrix \(\mQ\in\bbR^{n \times k}\) such that \(\sum_{i=1}^k \vq_i^\intercal\mA\vq_i \geq (1-\eps) \sum_{i=1}^k \sigma_i(\mA)\) requires at least \(\Omega(\frac{k \log n}{\sqrt{g_{k\rightarrow k}}})\) matrix-vector products when \(\eps = g_{k \rightarrow k} = \frac{\sigma_k - \sigma_{k+1}}{\sigma_{k}}\). In comparison, 
applying \Cref{thm:decay-dependence} with $\ell = k$ yields an upper bound of roughly $O(\frac{k \log (1/g_{min})}{\sqrt{g_{k\rightarrow k}}} + \frac{\log n}{\sqrt{g_{k\rightarrow k}}})$, which only does not violate the lower bound of  \cite{simchowitz2018tight} since \(\frac{1}{g_{min}} = \text{poly}(n)\) for their input. I.e., their input suffers from very small singular value gaps. 

When \(k=1\), the  intuition for the \(\log n\) dependence is that since a random start vector has \(\frac{1}{\text{poly}(n)}\) inner product with the top singular vector, it requires \(\log n\) iterations to converge.
The matching upper and lower bounds of \(\Theta(\frac{k \log n}{\sqrt{g_{k\rightarrow k}}})\) from \cite{MuscoMusco:2015} and \cite{simchowitz2018tight} suggest that the cost of rank-\(k\) approximation is simply \(k\) times the cost of rank-1 approximation, i.e., roughly \(k \log n\).
In fact, the situation is more nuanced.
Our work suggests that, unless gaps between singular values are very small, the cost of rank-\(k\) approximation is much actually cheaper.

\subsection{Improved Results for Frobenius Norm Low-Rank Approximation}
\label{sec:fast-frob}

\Cref{thm:decay-dependence} shows that single vector Krylov methods achieve strong spectrum-adaptive guarantees, converging at essentially the same rate as the best choice of block size, if not faster.
As a concrete application of this observation, we show that single vector Krylov methods automatically match a recent result of \cite{bakshi2022low} on Frobenius norm low-rank approximation.
\cite{bakshi2022low} propose an algorithm that combines the results of running block Krylov iteration twice, once with block size \(k\) for \(\tilde O(\frac{1}{\eps^{1/3}})\) iterations and once with block size \(O(\frac{k}{\eps^{1/3}})\) for \(\tilde O(1)\) iterations.
Their algorithm obtains a \((1+\eps)\) optimal low-rank approximation in the Frobenius norm using \(\tilde O(\frac{k}{\eps^{1/3}})\) matrix-vector products instead of \(\tilde O(\frac{k}{\sqrt\eps})\), as required by \Cref{thm:single-vec-guarantee}.
Since \Cref{thm:single-vec-guarantee} and \Cref{thm:decay-dependence} show that single vector Krylov methods match the convergence rates of both block size \(k\) and block size \(O(\frac{k}{\eps^{1/3}})\), they therefore match the result of \cite{bakshi2022low}. Formally we have:
\begin{theorem}
\label{thm:fast-frob}
For \(\mA \in\bbR^{n \times d}\), let \(g_{min} = \min_{i\in\{1,\ldots,\ell-1\}} \frac{\sigma_i - \sigma_{i+1}}{\sigma_{i+1}}\) where \(\ell = \Theta(\frac{k}{\eps^{1/3}})\).
For any \(\eps,\delta\in(0,1)\), \Cref{alg:single-vec-krylov} initialized with \(\vx\sim\cN(\vec0,\mI)\) and run for \(t = O(\frac{k}{\eps^{1/3}} \log(\frac1{g_{min}}) + \frac{1}{\eps^{1/3}} \log(\frac{n}{\delta \eps}))\) iterations returns an orthogonal \(\mQ\in\bbR^{d \times k}\) such that, with probability at least \(1-\delta\), we have
\[
	\normof{\mA-\mQ\mQ^\intercal\mA}_F \leq (1+\eps) \normof{\mA-\mQ\mQ^\intercal\mA}_F.
\]
\end{theorem}
We formally prove \Cref{thm:fast-frob} in \Cref{app:fast-frob}, by generalizing and formalizing an analysis  stated in the introduction of \cite{bakshi2022low}.
We note that, in concurrent work, \cite{Bakshi:2023uu} show a similar result which has no $g_{min}$ dependence and further removes the $\log n$ dependence.
However, their bound only applies to the special case of rank-$1$ approximation.

\subsection{Improved Rates for Schatten Norm Low-Rank Approximation}
\label{sec:fast-schatten}

The work of \cite{bakshi2022low} also gives bounds for low-rank approximation in general Schatten \(p\)-Norms for any \(p\geq1\).
They show that by running Krylov iteration 4 times, on both \mA and \(\mA^\intercal\) and with both a relatively small block size \(\ell=k\) and a relatively large block size \(\ell=\tilde O(\frac{k}{\eps^{1/3}})\), they can recover a low-rank approximation to \mA in the Schatten \(p\)-norm using \(\tilde O(\frac{kp^{1/6}}{\eps^{1/3}})\) matrix-vector multiplications.
As in the case of Frobenius norm low-rank approximation, we show that we can match this entire process with a single instantiation of single vector Krylov, yielding:

\begin{theorem}[Schatten-p Norm Low-Rank Approximation]
\label{thm:schatten_p}
For \(\mA \in\bbR^{n \times d}\) and \(p \geq 1\), let \(g_{min} = \min_{i\in\{1,\ldots,\ell-1\}} \frac{\sigma_i - \sigma_{i+1}}{\sigma_{i+1}}\) where \(\ell = \Theta(\frac{k}{\eps^{1/3}p^{1/3}})\).
For any \(\eps,\delta\in(0,1)\), let \(\mQ\in\bbR^{d \times k}\) be the result of running \Cref{alg:single-vec-krylov} on \(\mA^\intercal\) initialized with \(\vx\sim\cN(\vec0,\mI)\) and run for
\[
	t
	= O \left (\tsfrac{kp^{1/6}}{\eps^{1/3}} \log(\tsfrac{1}{g_{min}}) + (\sqrt{p}+\tsfrac{p^{1/6}}{\eps^{1/3}}) \log(\tsfrac{np}{\delta\eps}) \right)
	= \tilde O \left (\tsfrac{kp^{1/6}}{\eps^{1/3}} + \sqrt p \right)
\]
iterations.
Let \(\mZ\in\bbR^{n \times k}\) be an orthonormal basis for \(\mA\mQ\).
Then, with probability at least \(1-\delta\),
\[
	\normof{\mA-\mZ\mZ^\intercal\mA}_p \leq (1+\eps)\normof{\mA-\mA_k}_p,
\]
where \(\normof\mA_p \defeq (\sum_{i=1}^n\sigma_i(\mA)^p)^{1/p}\) is the Schatten \(p\)-norm.
\end{theorem}
We prove \Cref{thm:schatten_p} in \Cref{app:fast-schatten}.
Unlike our previous results, \Cref{thm:schatten_p} first uses single vector Krylov to compute an orthonormal basis \mQ, but then outputs \mZ, which is an orthonormal basis for \(\mA\mQ\).
We suspect that this two-step process is an artifact of the analysis in \cite{bakshi2022low}, and that simply returning the output of a single vector Krylov method suffices.

Note that since the same single vector Krylov method obtains the result of \Cref{thm:schatten_p} for any choice of $p$, and setting \(p = O(\frac{\log n}{\eps})\) closely approximates the Schatten-\(\infty\) norm (i.e., the spectral norm), we achieve the best known result for outputting a low-rank approximation \emph{simultaneously} in all Schatten \(p\)-norms:

\begin{corollary}
\label{corol:simul-schatten}
For \(\mA \in\bbR^{n \times d}\), let \(g_{min} = \min_{i\in\{1,\ldots,\ell-1\}} \frac{\sigma_i - \sigma_{i+1}}{\sigma_{i+1}}\) where \(\ell = \Theta(\frac{k}{\log^{1/3}(n)})\).
For any \(\eps,\delta\in(0,1)\), let \(\mQ\in\bbR^{d \times k}\) be the result of running \Cref{alg:single-vec-krylov} on \(\mA^\intercal\) initialized with \(\vx\sim\cN(\vec0,\mI)\) and run for
\[
	t
	= O \left( \tsfrac{k \log^{\frac{1}{6}\hspace{-0.05cm}}(n)}{\sqrt\eps} \log(\tsfrac{1}{g_{min}}) + \tsfrac{\sqrt{\log (n)}}{\sqrt{\eps}} \log(\tsfrac{n}{\delta\eps}) \right)
	= \tilde O \left( \tsfrac{k}{\sqrt\eps} \right)
\]
iterations.
Let \(\mZ\in\bbR^{n \times k}\) be an orthonormal basis for \(\mA\mQ\).
Then, for  \(\eps_p \defeq \eps \cdot \min\{1, \frac{\sqrt{p\eps}}{\log(n)}\} \leq \eps\), with probability at least \(1-\delta\), simultaneously for all \(p\geq1\),
\[
	\normof{\mA-\mQ\mQ^\intercal\mA}_p \leq (1+\eps_p)\normof{\mA-\mA_k}_p.
\]

\end{corollary}
A similar but weaker guarantee is available from \cite{bakshi2022low}.
They show that running four Block Krylov methods with block size choices depending on \(p\) suffices to obtain a low-rank approximation in the Schatten \(q\)-norm for all \(q \leq p\).
If we let \(p = O(\frac{\log n}{\eps})\), so that \(\normof{\mA}_p\) approximates the spectral norm, 
then, their result shows that \(\tilde O(\frac{k}{\sqrt\eps})\) matrix-vector products suffice to output a \((1+\eps)\) relative error low-rank approximation in all Schatten norms, including the Frobenius norm.
In contrast, \Cref{thm:fast-frob} shows that running single vector Krylov for \(\tilde O(\frac{k}{\sqrt\eps})\) iterations actually gives a better relative error of \((1 + \eps^{3/2})\) in the Frobenius norm.

\subsection{General Bounds for Small-Block Methods}
\label{sec:beyond-single-vec}

Our approach to analyzing  single vector Krylov also extends to `small-block' Krylov methods, which use block size \(1 < b < k\).
Such methods are common in practice as they help avoid slow convergence due to nearby singular values:
intuitively, we expect a block size of \(b\) to be effective even when the input \mA has clusters of at most \(b\) very close singular values.
Additionally, parallelism often lets us compute multiple matrix-vector products with \mA just as quickly as computing a single matrix-vector product, with incentives the use of a block size $> 1$.

We outline results for small block methods in this section, but defer proofs to \Cref{app:small-block-analysis}.
To start, we generalize the notion of a gap between singular values:

\begin{definition}
\label{def:b-order-gap}
Fix block size \(b \in [k]\).
For each \(i\in[k]\), we let \(\cN_i \subset [k] \setminus \{i\}\) be the indices of the \(b-1\) singular values other than \(i\) that minimize \(\abs{\frac{\sigma_i-\sigma_j}{\sigma_j}}\).
Then let \(g_{min,b}\) be the \emph{\(b^{th}\)-order gap} of \mA:
\[
	g_{min,b}
	= \min_{i\in[k]} \min_{j\in[k] \setminus \cN_i, j \neq i} \left|\frac{\sigma_i - \sigma_j}{\sigma_j}\right|.
\]
If \(b=k\), then the sets \(\cN_i\) have no terms, and we define \(g_{min,b}\defeq1\).
\end{definition}
If \(b=1\), then we see that the sets \(\cN_i\) are empty, and we recover \(g_{min,b}=g_{min}\) as defined in \Cref{thm:single-vec-guarantee}.
If \(b=k\), then \(g_{min,b}\) is just \(1\), which matches the fact that block size \(k\) does not require a gap dependence \cite{MuscoMusco:2015}.
If e.g., \(b=2\) and \(\mA\) has two identical singular values, then we may still have \(g_{min,b}>0\), and so  an algorithm can depend on \(\frac1{g_{min,b}}\) without risking total failure.
With this characterization of small-block gaps, we prove the following generalization of \Cref{thm:low-rank-approx}  in \Cref{app:small-block-analysis}:
\begin{theorem}
\label{thm:block-low-rank-approx}
Fix any PSD  \(\mA\in\bbR^{n \times n}\) and block size $b \in \{1,\ldots,k\}$.
Let $\mG\in\bbR^{n \times b}$ be a matrix with i.i.d. \(\cN(0,1)\) entries, and let $\mS_r = \bmat{\mG & \mA^2\mG & \mA^4\mG & \ldots & \mA^{2(r-1)}\mG}$, where \(r = k-b+1\).
For any $\delta \in (0,1)$, with probability at least $(1-\delta)$, there exists a matrix \(\mQ\in\bbR^{n \times k}\) that lies in the span of \(\mS_r\) and is a $(k,L)$-good starting matrix for $\mA$ for $L = \frac{cb^2k^2n \log(\nicefrac 1\delta)}{\delta^2 g_{min,b}^{4(k-b)}}$.
Here, \(c\) is a fixed constant.
\end{theorem}
Above, the starting block \(\mS_r\) has \(rb \approx b(k-b)\) columns, which can be larger than \(k\). \Cref{thm:block-low-rank-approx} shows that
there exists a matrix $\mQ \in \R^{n \times k}$ in the span of \(\mS_r\) which satisfies the \((k,L)\)-good property of \Cref{def:lgood}.
Since the Krylov subspace generated by \mQ lies within the Krylov subspace generated by \(\mS_r\), we can thus use \Cref{impthm:poly-eps-low-rank} and \Cref{impthm:spectral-decay-low-rank} to obtain guarantees for the subspace generated by \(\mS_r\). For a formal argument, see \Cref{app:small-block-analysis}. Overall, by plugging in the value of \(L=\text{poly}(\frac{n}{\delta g_{min,b}^{k-b}})\) into these theorems, we achieve similar guarantees as for single vector Krylov, but with a dependence on $g_{min,b}$.
For instance, we generalize \Cref{thm:single-vec-guarantee}, obtaining:
\begin{theorem}
\label{thm:block-vec-guarantee}
For \(\mA \in\bbR^{n \times d}\) and \(b \leq k\), let \(g_{min,b}\) as in \Cref{def:b-order-gap}.
For any \(\eps,\delta\in(0,1)\), \Cref{alg:block-krylov} initialized with i.i.d. \(\cN(0,1)\) matrix \(\mG\in\bbR^{n \times b}\) and run for \(t = O(\frac{k-b}{\sqrt{\eps}} \log(\frac1{g_{min,b}}) + \frac{1}{\sqrt{\eps}}\log(\frac{n}{\delta\eps}))\) iterations returns an orthogonal \(\mQ\in\bbR^{n \times k}\) such that, with probability at least \(1-\delta\),
\begin{align*}
	\normof{\mA-\mQ\mQ^\intercal\mA}_{\xi} &\leq (1+\eps) \normof{\mA-\mA_k}_{\xi} & &\text{and} & \left| \vq_i^\intercal \mA\mA^\intercal \vq_i -\sigma_i(\mA)^2 \right| &\leq \eps\sigma_{k+1}(\mA)^2.
\end{align*}
In particular, this requires \(O(\frac{b(k-b)}{\sqrt{\eps}} \log(\frac1{g_{min,b}}) + \frac{b}{\sqrt{\eps}}\log(\frac{n}{\delta\eps}))\) matrix-vector products with \mA.
\end{theorem}
For constant \(b = O(1)\), the above theorem recovers the same asymptotic matrix-vector complexity as \Cref{thm:single-vec-guarantee} but with an improved dependence on singular value gaps.
For \(k-b = O(1)\), it nearly matches the matrix-vector complexity of block Krylov with block size $k$, but with a very mild gap dependence.
When \(b = \frac{k}{2}\), we get the worst of both worlds, needing \(m = \tilde O(\frac{k^2}{\sqrt\eps})\) matrix-vector products.
This may be a limitation of our proof techniques: it could be possible to show that the matrix-vector complexity scales linearly in \(k\) for any block size.

Further, observe that by using \Cref{thm:block-low-rank-approx}, the spectrum-adaptive results of \Cref{sec:spectral-decay}, the fast Frobenius results of \cref{sec:fast-frob}, and the fast Schatten norm results of \Cref{sec:fast-schatten} also apply in the general block size $b \le k$ case.
Wherever those theorems have \(g_{min}\), we replace this with \(g_{min,b}\), where \(g_{min,b}\) is defined now across the top \(\ell-b\) singular values instead of the top \(k-b\).


\section{Random Perturbation Analysis}
\label{sec:perturb}

\Cref{thm:single-vec-guarantee,thm:decay-dependence,thm:schatten_p,thm:block-vec-guarantee} all show that single vector or small-block Krylov methods match or improve the existing bounds for large block methods, up to a factor of \(\log(\frac1{g_{min}})\).
Single vector methods cannot avoid this dependence in general: if one of \mA's top singular values is repeated, so that \(g_{min} = 0\), then the Krylov subspace can be rank deficient and fail to converge to the top subspace of \mA.
To address this possible point of failure, we take a smoothed-analysis approach, showing that if our input instance is subjected to a small random perturbation, the dependence on $g_{min}$ can be removed. The key tool we leverage is a line of work from random matrix theory called \emph{eigenvalue repulsion}, which shows that small eigenvalue gaps are  brittle: by adding a tiny amount of random noise to any matrix, we can ensure that its eigenvalues are well separated \cite{minami1996local,nguyen2017random,beenakker1997random}. 

\subsection{Gap-Independent Bounds for PSD Matrices}

Our first result shows that we can remove the dependence on $g_{min}$ when our input matrix $\bv A$ is PSD.
In \Cref{sec:asymetric-gap-free}, we  show that this result gives some bounds for non-PSD matrices as well.
Our result leverages the following eigenvalue repulsion bound, which we derive in \Cref{app:symmetric-repulse} using a result of \cite{minami1996local}.
\begin{lemma}
\label{lem:diagonal-perturb-gaps}
Fix symmetric matrix \(\mA\in\bbR^{n \times n}\), \(\delta \in (0,1)\), and \(\Delta \leq \normof{\mA}_2\).
Let \(\mD \in \bbR^{n \times n}\) be a diagonal matrix whose entries are uniformly distributed in \([-\Delta,+\Delta]\).
Then, letting \(\tilde\mA = \mA + \mD\) and letting $C$ denote some universal constant, with probability at least \(1-\delta\),
\[
	\min_{i\in[n-1]} \frac{\abs{\lambda_i(\tilde\mA) - \lambda_{i+1}(\tilde\mA)}}{\abs{\lambda_{i+1}(\tilde\mA)}} \geq \frac{\delta}{Cn^2} \cdot \frac{\Delta^2}{\normof{\mA}_2^2}.
\]
\end{lemma}
That is, by adding a small amount of noise to the diagonal of \mA, we can ensure its eigenvalue gaps are polynomially large in the problem parameters. While \Cref{lem:diagonal-perturb-gaps} holds for general symmetric matrices, we will apply it specifically to PSD  \mA in our analysis, since for indefinite matrices, having large {eigenvalue gaps} does not necessarily imply having  \emph{singular value gaps}, which are required for our convergence results for single vector Krylov methods. It would be interesting to prove an analogous result to \Cref{lem:diagonal-perturb-gaps} that applies directly to singular values and use it to generalize our  results beyond PSD matrices.

Naturally, the larger the perturbation parameter $\Delta$ in \Cref{lem:diagonal-perturb-gaps}, the larger the resulting gaps. Thus, $\Delta$  gives  a tradeoff between runtime and accuracy: higher $\Delta$ leads to larger gaps and thus faster convergence. However, it will also make the result of our algorithms less accurate.
In \Cref{app:perturbed-correctness} we show that picking \(\Delta\) such that \(\normof{\mD}_2 \leq \frac{\eps\sigma_{k+1}(\mA)}{n}\) suffices to guarantee that a near-optimal low-rank approximation of $\bv{\tilde A}$ is also near optimal for $\bv{A}$:
\begin{lemma}
\label{lem:perturbed-correctness}
Let \(\tilde\mA = \mA+\mD\) where \(\norm{\mD}_2 \leq \frac{\eps}{3n}\sigma_{k+1}(\mA)\) and \(\eps\in(0,1)\).
Fix any \(\mQ\in\bbR^{n \times k}\) with orthonormal columns \(\vq_1,\ldots,\vq_k\).
Then, with probability at least \(1-\delta\),
\begin{enumerate}
	\item If \(\abs{\vq_i^\intercal\tilde\mA\tilde\mA^\intercal\vq_i - \sigma_i(\tilde\mA)^2}\leq\eps\sigma_{k+1}(\tilde\mA)^2\), then \(\abs{\vq_i^\intercal\mA\mA^\intercal\vq_i - \sigma_i(\mA)^2}\leq 8\eps\sigma_{i}(\mA)^2.\)
	\item If \(\normof{\tilde\mA-\mQ\mQ^\intercal\tilde\mA}_2 \leq (1+\eps)\normof{\tilde\mA-\tilde\mA_k}_2\), then \(\normof{\mA-\mQ\mQ^\intercal\mA}_2 \leq (1+2\eps)\normof{\mA-\mA_k}_2.\)
	\item If \(\normof{\tilde\mA-\mQ\mQ^\intercal\tilde\mA}_F \leq (1+\eps)\normof{\tilde\mA-\tilde\mA_k}_F\), then \(\normof{\mA-\mQ\mQ^\intercal\mA}_F \leq (1+4\eps)\normof{\mA-\mA_k}_F.\)
\end{enumerate}
\end{lemma}
In particular, since we pick \(\mD\) to be diagonal, we have \(\normof{\mD}_2 \leq \Delta \leq \frac{\eps\sigma_{k+1}(\mA)}{3n}\).
\Cref{lem:diagonal-perturb-gaps,lem:perturbed-correctness} then together imply the following gap-independent variant of \Cref{thm:single-vec-guarantee} for PSD matrices.
\begin{corollary}[Gap-Independent Convergence]
\label{corol:perturbed-svk-symmetric}
For PSD \(\mA \in\bbR^{n \times n}\), let \(\kappa_k = \frac{\sigma_1}{\sigma_k}\) and \(\Delta = \frac{\eps\sigma_{k+1}}{3n}\).
For any \(\eps,\delta\in(0,\frac12)\), let \(\tilde\mA = \mA+\mD\) where \(\mD\in\bbR^{n \times n}\) is a diagonal matrix with entries drawn uniformly and i.i.d. from \([-\Delta,\Delta]\).
Then, \Cref{alg:single-vec-krylov} run on \(\tilde\mA\) initialized with \(\vx\sim\cN(\vec0,\mI)\) and run for \(t = O(\frac{k}{\sqrt\eps} \log(\frac{n\kappa_k}{\delta\eps}))\) iterations returns orthogonal \(\mQ\in\bbR^{n \times k}\) such that, with probability at least \(1-\delta\),
\begin{align*}
	\normof{\mA-\mQ\mQ^\intercal\mA}_{\xi} &\leq (1+\eps) \normof{\mA-\mA_k}_{\xi} & &\text{and} & \left| \vq_i^\intercal \mA\mA^\intercal \vq_i -\sigma_i(\mA)^2 \right| &\leq \eps\sigma_{k+1}(\mA)^2.
\end{align*}
\end{corollary}
\begin{proof}
Proving this result simply requires showing that \Cref{lem:diagonal-perturb-gaps} implies gaps between \mA's singular values. This is not immediate since, even if we assume \mA is PSD, $\tilde{\mA}$ might not be, so could have negative eigenvalues. An eigenvalue at \(\eta\) and another eigenvalue at \(-\eta\), which would give a singular value gap of 0, which we need to rule out.
To do so, note that \(\normof{\mD}_2 = \max_i \abs{\mD_{i,i}} \leq \Delta < \eps\sigma_{k+1}(\mA)\).
So by assuming that \mA is PSD, we know that for any negative eigenvalue \(\lambda_i(\tilde\mA)\) of $\tilde{\mA}$,
\[
	\abs{\lambda_i(\tilde\mA)}
	\leq \abs{\lambda_i(\tilde\mA) - \lambda_i(\mA)}
	\leq \normof{\mD}_2
	< \eps\sigma_{k+1}(\mA)
	\leq \sigma_{k+1}(\tilde\mA).
\]
The last inequality assumes \(\eps \leq \frac12\) to say that \(\sigma_{k+1}(\tilde\mA) \geq \sigma_{k+1}(\mA) - \normof{\mD}_2 \geq (1-\eps)\sigma_{k+1}(\mA) \geq \eps\sigma_{k+1}(\mA)\).
So, we know that if \(\lambda_i(\tilde\mA) < 0\) then any singular value associated with \(\lambda_i(\tilde\mA)\) is not one of the top \(k\) singular values of \(\tilde\mA\).
So, the top \(k\) singular values of \(\tilde\mA\) must all be associated with nonnegative eigenvalues.
That is \(\sigma_i(\tilde\mA)=\lambda_i(\tilde\mA)\) for \(i\in[k]\).
And so, we find that
\[
	g_{min}
	= \min_{i\in\{1,\ldots,k-1\}} \frac{\abs{\sigma_i(\tilde\mA) - \sigma_{i+1}(\tilde\mA)}}{\sigma_{i+1}(\tilde\mA)}
	= \min_{i\in\{1,\ldots,k-1\}} \frac{\abs{\lambda_i(\tilde\mA) - \lambda_{i+1}(\tilde\mA)}}{\abs{\lambda_{i+1}(\tilde\mA)}}
	\geq \frac{\delta}{Cn^2} \cdot \frac{\Delta^2}{\normof{\mA}_2^2}.
\]
To complete the theorem, we then plug in \(\Delta = \frac{\eps\sigma_{k+1}}{3n}\), getting
\[
	g_{min}
	\geq \frac{\delta}{Cn^2} \cdot \frac{\frac{\eps^2\sigma_{k+1}^3}{9n^2}}{\sigma_1^2}
	= \frac{\delta\eps^2}{9Cn^4\kappa_k^2}.
\]
We then appeal to \Cref{thm:single-vec-guarantee} to get the final iteration complexity of 
\(t = O(\tsfrac{k}{\sqrt \eps} \log(\frac1{g_{min}}) + \frac1{\sqrt\eps} \log(\tsfrac{n}{\eps\delta})) = O(\frac{k}{\sqrt\eps}\log(\frac{n\kappa_k}{\delta\eps}))\).
\end{proof}
Up to a logarithmic dependence on $\kappa_k$, \Cref{corol:perturbed-svk-symmetric} exactly matches the gap-independent low-rank approximation for the block Krylov method \cite{MuscoMusco:2015}.
The same approach can be generalized to give analogs to \Cref{thm:decay-dependence}, \Cref{thm:fast-frob}, \Cref{thm:schatten_p}, and \Cref{corol:simul-schatten} with a dependence on \(\kappa_k\) instead of \(g_{min}\).

\Cref{corol:perturbed-svk-symmetric} can be interpreted in several ways.
In practice, it is unlikely that adding random noise is in fact needed to break small singular value gaps.
Noise inherent in the input matrix or due to roundoff error will generally suffice to rule out the existence of tiny singular value gaps.
Thus, the corollary can be thought of as a {smoothed-analysis} result \cite{Spielman:2004vw,sankar2006smoothed}, showing that single vector Krylov methods display gap-independent convergence even on input instances which are tiny random perturbations of potentially worst-case instances.
Alternatively, when rigorous worst-case guarantees are required, we could actually run \Cref{alg:single-vec-krylov} on the true input matrix $\mA$ with a random diagonal perturbation added.
Since \mD is diagonal, the runtime of matrix-vector products with \(\tilde\mA\) will generally be dominated by the runtime of matrix-vector products with \(\mA\), so this is very efficient.
We experimentally explore the convergence of this perturbed iteration on a PSD matrix in \Cref{sec:experiment-perturb}.

\subsection{Gap-Independent Bounds for Rectangular and Indefinite Inputs}\label{sec:asymetric-gap-free}

We next observe that \Cref{corol:perturbed-svk-symmetric} gives results for non-PSD matrices as well, at least for low-rank approximation with respect to the {spectral norm}. In this case, we can run a single vector Krylov method on a perturbation of PSD matrix \(\mA\mA^\intercal\).
This suffices because the spectral norm guarantee has the property that a near-optimal basis for approximating \(\mA\mA^\intercal\) is also a near-optimal for \(\mA\):
\begin{lemma}
\label{lem:spectral-square-guarantee}
Let \(\mA\in\bbR^{n \times d}\).
Let \(\mQ\in\bbR^{n \times k}\) be a matrix with orthonormal columns such that  \(\normof{\mA\mA^\intercal-\mQ\mQ^\intercal\mA\mA^\intercal}_2 \leq (1+\eps) \normof{\mA\mA^\intercal - (\mA\mA^\intercal)_k}_2\).
Then, \(\normof{\mA-\mQ\mQ^\intercal\mA}_2\leq(1+\eps)\normof{\mA-\mA_k}_2\).
\end{lemma}
\begin{proof}
Let \(\mP\defeq\mI-\mQ\mQ^\intercal\).
Observe that \(\normof{\mA\mA^\intercal-(\mA\mA^\intercal)_k}_2 = \sigma_{k+1}(\mA)^2\), so we are given that
\(
	\normof{\mP\mA\mA^\intercal}_2 \leq (1+\eps) \ \sigma_{k+1}(\mA)^2
\).
Next note that \(\mP\) is a projection matrix, which can only decrease spectral norms, so we have
\[
	\normof{\mP\mA}_2^2 = \normof{(\mP\mA)(\mP\mA)^\intercal}_2 = \normof{\mP\mA\mA^\intercal\mP}_2 \leq \normof{\mP\mA\mA^\intercal}_2 \leq (1+\eps) \ \sigma_{k+1}(\mA)^2.
\]
Taking the square root of both sides, and noting that \(\sqrt{1+\eps} \leq 1+\eps\), we conclude that \(\normof{\mA-\mQ\mQ^\intercal\mA}_2 \leq (1+\eps) \sigma_{k+1}(\mA) = (1+\eps) \normof{\mA-\mA_k}_2\).
\end{proof}
Combining \Cref{lem:spectral-square-guarantee} with \Cref{lem:spectral-square-guarantee} we obtain the following result: 
\begin{corollary}
\label{corr:rect}
For \(\mA \in\bbR^{n \times d}\), let \(\kappa_k = \frac{\sigma_1}{\sigma_k}\) and \(\Delta = \frac{\eps\sigma_{k+1}^2}{3n}\).
For any \(\eps,\delta\in(0,\frac12)\), let \(\tilde\mA = \mA\mA^\intercal+\mD\) where \(\mD\in\bbR^{n \times n}\) is a diagonal matrix with entries drawn uniformly and i.i.d. from \([-\Delta,\Delta]\).
Then, \Cref{alg:single-vec-krylov} run on \(\tilde\mA\) initialized with \(\vx\sim\cN(\vec0,\mI)\) and run for \(t = O(\frac{k}{\sqrt\eps} \log(\frac{n\kappa_k}{\delta\eps}))\) iterations returns an orthogonal \(\mQ\in\bbR^{n \times k}\) such that, with probability at least  \(1-\delta\),
\[
	\normof{\mA-\mQ\mQ^\intercal\mA}_2 \leq (1+\eps) \normof{\mA-\mQ\mQ^\intercal\mA}_2.
\]
\end{corollary}
Note that it is not clear that \Cref{lem:spectral-square-guarantee} extends beyond the spectral norm, e.g., to the Frobenius norm. Nevertheless, we suspect that a result comparable to \Cref{corr:rect} should hold for all other error metrics considered in this paper.


\section{Numerical Experiments}

\label{sec:experiments}
In this section we validate the core findings of our theoretical results with numerical experiments. We focus on four key findings. In \Cref{sec:experiment-gaps}, we verify that  the dependence of single vector Krylov on the sequential gap size \(g_{min}\) is in fact logarithmic, matching the theoretical bounds of \Cref{sec:gap-dep-analysis,sec:applications}. In \Cref{sec:experiment-block-size}, we show that using a small block size $b > 1$ can ameliorate this gap dependence, by replacing $\log(1/g_{min})$ with $\log(1/g_{min,b})$ as shown in \Cref{thm:block-vec-guarantee}. Relatedly, in \Cref{sec:experiment-perturb} we show that a small random perturbation of the input matrix can break up overlapping singular values and lead to much faster convergence of the single vector Krylov method, matching our theoretical findings from \Cref{sec:perturb}. In \Cref{sec:experiment-grid} we compare single vector and large block Krylov methods. We find that for a wide range of matrices, single vector Krylov methods significantly outperform larger block methods. However, for some very specific worst case instances, large block methods can perform better.

Finally, while our theoretical bounds ignore issues of numerical stability, in \Cref{sec:experiment-ortho}, we observe empirically that small block methods tend to have significantly more issues with stability than large block methods. Exploring this issue further in future work would be very interesting. 

\subsection{Experimental Set Up}

To control against stability issues in our primary experiments, we implement algorithms \Cref{alg:single-vec-krylov} and \Cref{alg:block-krylov} using a full reothogonalization strategy to keep the Krylov subspace close to orthogonal. At every iteration of the (block) Krylov iteration, we orthogonalize the most recently generated column (resp. block) of the Krylov subspace against all previous columns (resp. blocks) using the modified Gram-Schmidt process. At the next iteration, this column (resp. block) is multiplied by $\mA$ to produce a new column (resp. block) of the Kyrlov subspace. See  \Cref{sec:experiment-ortho} for more details or our code, which is available on GitHub\footnote{\href{https://github.com/RaphaelArkadyMeyerNYU/SingleVectorKrylov}{https://github.com/RaphaelArkadyMeyerNYU/SingleVectorKrylov}}. 

We also note that, like most standard implementations, including those based on the Lanczos recurrence, our code implements \Cref{alg:single-vec-krylov} run for $t$ iterations using $t+1$ matrix-vector products with $\mA\mA^\intercal$. To see why this is possible, note that the matrix $\mA\mA^\intercal \mZ$ computed on Line 2 of the algorithm can be formed ``on-the-fly'' as we generate the Krylov subspace. Let $\vz_i$ be the $i^\text{th}$ column in $\mZ$. At each iteration we already compute $\mA\mA^\intercal \vz_i$ for column $\vz_i$ to form column $\vz_{i+1}$, so can just store this result to form $\mA\mA^\intercal\mZ$. Similarly, implementing  \Cref{alg:block-krylov}  requires $(t+1)\ell$ matrix-vector products with $\mA\mA^\intercal$.  

Throughout our experiments, we only report low-rank approximation error in terms of the Frobenius norm, as we found convergence in the spectral and Frobenius norms typically matched quite closely. In particular, letting $\bv Q$ be the output of  \Cref{alg:single-vec-krylov} or \Cref{alg:block-krylov}, we report
\[
	\eps_{empirical} 
	\defeq
	\frac{\normof{\mA-\mQ\mQ^\intercal\mA}_F - \normof{\mA-\mA_k}_F}{\normof{\mA-\mA_k}_F}.
\]
Our theory describes how \(\eps_{empiricial}\) should change as a function of the number of iterations, the block size, and the singular value gaps of $\bv A$.

Lastly, we note that, since Krylov methods initialized with random Gaussian vectors are invariant to rotation, without loss of generality we test on diagonal input matrices for all synthetic data experiments. Each matrix's diagonal entries correspond to its singular values.

\subsection{Verifying Gap Dependence}
\label{sec:experiment-gaps}
We first empirically show that single vector Krylov has a logarithmic dependence on the minimum sequential gap size \(g_{min} = \min_{i\in[k-1]} \frac{{\sigma_i-\sigma_{i+1}}}{\sigma_{i+1}}\), as predicted by the  bounds of \Cref{sec:gap-dep-analysis,sec:applications}.
We consider an exponentially  decaying spectrum with parameter \(\alpha=1.1\) whose singular values are all nearly repeated, with  gap sizes varying between \(g_{min}\in[10^{-10},1]\).
That is, letting our vector of singular values be denoted $\vsigma = [\sigma_1,\ldots,\sigma_n]$ so that $\bv A = \diag(\vsigma)$, and fixing $n = 1000$, we let
\begin{align}\label{eq:sigma}
	\vsigma &= \left[
		1, \hspace{1em} \tsfrac{1}{1+g_{min}}, \hspace{1em}
		\alpha^{-1}, \hspace{1em} \tsfrac{\alpha^{-1}}{1+g_{min}}, \hspace{1em}
		\alpha^{-2}, \hspace{1em} \tsfrac{\alpha^{-2}}{1+g_{min}}, \hspace{1em}
		\ldots \hspace{1em}
		\alpha^{-499}, \hspace{1em} \tsfrac{\alpha^{-499}}{1+g_{min}}
	\right].
\end{align}

\begin{figure}[t]
	\centering
	\includegraphics[height=0.42\columnwidth]{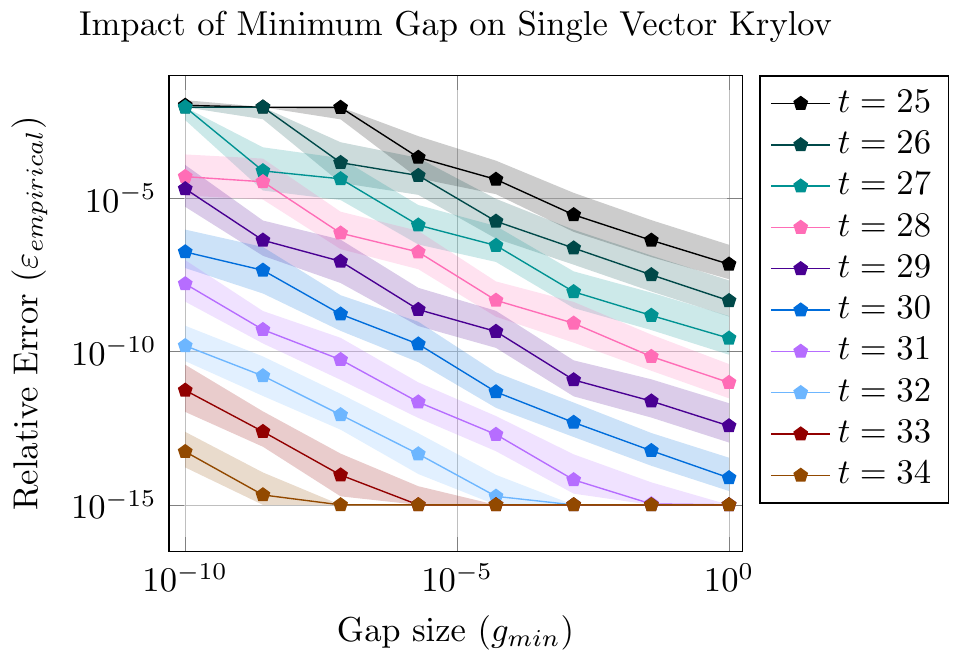}
	\caption{
		Low-rank approximation error vs. minimum gap size for diagonal \(\bv A\in\bbR^{1000 \times 1000}\) with singular values \(\vsigma\) as described in \Cref{eq:sigma}.
		For 8 different gap sizes logarithmically spaced in \([10^{-10},1]\), we run single vector Krylov for  \(t=25,26,\ldots,34\) iterations with target rank \(k=10\).
		The median of \(\eps_{empirical}\) over $500$ independent trials is plotted, with the \(25^{th}\) and \(75^{th}\) quartiles shaded in.
		When \(\eps_{empirical}<10^{-15}\), we plot it at \(10^{-15}\) so that the log/log plot does not degenerate. As expected given the theoretical bound of \Cref{thm:decay-dependence}, we see a linear relationship between the log relative error and log gap size.
	}
	\label{fig:gap-size}
\end{figure}

Since this matrix has a fast decay of singular values, we expect the performance of single vector Krylov to follow the spectral decay rate of \Cref{thm:decay-dependence}. That is, fixing the dimension $d$ and failure probability $\delta$, we should expect the number of iterations to scale as:
\[
	t \propto \tsfrac{\ell}{\sqrt{g_{k\rightarrow \ell}}} \log\left(\tsfrac1{g_{min}}\right) + \tsfrac{1}{\sqrt{g_{k\rightarrow \ell}}} \log\left(\tsfrac{1}{\eps}\right).
\]
Rearranging this expression, we can equivalently expect
\begin{alignat*}{3}
	\log(\eps)
	&=  - t\sqrt{g_{k\rightarrow \ell}} \ &-\ \ell \log(g_{min}) \\
	&=  ~~- C_t \  &-\  \ell \log(g_{min})
\end{alignat*}
where \(C_t \defeq t\sqrt{g_{k\rightarrow \ell}}\) is independent of \(g_{min}\).
So, if we plot \(\eps_{empirical}\) versus \(g_{min}\) on a log/log plot for a fixed value of \(t\), we should see a line with negative slope.
Further, since the vertical offset of these lines are \(C_t \propto t\), we should expect that increasing \(t\) should shift these lines downwards and proportionally to \(t\).
We see this behavior exactly in \Cref{fig:gap-size}.

\subsection{Verifying the Effect of Block Size on Gap Dependence}
\label{sec:experiment-block-size}

We next show that when $\mA$ has very small singular value gaps (or even exactly overlapping singular values), the dependence on $\log(1/g_{\min})$ can be avoided by using a small constant block size $b$.
This lets us instead depend on $\log(1/g_{min,b})$, as in the analysis of \Cref{thm:block-vec-guarantee}.

We focus on when \mA has pairs, but not triplets, of exactly overlapping singular values.
In this case, block Krylov with block size \(b=2\) should perform well, since it should not suffer due to the overlapping singular values.
Further, it should match or outperform larger block methods.
To show this, we construct an exponentially decaying spectrum with parameter \(\alpha=1.005\) and whose top \(k = 50\) singular values are each repeated, with sequential gap size \(g_{min}=0\).
Formally, we choose $1000$ singular values as follows:
\begin{align}\label{eq:sigma2}
	\vsigma_A &= \left[
		1 \hspace{1em} 1 \hspace{1em}
		\alpha^{-1} \hspace{1em} \alpha^{-1} \hspace{1em}
		\alpha^{-2} \hspace{1em} \alpha^{-2} \hspace{1em}
		\ldots \hspace{1em}
		\alpha^{-25} \hspace{1em} \alpha^{-25}
	\right] \nonumber \\
	\vsigma_B &= \left[
		\alpha^{-26} \hspace{1em} \alpha^{-27} \hspace{1em}
		\alpha^{-28} \hspace{1em} \ldots \hspace{1em}
		\alpha^{-975}
	\right] \nonumber \\
	\vsigma &= \left[ \vsigma_A \hspace{1em} \vsigma_B \right].
\end{align}
In theory, single vector Krylov should completely fail in this case, only capturing a $k/2$-dimensional subspace of the span of the top $k$ singular vectors.
Due to finite precision roundoff, the method nevertheless converges.
However, it is still significantly handicapped by the repeated singular values.

\begin{figure}[t]
	\centering
	\includegraphics[width=\textwidth]{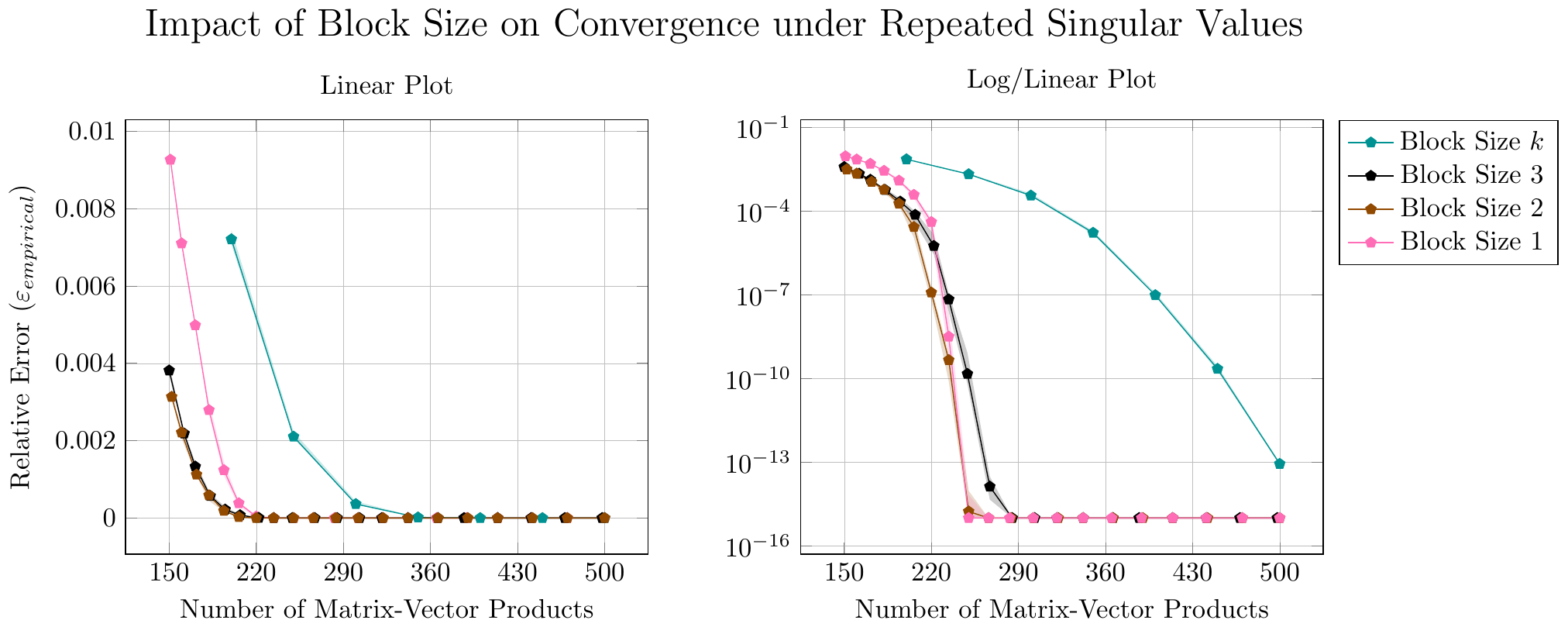}
	\caption{
		Low-rank approximation error vs. number of matrix-vector products for diagonal \(\bv A\in\bbR^{1000 \times 1000}\) with repeated top singular values as in \Cref{eq:sigma2}. We run Krylov iteration with target rank \(k=50\) for block sizes \(b=1,2,3,50\).
		The median of \(\eps_{empirical}\) over $10$ independent trials  is plotted, with the \(25^{th}\) and \(75^{th}\) quartiles shaded in. Note that these quartiles are very close together in this example.
		When \(\eps_{empirical}<10^{-15}\), we plot it at \(10^{-15}\) so that the plot does not degenerate.
		The plot on the left has a linear y-axis and highlights the performance for early iterations, while the plot on the right uses a logarithmic y-axis and highlights the performance for later iterations. Overall, we see that block size $b=2$, which is just large enough to avoid the repeated pairs of singular values in $\bv{A}$'s spectrum, performs best. Single vector Krylov is handicapped by the repeated singular values, especially at early iterations. Large block Krylov converges at a significantly slower rate than the small block variants.
	}
	\label{fig:block-size}
\end{figure}

In \Cref{fig:block-size} we plot the low-rank approximation error vs. number of matrix-vector products of single vector Krylov and block Krylov with block sizes $2,3$, and $50$, for target rank $k = 50$.
We show both y-linear and y-logarithmic plots to highlight the performance at early and later iterations.
We see that block size 2 performs the best across the board, and that block size 3 is only mildly worse.
Due to the repeated singular values, single vector Krylov performs worse, especially for the early iterations. It becomes competitive with block size 3 eventually.
In contrast, the full block size \(k\) method converges much more slowly.

\subsection{Verifying the Effect of Random Perturbations on Gap Dependence}
\label{sec:experiment-perturb}
\begin{figure}[t]
	\centering
	\includegraphics[width=\textwidth]{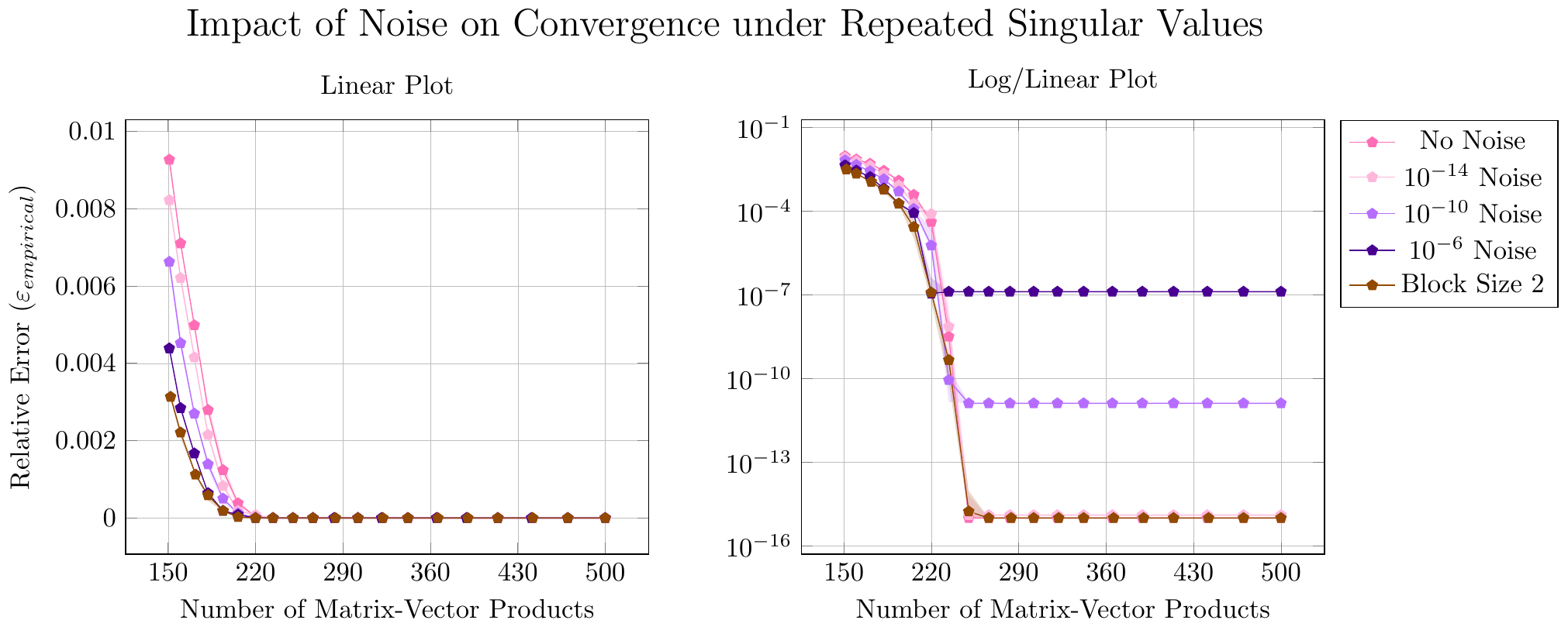}
	\caption{
		Low-rank approximation error vs. number of matrix-vector products for diagonal \(\bv A\in\bbR^{1000 \times 1000}\) with repeated top singular values as described in \Cref{eq:sigma2}. We run the single vector Krylov method with target rank $k = 50$, with varying levels of noise added to $\bv A$ to separate its repeated singular values. In particular, following \Cref{corol:perturbed-svk-symmetric}, we run the method on \(\mA+\mD\) where \(\mD\in\bbR^{1000\times1000}\) is a random diagonal matrix with entries drawn uniformly on \([-\Delta,\Delta]\), for \(\Delta \in \{10^{-6},10^{-10},10^{-14}, 0\}\). Low-rank approximation error is measured with respect to the original input \mA. For comparison, we also show performance of block Krylov with block size $b=2$ with no noise added. This was the optimal block size for this spectrum as seen in \Cref{fig:block-size}. The median of \(\eps_{empirical}\) over $10$ independent trials  is plotted, with the \(25^{th}\) and \(75^{th}\) quartiles shaded in. Note that these quartiles are very close together in this example.
		When \(\eps_{empirical}<10^{-15}\), we plot it at \(10^{-15}\) so that the plot does not degenerate.
		The plot on the left has a linear y-axis and highlights the performance for early iterations, while the plot on the right uses a logarithmic y-axis and highlights the performance for later iterations. We can see that by adding a small random perturbation, we can improve the convergence of single vector Krylov to nearly match that of block Krylov with $b = 2$. Larger noise leads to faster convergence but also larger final error.
	}
	\label{fig:perturb}
\end{figure}

Next, we show that adding a small amount of random noise to break up small singular value gaps can also make single vector Krylov converge more quickly, verifying the results of \Cref{sec:perturb}.
We use the same matrix as in \Cref{sec:experiment-block-size}, with spectrum given in \Cref{eq:sigma2}.
In \Cref{fig:perturb}, we show that adding noise to $\bv{A}$ the order of \(10^{-6}\) leads to single vector Krylov converging nearly as quickly as the optimal \(b=2\) block Krylov method as seen in \Cref{sec:experiment-block-size}. This noise does limit our eventual accuracy at convergence, which can be seen clearly in our logarithmic error plot. 
Changing the magnitude of the noise lets us interpolate between fast convergence and high accuracy.

\subsection{Effect of Block Size in Convergence}
\label{sec:experiment-grid}
\begin{figure}
	\centering
	\includegraphics[width=0.945\columnwidth]{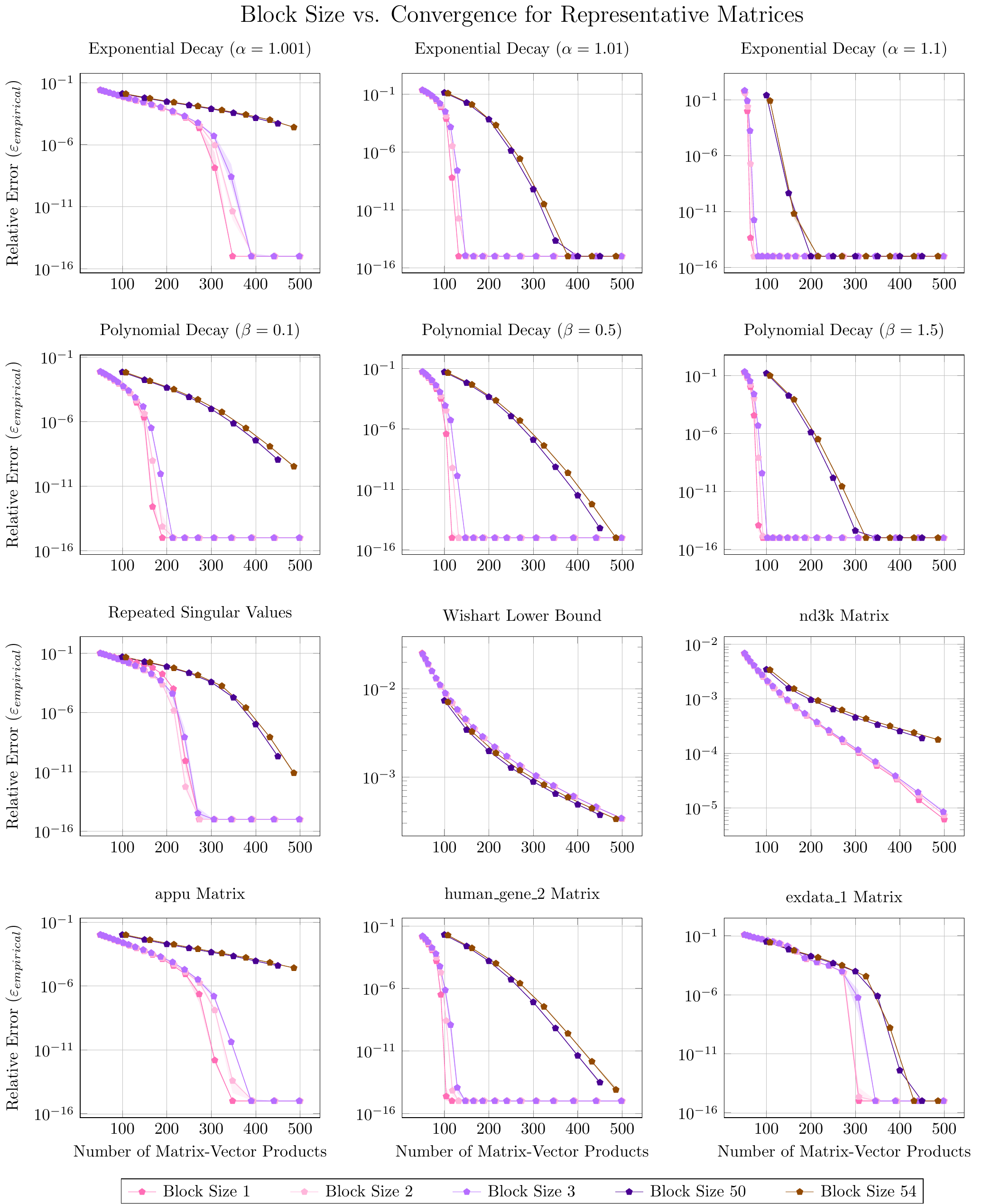}
	\caption{Low-rank approximation error vs. number of matrix-vector products for Krylov iteration with various block sizes on synthetic and real-world input matrices as described in \Cref{sec:experiment-grid}. In all cases, the target rank is set to $k = 50$. 
	The median of \(\eps_{empirical}\) is plotted over 10 independent trials, with the \(25^{th}\) and \(75^{th}\) quartiles shaded in. Note that these quartiles are very close together in most plots.
	When \(\eps_{empirical}<10^{-15}\), we plot it at \(10^{-15}\) so that the plot does not degenerate.
	}
	\label{fig:block-size-grid}
\end{figure}

We next present a wider comparison of how the choice of block size for Krylov iteration effects convergence to a near-optimal low-rank approximation. We fix target rank $k = 50$ and compare  block sizes 1, 2, 3, 50, and 54. Block size $1$ corresponds to the single vector Krylov method.
Block sizes 2 and 3 should be more resilient to a pairs or triplets of very close singular values, respectively.
Block sizes \(k\) and \(k+4\) are recommended by prior theoretical work on Krylov Iteration for low-rank approximation \cite{MuscoMusco:2015,tropp2018analysis}.
We consider eight input matrices. All synthetic inputs are $1000 \times 1000$ and diagonal.
\begin{enumerate}
	\item \textbf{Exponential Decay}: \(\sigma_i = \alpha^{-i}\) for \(\alpha\in\{1.001,\ 1.01,\ 1.1\}\)
	\item \textbf{Polynomial Decay}: \(\sigma_i = i^{-\beta}\) for \(\beta\in\{0.1,\ 0.5,\ 1.5\}\)
	\item \textbf{Repeated Singular Values}: A matrix with each of its top $k$ singular values repeated, as defined in \Cref{eq:sigma2}. 
	\item \textbf{Wishart Lower Bound}:  \(\sigma_i = \sqrt{1-(\frac{i}{1000})^2}\).
		  This is an approximation of the spectrum of \(\mI - \frac1{5n}\mG^\intercal\mG\) where \(\mG^{1000\times1000}\) has i.i.d \(\cN(0,1)\) entries.
		  This matrix is used as a lower bound instance for rank-1 low-rank approximation in \cite{bakshi2022low}.
  	\item \textbf{nd3k, appu, human\_gene\_2, exdata\_1}: Various real-world matrices arbitrarily chosen from SuiteSparse \cite{sparseSuite}.
\end{enumerate}
We can see the results of these experiments in \Cref{fig:block-size-grid}.
We see that for all except the repeated singular value and Wishart lower bound matrices, single vector Krylov dominates.
For the repeated singular value matrix, as in \Cref{sec:experiment-block-size}, we see that block size 2 dominates again.
For the Wishart lower bound matrix from \cite{bakshi2022low}, we see that large block methods marginally (though consistently) outperform small block methods.
This lower bound instance is designed to force Krylov methods to converge at a rate of \(\frac{1}{\eps^{1/3}}\), instead of at the spectral decay rate \(\frac{1}{\sqrt{g_{k\rightarrow\ell}}}\).
This seems to makes the rate of convergence of single vector Krylov slower than block Krylov, since we pay a \(\log(1/g_{min})\) dependence, while only benefiting a small amount from separating the $k$ and $\log(n/\eps)$ dependence (see \Cref{thm:single-vec-guarantee}).
In contrast, the other figures show matrices where the spectral decay rate controls convergence, where single vector Krylov still pays \(\log(1/g_{min})\) but seems to see performance gains from being able to simulate general block sizes and from separating the $\log(n/\eps)$ dependence from the (simulated) block size dependence (see \Cref{thm:decay-dependence}).

\subsection{Block Size and Numerical Stability}
\label{sec:experiment-ortho}
It is well known that Krylov methods can suffer from numerical stability issues \cite{golub2013matrix,meurant2006lanczos}.
In particular, the iterates \((\mA\mA^\intercal)^t\vg\) approach the same vector (the top singular vector of $\bv A$) as \(t\) grows large.
So, \mK becomes ill-conditioned.
So far, we have focused on convergence guarantees and ignored numerical stability.
As discussed, our implementations use orthogonalization to keep \mK well-conditioned at all iterations.
That is, for single vector Krylov, at every iteration, we compute \((\mA\mA^\intercal)\vk_{t-1}\) where \(\vk_{t-1}\) is the last column in the Krylov matrix \mK.
Then we project \((\mA\mA^\intercal)\vk_{t-1}\) away from all of the previous columns \(\vk_1,\ldots,\vk_{t-1}\) via modified Gram-Schmidt, and store the resulting vector as \(\vk_t\).
We do the same for block Krylov, where we compute \((\mA\mA^\intercal)\bmat{\vk_{t-b-1} & \ldots & \vk_{t-1}}\) and add the resulting \(b\) columns to \mK iteratively via modified Gram-Schmidt.

In practice, Krylov implementations typically spend less effort orthogonalizing at each step.
For example, they are commonly implemented via the Lanczos method, where \(\vk_t\) is only projected away from \(\vk_{t-1}\) and \(\vk_{t-2}\).
In infinite precision, this is equivalent to projecting away from all previous columns  \cite{golub2013matrix}.
Similar ideas can be applied to block Krylov methods \cite{RokhlinSzlamTygert:2009,Saad:1980}.
While such methods are highly efficient, when using them, $\bv{K}$ can lose orthogonality.
This can lead to slower convergence or necessitate modifications such as restarts or reorthogonalization \cite{calvetti1994implicitly,paige1972computational,parlett1998symmetric}.

\begin{figure}
	\centering
	\includegraphics[height=0.42\columnwidth]{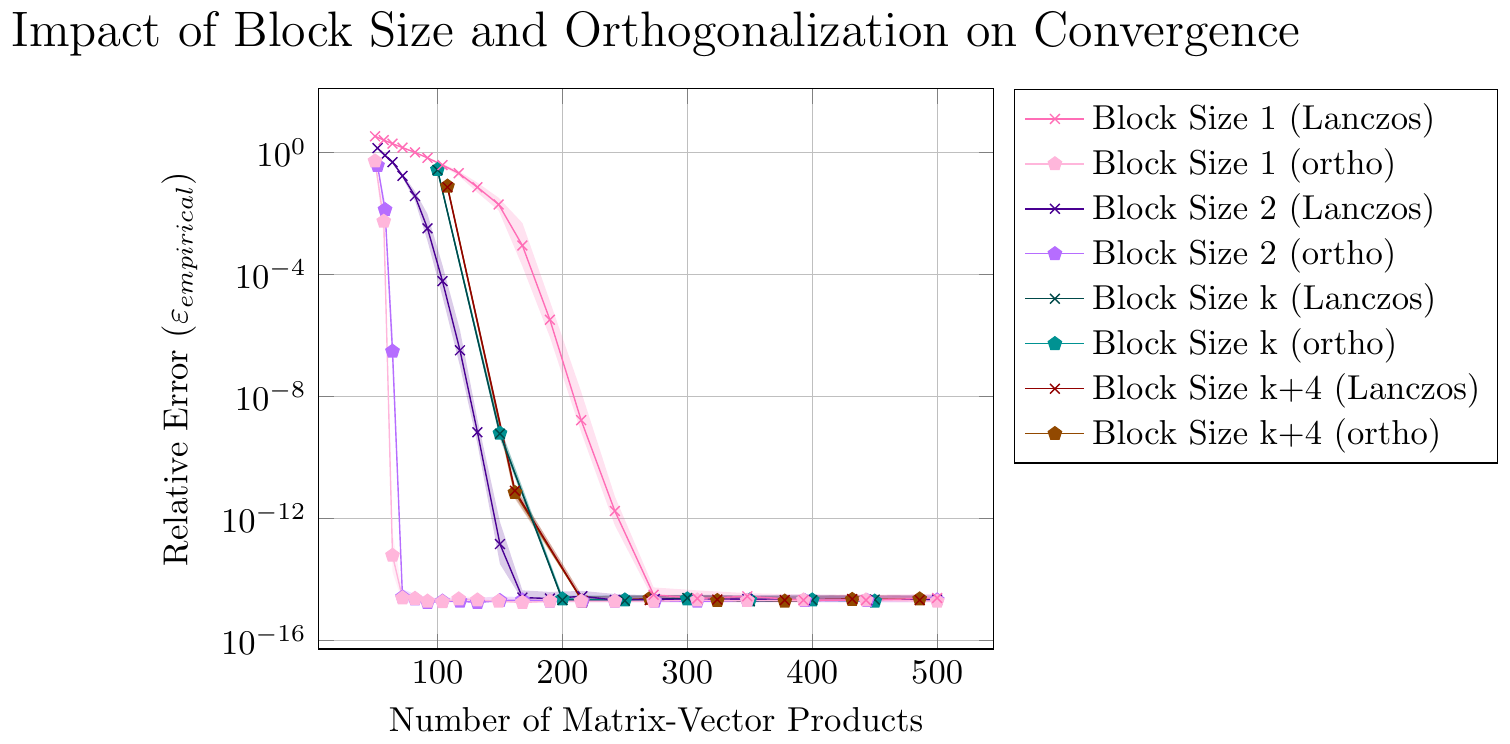}
	\caption{Low-rank approximation error vs. number of matrix-vector products for diagonal \(\bv A\in\bbR^{1000 \times 1000}\) with singular values \(\sigma_i=1.1^{-i}\) and target rank $k = 50$. We run for block sizes \(b\in\{1,2,k,k+4\}\) both with Lanczos and with full orthogonalization.
		The median of \(\eps_{empirical}\) over $100$ independent trials  is plotted, with the \(25^{th}\) and \(75^{th}\) quartiles shaded in.
				Notice that for this experiment, single vector Krylov converges the fastest with full orthogonalization and slowest with Lanczos, and that large block methods have no real gap between full orthogonalization and Lanczos. I.e., large block methods are much more stable even with partial orthogonalization.
	}
	\label{fig:ortho}
\end{figure}

Intuitively, comparing single vector or small block Krylov to  large block Krylov with a fixed size Krylov subspace \mK, we expect that single vector and small block Krylov will be more susceptible to conditioning issues, since they require more iterations to reach the same sized subspace. Thus, we should expect partial orthogonalization methods like Lanczos to lead to slower convergence for these methods as compared to large block methods.
With full orthogonalization, we should instead expect to see small block methods dominate.
We see this trend exactly in \Cref{fig:ortho}. An interesting extension to our work would be to more closely study the stability of Krylov methods for low-rank approximation, and to develop a more clear theoretical understanding of the advantages of large block methods in this regard.

\section*{Acknowledgements}
This work was supported by NSF Awards 2046235 and 2045590. 
Cameron Musco thanks Alex Breuer for helpful conversations in 2016 that helped inspire the initial idea behind this work.
Raphael Meyer was partially supported by a GAANN fellowship from the US Department of Education.
He thanks Tyler Chen, Apoorv Singh, and Axel Elaldi for help designing the figures.
He also thanks Kathryn Lund, Kirk Soodhalter, and David Persson for helpful conversations contextualizing this work.
He further thanks David Persson for help formalizing the results on single vector simultaneous iteration.

\clearpage

\bibliographystyle{apalike}
\bibliography{singleVector}

\appendix
\section{Reduction to the Positive Semidefinite Case}

\label{app:reduct_to_psd_case}
In our analysis, we can assume without loss of generality that the input \(\mA\) is a square PSD matrix. To see why, for any $\mC \in \R^{n \times d}$, let $\mA = (\mC \mC^\intercal)^{1/2}$. Observe that $\mA \in \R^{n \times n}$ is PSD. Further, observe that since $\mA^2 = \mA \mA^\intercal =\mC\mC^\intercal$, \Cref{alg:single-vec-krylov} and \Cref{alg:block-krylov} yield identical outputs for $\mA$ and $\mC$.

Expanding the SVD \(\mC=\mU\mSigma\mV^\intercal\), we can have \(\mA = (\mC\mC^\intercal)^{\nicefrac12} = \mU\mSigma\mU^\intercal\). Thus, $\mA$ and $\mC$ have identical singular values and $\normof{\mA-\mA_k}_\xi = \normof{\mC- \mC_k}_\xi$ for any unitarily invariant norm $\norm{\cdot }_{\xi}$ (including the spectral and Frobenius norms) and any $k$. Additionally, for any $\bv{Q} \in \R^{n \times k}$,
\[
\normof{\mA-\mQ\mQ^\intercal\mA}_\xi = \normof{\mU\mSigma-\mQ\mQ^\intercal\mU\mSigma}_\xi =\normof{\mC-\mQ\mQ^\intercal\mC}_\xi.
\]
Thus, any bound on $\normof{\mA-\mQ\mQ^\intercal\mA}_\xi$ in terms of $\normof{\mA-\mA_k}_\xi $ holds identically for $\mC$. Finally, for any $\vq \in \R^n$, $\vq^\intercal \mC\mC^\intercal \vq= \vq^\intercal \mA^2 \vq$. Thus, any bound on $\vq^\intercal \mA^2 \vq$ in terms of $\sigma_i(\mA)$ holds identically for $\mC$.

\section{Frobenius Low-Rank Approximation with \texorpdfstring{\(\eps^{-1/3}\)}{eps 1/3} Dependence}
\label{app:fast-frob}

In this section, we prove \Cref{thm:fast-frob}.
This analysis closely follows the intuition given in the introduction of \cite{bakshi2022low}.
We have the following:
\begin{reptheorem}{thm:fast-frob}
For \(\mA \in\bbR^{n \times d}\), let \(g_{min} = \min_{i\in\{1,\ldots,\ell-1\}} \frac{\sigma_i - \sigma_{i+1}}{\sigma_{i+1}}\) where \(\ell = \Theta(\frac{k}{\eps^{1/3}})\).
For any \(\eps,\delta\in(0,1)\), \Cref{alg:single-vec-krylov} initialized with \(\vx\sim\cN(\vec0,\mI)\) and run for \(t = O(\frac{k}{\eps^{1/3}} \log(\frac1{g_{min}}) + \frac{1}{\eps^{1/3}} \log(\frac{n}{\delta \eps}))\) iterations returns an orthogonal \(\mQ\in\bbR^{n \times k}\) such that, with probability at least \(1-\delta\),
\[
	\normof{\mA-\mQ\mQ^\intercal\mA}_F \leq (1+\eps) \normof{\mA-\mQ\mQ^\intercal\mA}_F.
\]
\end{reptheorem}
\begin{proof}
We first recall that we have two different guarantees for \Cref{alg:single-vec-krylov}.
Fix some \(\gamma > 0\).
By \Cref{thm:single-vec-guarantee}, if we run \Cref{alg:single-vec-krylov} for \(t = O(\tsfrac k{\sqrt\gamma} \log(\tsfrac1{g_{min}}) + \tsfrac1{\sqrt\gamma} \log(\tsfrac{n}{\delta\gamma}))\) iterations, then with probability at least \(1-\delta\),
\[
	\abs{\vq_i^\intercal\mA\mA^\intercal\vq_i - \sigma_i^2} \leq \gamma \sigma_{k+1}^2.
\]
Fix some \(\ell \geq k\), and let \(g_{k\rightarrow \ell} = \frac{\sigma_k - \sigma_{\ell+1}}{\sigma_k}\).
Then, by \Cref{thm:decay-dependence}, if we run \Cref{alg:single-vec-krylov} for \(t = O(\tsfrac{\ell}{\sqrt{g_{k\rightarrow \ell}}} \log(\tsfrac1{g_{min}}) + \tsfrac{1}{\sqrt{gap_{k\rightarrow \ell}}} \log (\tsfrac{n}{\delta\eta}))\) iterations, then with probability at least \(1-\delta\) we again have
\[
	\abs{\vq_i^\intercal\mA\mA^\intercal\vq_i - \sigma_i^2} \leq \eta \sigma_{k+1}^2.
\]
We will show that we can get error \eps\xspace in Frobenius norm by taking \(\gamma = \eps^{2/3}\) and \(\eta = \frac\eps k\).
In particular, we run a case-analysis between either large-tailed or small-tailed spectra of \mA.

\medskip
\noindent \textbf{Small Tailed Case:}
Suppose \(\normof{\mA-\mA_k}_F^2 \leq \frac{k}{\eps^{1/3}} \sigma_{k+1}^2\).
Then \mA must have a fast spectral decay.
In particular, let \(\ell = (1+\frac{4}{\eps^{1/3}})k = O(\frac k{\eps^{1/3}})\).
Then \(\sigma_{\ell}\) is substantially smaller than \(\sigma_k\):
\[
	\frac{k}{\eps^{1/3}}\sigma_{k+1}^2
	\geq \normof{\mA-\mA_k}_F^2
	= \sum_{i=k+1}^n \sigma_i^2
	\geq (\ell-k)\sigma_\ell^2
	= \frac{4k}{\eps^{1/3}} \sigma_\ell^2.
\]
That is, \(\sigma_\ell \leq \frac{\sigma_{k+1}}{2}\), so that \(\sqrt{g_{k\rightarrow \ell}} = \sqrt{\frac{\sigma_k-\sigma_{\ell+1}}{\sigma_k}} \geq \frac1{\sqrt2}\), so the spectral-decay analysis of \Cref{thm:decay-dependence} says that \(t = O(\frac{k}{\eps^{1/3}} \log(\tsfrac1{g_{min}}) + \log (\tsfrac{n}{\delta\eta}))\) iterations suffice to get the singular value guarantee \(\normof{\vq_i^\intercal\mA}_2^2 \in \sigma_i^2 \pm \eta \sigma_{k+1}^2\).
Since \(\mQ\mQ^\intercal=\sum_{i=1}^k \vq_i\vq_i^\intercal\) is a sum of orthogonal projection matrices,
\begin{align*}
	\normof{\mA-\mQ\mQ^\intercal\mA}_F^2
	&= \normof{\mA}_F^2 - \sum_{i=1}^k \normof{\vq_i\vq_i^\intercal\mA}_F^2  \tag{Matrix Pythagoras}\\
	&= \normof{\mA}_F^2 - \sum_{i=1}^k \normof{\vq_i^\intercal\mA}_2^2 \\
	&\leq \normof{\mA}_F^2 - \sum_{i=1}^k \sigma_i^2 + \eta k \sigma_{k+1}^2 \tag{Singular Value Guarantee} \\
	&= \normof{\mA-\mA_k}_F^2 + \eta k \sigma_{k+1}^2 \\
	&\leq \normof{\mA-\mA_k}_F^2 + \eta k \normof{\mA-\mA_k}_F^2 \\
	&= (1+\eta k)\normof{\mA-\mA_k}_F^2.
\end{align*}
So, taking \(\eta = \frac{\eps}{k}\) for a total iteration count of \(t = O(\frac{k}{\eps^{1/3}} \log(\tsfrac1{g_{min}}) + \log (\tsfrac{n}{\delta\eps}))\) suffices in this case.

\medskip
\noindent
\textbf{Large Tailed Case:} Suppose \(\normof{\mA-\mA_k}_F^2 > \frac{k}{\eps^{1/3}} \sigma_{k+1}^2\).
Since the tail is large, even a low-accuracy singular value guarantee still ensures a good Frobenius norm guarantee.
In particular, we take the gap-independent analysis of \Cref{thm:single-vec-guarantee} with \(\gamma=\eps^{2/3}\), so that \(\normof{\vq_i^\intercal\mA}_2^2 \in \sigma_i^2 \pm \eps^{2/3}\sigma_{k+1}^2\), and we get
\begin{align*}
	\normof{\mA-\mQ\mQ^\intercal\mA}_F^2
	&= \normof{\mA}_F^2 - \sum_{i=1}^k \normof{\vq_i\vq_i^\intercal\mA}_F^2 \tag{Matrix Pythagoras} \\
	&= \normof{\mA}_F^2 - \sum_{i=1}^k \normof{\vq_i^\intercal\mA}_2^2 \\
	&\leq \normof{\mA}_F^2 - \sum_{i=1}^k \sigma_i^2 + \eps^{2/3} k\sigma_{k+1}^2 \tag{Singular Value Guarantee}\\
	&= \normof{\mA-\mA_k}_F^2 + \eps^{2/3} k\sigma_{k+1}^2 \\
	&\leq \normof{\mA-\mA_k}_F^2 + \eps \normof{\mA-\mA_k}_F^2 \tag{\(\sigma_{k+1}^2 < \tsfrac{\eps^{1/3}}{k} \normof{\mA-\mA_k}_F^2\)}\\
	&= (1+\eps)\normof{\mA-\mA_k}_F^2.
\end{align*}
So, since \(\gamma=\eps^{2/3}\) here, we achieve error \(\eps\) under Frobenius norm with \(t = O(\tsfrac k{\eps^{1/3}} \log(\tsfrac1{g_{min}}) + \tsfrac1{\eps^{1/3}} \log(\tsfrac{n}{\delta\eps}))\).

\medskip

\noindent
\textbf{Putting it Together. }
So, in either case, running \(t = \tilde O(\frac{k}{\eps^{1/3}})\) iterations suffices to obtain a $(1+\eps)$ optimal low-rank approximation in the Frobenius norm.
Further, the algorithm used in the two cases is identical, so (unlike \cite{bakshi2022low}) we do not have to detect which case we are in and alter the algorithm accordingly.
We simply run single vector Krylov.
\end{proof}


\section{Schatten Norm Low-Rank Approx. with \texorpdfstring{\(\eps^{-1/3}\)}{eps 1/3} Dependence}
\label{app:fast-schatten}

In this section, we prove \Cref{thm:schatten_p} by arguing that running \Cref{alg:single-vec-krylov} once effectively simulates running Algorithm 5.4 from \cite{bakshi2022low}.
We have the following:

\begin{reptheorem}{thm:schatten_p}
For \(\mA \in\bbR^{n \times d}\) and \(p \geq 1\), let \(g_{min} = \min_{i\in\{1,\ldots,\ell-1\}} \frac{\sigma_i - \sigma_{i+1}}{\sigma_{i+1}}\) where \(\ell = \Theta(\frac{k}{\eps^{1/3}p^{1/3}})\).
For any \(\eps,\delta\in(0,1)\), let \(\mQ\in\bbR^{d \times k}\) be the result of running \Cref{alg:single-vec-krylov} on \(\mA^\intercal\) initialized with \(\vx\sim\cN(\vec0,\mI)\) and run for
\[
	t
	= O \left (\tsfrac{kp^{1/6}}{\eps^{1/3}} \log(\tsfrac{1}{g_{min}}) + (\sqrt{p}+\tsfrac{p^{1/6}}{\eps^{1/3}}) \log(\tsfrac{np}{\delta\eps}) \right)
	= \tilde O \left (\tsfrac{kp^{1/6}}{\eps^{1/3}} + \sqrt p \right)
\]
iterations.
Let \(\mZ\in\bbR^{n \times k}\) be an orthonormal basis for \(\mA\mQ\).
Then, with probability at least \(1-\delta\),
\[
	\normof{\mA-\mZ\mZ^\intercal\mA}_p \leq (1+\eps)\normof{\mA-\mA_k}_p,
\]
where \(\normof\mA_p \defeq (\sum_{i=1}^n\sigma_i(\mA)^p)^{1/p}\) is the Schatten \(p\)-norm.
\end{reptheorem}
\begin{proof}
Note that \cite{bakshi2022low} outputs orthonormal $\bv{Q} \in \R^{d \times k}$ with \(\normof{\mA-\mA\mQ\mQ^\intercal}\) bounded, rather than $\bv{Q} \in \R^{n \times k}$ with \(\normof{\mA-\mQ\mQ^\intercal\mA}\) bounded. However, we can translate their analysis to the later case simply by running their algorithms on $\bv{A}^\intercal$. Thus we consider this case going forward.
The first two lines of Algorithm 5.4 in \cite{bakshi2022low} run Block Krylov Iteration twice on \(\mA^\intercal\). 
First, they let \(\mW_1 \in \bbR^{d \times k}\) be the result of using block size \(\ell_1 = k\) and running until the gap-independent rate gives a singular value guarantee (i.e. \Cref{eq:per_vec_gen}) with relative error at most \(\gamma_1 = \frac{\eps^{2/3}}{p^{1/3}}\).
For single vector Krylov, by \Cref{thm:single-vec-guarantee}, this takes
\[
	O\left(\tsfrac{k}{\sqrt{\gamma_1}} \log(\tsfrac{1}{g_{min}}) + \tsfrac{1}{\sqrt{\gamma_1}} \log(\tsfrac{n}{\gamma_1 \delta})\right)
	= O\left(\tsfrac{kp^{1/6}}{\eps^{1/3}} \log(\tsfrac{1}{g_{min}}) + \tsfrac{p^{1/6}}{\eps^{1/3}} \log(\tsfrac{np}{\eps \delta})\right)
\]
iterations.
Second, they let \(\mW_2\in\bbR^{d \times k}\) be the result of running with block size \(\ell_2 = O(\frac{k}{\eps^{1/3}p^{1/3}})\) for enough iterations so that if \(g_{k \rightarrow \ell_2} \geq \frac1p\), then block Krylov would achieve error \(\gamma_2 = \poly(\frac \eps n)\)\footnote{They write \(\gamma_2 = \eps\) in the algorithm but in Equation (5.21) we can see they actually want this smaller error. Since gap-dependent rate depends on \(\log(\frac{n}{\gamma_2})\), shrinking \(\gamma_2\) from \(\eps\) to \(\text{poly}(\frac{\eps}{n})\) does not change the asymptotic complexity.}.
For single vector Krylov, by \Cref{thm:decay-dependence}, this takes
\[
	O\left(\tsfrac{\ell_2}{\sqrt{g_{k\rightarrow \ell_2}}} \log(\tsfrac{1}{g_{min}}) + \tsfrac{1}{\sqrt{g_{k\rightarrow \ell_2}}} \log(\tsfrac{n}{\gamma_2 \delta})\right)
	= O\left(\tsfrac{kp^{1/6}}{\eps^{1/3}} \log(\tsfrac{1}{g_{min}}) + \sqrt{p} \log(\tsfrac{n}{\eps \delta})\right)
\]
iterations.
Note that \Cref{alg:single-vec-krylov} outputs a single matrix \mQ that achieves the guarantees needed by both \(\mW_1\) and \(\mW_2\).

Next, we consider the third and fourth lines of Algorithm 5.4 in \cite{bakshi2022low}.
The third line runs block Krylov on \mA directly to estimate several of its singular values.
The fourth line uses those estimated singular values to determine if we should return an orthogonal basis for \(\mW_1^\intercal\mA^\intercal\) or \(\mW_2^\intercal\mA^\intercal\).\footnote{In a personal communication with the authors of \cite{bakshi2022low}, we confirmed there is a typo in the current arXiv version of the paper, where the algorithm says to return a matrix \(\mZ_1\). It should return an orthonormal basis for \(\mW_1^\intercal\mA^\intercal\) instead.}
Since we have \(\mW_1 = \mW_2 = \mQ\), we can ignore the tests in the third and fourth lines, and just always return a basis for \(\mQ^\intercal\mA^\intercal\). 
So, overall, we compute a matrix \mZ with the exact same guarantees as the \cite{bakshi2022low} by only using a one instance of single vector Krylov.
\end{proof}


\section{Eigenvalue Repulsion Corollaries}

This appendix covers the proofs needed for \Cref{corol:perturbed-svk-symmetric}.
First, we take a result of \cite{minami1996local} and use it to prove a gap on the eigenvalues of symmetric matrices.
Second, we show that a optimal projection matrix that achieves near-optimal low-rank approximation on the perturbation of \mA must also achieve near-optimal low-rank approximation on \mA itself.

\subsection{Proof of \texorpdfstring{\Cref{lem:diagonal-perturb-gaps}}{Lemma \ref*{lem:diagonal-perturb-gaps}}}
\label{app:symmetric-repulse}

We first import a result of \cite{minami1996local}, originally studied in relation to the Wegner Estimate \cite{wegner1981bounds}.
This result is also given as Equation (1.11) in \cite{aizenman2017matrix}:
\begin{importedtheorem}
\label{impthm:minami-bound}
Let \(\mA\in\bbR^{n \times n}\) be symmetric, and let \(\mD\in\bbR^{n \times n}\) be diagonal, with entries  drawn i.i.d. from a distribution with pdf \(p(\cdot)\).
Then, for any interval \(\cI \subset \bbR\),
\[
	\Pr[\mA+\mD \text{ has at least 2 eigenvalues in } \cI] \leq C (\normof{p}_{\infty} \abs{\cI} n)^2,
\]
for some universal constant \(C\), where \(\abs{\cI}\) is the length of \cI, and where \(\normof{p}_{\infty} = \max_{t\in\bbR} p(t)\).
\end{importedtheorem}

\begin{replemma}{lem:diagonal-perturb-gaps}
Fix symmetric matrix \(\mA\in\bbR^{n \times n}\), \(\delta \in (0,1)\), and \(\Delta \leq \normof{\mA}_2\).
Let \(\mD \in \bbR^{n \times n}\) be a diagonal matrix whose entries are uniformly distributed in \([-\Delta,+\Delta]\).
Then, letting \(\tilde\mA = \mA + \mD\) and letting $C$ denote some universal constant, with probability at least \(1-\delta\),
\[
	\min_{i\in[n-1]} \frac{\abs{\lambda_i(\tilde\mA) - \lambda_{i+1}(\tilde\mA)}}{\abs{\lambda_{i+1}(\tilde\mA)}} \geq \frac{\delta}{Cn^2} \cdot \frac{\Delta^2}{\normof{\mA}_2^2}.
\]
\end{replemma}
\begin{proof}
Let \(R \defeq 2\normof{\mA}_2\).
Since \(\Delta < \normof{\mA}_2\), we know that \(\normof{\tilde\mA}_2 \leq \normof{\mA}_2 + \normof{\mD}_2 \leq 2 \normof{\mA}_2 = R\).
Let \(\gamma > 0\) be a number to be fixed later.
Then define
\[
	\cI_{i} \defeq (-R+i\gamma) \pm \gamma = [(-R+\gamma i) - \gamma, (-R+\gamma i) + \gamma]
\]
for \(i=1,\ldots,m\), where \(m = \frac{4\normof{\mA}_2}{\gamma}-1\).
These are intervals of width \(2\gamma\) that overlap and cover the range \([-R, R]\).
For instance, we have \(\cI_1 = [-R, -R+2\gamma]\), \(\cI_2 = [-R+\gamma, -R+2\gamma]\), and \(\cI_3 = [-R + 2\gamma, R+4\gamma]\), so that \(\cI_2\) overlaps with \(\cI_1\) and \(\cI_3\).
In particular, if \(\tilde\mA\) has two eigenvalues that are additively \(\gamma\) close, so that \(\abs{\lambda_i(\tilde\mA) - \lambda_{i+1}(\tilde\mA)} \leq \gamma\), then we know that \(\lambda_i(\tilde\mA)\) and \(\lambda_{i+1}(\tilde\mA)\) both lie in some \(\cI_j\).
Therefore, we can write
\begin{align*}
	\Pr[\exists i:\, \abs{\lambda_i(\tilde\mA) - \lambda_{i+1}(\tilde\mA)} < \gamma]
	&\leq \Pr[\text{at least two eigenvalues of } \tilde\mA \text{ lie in some } \cI_{j}] \\
	&\leq \sum_{j=1}^m \Pr[\text{at least two eigenvalues of } \tilde\mA \text{ lie in } \cI_{j}] \\
	&\leq Cm \left( \frac{1}{2\Delta} \cdot 2\gamma \cdot n \right)^2 \tag{\Cref{impthm:minami-bound}} \\
	&\leq \frac{4C\gamma n^2\normof{\mA}_2}{\Delta^2} \tag{\(\text{using that } m \leq \tsfrac{4\normof{\mA}_2}{\gamma}\)} \\
	&= \delta,
\end{align*}
where the last line holds if we fix \(\gamma = \frac{\Delta^2\delta}{4Cn^2\normof{\mA}_2}\).
That is, with probability at least \(1-\delta\), we know that \(\abs{\lambda_i(\tilde\mA) - \lambda_{i+1}(\tilde\mA)} \geq \frac{\Delta^2\delta}{4Cn^2\normof{\mA}_2}\) for all \(i=1,\ldots,n-1\).
Lastly, we take
\[
	\frac{\abs{\lambda_i(\tilde\mA) - \lambda_{i+1}(\tilde\mA)}}{\abs{\lambda_{i+1}(\tilde\mA)}}
	\geq \frac{\abs{\lambda_i(\tilde\mA) - \lambda_{i+1}(\tilde\mA)}}{\normof{\tilde\mA}_2}
	\geq \frac{\frac{\Delta^2\delta}{4Cn^2\normof{\mA}_2}}{2\normof{\mA}_2}
	= \frac{\Delta^2\delta}{8Cn^2\normof{\mA}_2^2},
\]
which completes the proof.
\end{proof}

\subsection{Proof of \texorpdfstring{\Cref{lem:perturbed-correctness}}{\ref*{lem:perturbed-correctness}}}
\label{app:perturbed-correctness}
We next show that a small enough perturbation of $\bv{A}$ suffices to give approximate SVD results for $\bv A$ itself.
\begin{replemma}{lem:perturbed-correctness}
Let \(\tilde\mA = \mA+\mD\) where \(\norm{\mD}_2 \leq \frac{\eps}{3n}\sigma_{k+1}(\mA)\) and \(\eps\in(0,1)\).
Fix any \(\mQ\in\bbR^{n \times k}\) with orthonormal columns \(\vq_1,\ldots,\vq_k\).
Then, with probability at least \(1-\delta\),
\begin{enumerate}
	\item If \(\abs{\vq_i^\intercal\tilde\mA\tilde\mA^\intercal\vq_i - \sigma_i(\tilde\mA)^2}\leq\eps\sigma_{k+1}(\tilde\mA)^2\), then \(\abs{\vq_i^\intercal\mA\mA^\intercal\vq_i - \sigma_i(\mA)^2}\leq 8\eps\sigma_{i}(\mA)^2.\)
	\item If \(\normof{\tilde\mA-\mQ\mQ^\intercal\tilde\mA}_2 \leq (1+\eps)\normof{\tilde\mA-\tilde\mA_k}_2\), then \(\normof{\mA-\mQ\mQ^\intercal\mA}_2 \leq (1+2\eps)\normof{\mA-\mA_k}_2.\)
	\item If \(\normof{\tilde\mA-\mQ\mQ^\intercal\tilde\mA}_F \leq (1+\eps)\normof{\tilde\mA-\tilde\mA_k}_F\), then \(\normof{\mA-\mQ\mQ^\intercal\mA}_F \leq (1+4\eps)\normof{\mA-\mA_k}_F.\)
\end{enumerate}
\end{replemma}
\begin{proof}~

\medskip
\noindent \textbf{Singular Value Guarantee.~}
First note that for any real \(a,b,c\) such that \(\abs{a-b} \leq c\) and \(c < b\), we have \(\abs{a^2-b^2}\leq3bc\).
This follows from expanding \((b-c)^2 < a^2 < (b+c)^2\) and applying the AMGM inequality.
Then note that
\(
	\abs{\sigma_i(\tilde\mA)-\sigma_i(\mA)}\leq\normof{\tilde\mA-\mA}_2=\normof{\mD}_2
\).
We then find that for $i \le k+1$,
\[
	\abs{\sigma_i^2(\tilde\mA) - \sigma_i^2(\mA)}
	\leq 3\sigma_i(\mA)\normof{\mD}_2
	\leq \eps\sigma_{i}(\mA)\sigma_{k+1}(\mA)
	\leq \eps\sigma_i(\mA)^2.
\]
Similarly note that \(\abs{\normof{\tilde\mA\vq_i}_2 - \normof{\mA\vq_i}_2} \leq \normof{\mD}_2\) and \(\normof{\tilde\mA\vq_i}_2 \leq 2 \sigma_i(\tilde\mA) \leq 4 \sigma_i(\mA)\), so we have
\begin{align*}
	\abs{\normof{\tilde\mA\vq_i}_2^2 - \normof{\mA\vq_i}_2^2}
	\leq 3\normof{\tilde\mA\vq_i}_2\normof{\mD}_2
	\leq 4\eps\sigma_i(\mA)^2.
\end{align*}
Which completes this part by triangle inequality:
\begin{align*}
	\abs{\vq_i^\intercal\mA\mA^\intercal\vq_i - \sigma_i(\mA)^2}
	&\leq \abs{\vq_i^\intercal\tilde\mA\tilde\mA^\intercal\vq_i - \sigma_i(\mA)^2} + 4\eps\sigma_i(\mA)^2 \\
	&\leq \abs{\vq_i^\intercal\tilde\mA\tilde\mA^\intercal\vq_i - \sigma_i(\tilde\mA)^2} + 7\eps\sigma_i(\mA)^2 \\
	&\leq 8\eps\sigma_i(\mA)^2.
\end{align*}

\medskip \noindent
\textbf{Spectral Norm Guarantee.~}
Here, first note that \(\normof{\tilde\mA-\tilde\mA_k}_2 = \sigma_{k+1}(\tilde\mA) \leq (1+\eps)\sigma_{k+1}(\mA)\) by the prior analysis on the singular value guarantee. 
Next, we use the fact that \(\mI-\mQ\mQ^\intercal\) is a projection to simplify
\[
	\normof{\tilde\mA-\mQ\mQ^\intercal\tilde\mA}_2
	= \normof{(\mI-\mQ\mQ^\intercal)(\mA+\mD)}_2
	\geq \normof{(\mI-\mQ\mQ^\intercal)\mA}_2 - \normof{\mD}_2,
\]
and since \(\normof{\mD}_2 \leq \eps \sigma_{k+1}(\mA)\), we have
\(
	\normof{\mA-\mQ\mQ^\intercal\mA}_2
	\leq \normof{\tilde\mA-\mQ\mQ^\intercal\tilde\mA}_2 + \normof{\mD}_2
	\leq (1+2\eps)\sigma_{k+1}(\mA),
\)
which completes this part of the lemma.

\medskip \noindent
\textbf{Frobenius Norm Guarantee.~}
Here, first note that \(\normof{\tilde\mA-\tilde\mA_k}_F^2 \leq (1+\eps)\normof{\mA-\mA_k}_F^2\), since
\[
	\normof{\tilde\mA-\tilde\mA_k}_F^2
	= \sum_{i=k+1}^n \sigma_i^2(\tilde\mA)
	\leq \sum_{i=k+1}^n \left(\sigma_i(\mA) + \normof{\mD}_2\right)^2
	\leq \normof{\mA-\mA_k}_F^2 + \sum_{i=k+1}^n (\normof{\mD}_2^2 + 2\sigma_i(\mA)\normof{\mD}_2),
\]
where we can further upper bound
\[
	\sum_{k+1}^n (\normof{\mD}_2^2 + 2\sigma_i(\mA)\normof{\mD}_2)
	\leq n(\tsfrac{\eps\sigma_{k+1}(\mA)}{3n})^2 + 2n\sigma_{k+1}(\mA)\tsfrac{\eps\sigma_{k+1}(\mA)}{3n}
	\leq \eps\sigma_{k+1}^2(\mA)
	\leq \eps\normof{\mA-\mA_k}_F^2.
\]
Next, using the fact that \(\mP=\mI-\mQ\mQ^\intercal\) is a projection matrix, we simplify
\[
	\normof{\tilde\mA-\mQ\mQ^\intercal\tilde\mA}_F
	= \normof{\mP(\mA+\mD)}_F
	\geq \normof{\mP\mA}_F - \normof{\mP\mD}_F,
\]
and since \(\normof{\mP\mD}_F \leq \normof{\mP}_F\normof{\mD}_2 \leq \sqrt d \frac{\eps}{3n}\sigma_{k+1}(\mA) \leq \eps\normof{\mA-\mA_k}_F\), we have
\[
	\normof{\mA-\mQ\mQ^\intercal\mA}_F
	\leq \normof{\tilde\mA-\mQ\mQ^\intercal\tilde\mA}_F + \normof{\mP\mD}_F
	\leq ((1+\eps)^2+\eps) \normof{\mA-\mA_k}_F,
\]
which completes the proof since \(((1+\eps)^2+\eps) \leq 1+4\eps\) for \(\eps\in(0,1)\).
\end{proof}

\section{Krylov Analysis with Small Blocks}
\label{app:small-block-analysis}

This section proves \Cref{thm:block-vec-guarantee}.
We first define the starting matrix that we will simulate block Krylov iteration on, analogously to what we use in \Cref{sec:gap-dep-analysis}:
\[
	\mS_r \defeq \bmat{\mG & \mA^2\mG & \mA^4\mG & \ldots & \mA^{2(r-1)}\mG}
	\hspace{1cm}
	\mK \spaneq \bmat{\mS_r & \mA^2\mS_r & \mA^4\mS_r & \ldots & \mA^{2q}\mS_r}
\]
where \(\ell = r b\) is the simulated block size (so we assume the integer \(r\) has \(rb > k\)) and where \(q=t-r+1\) denotes the number of simulated block-Krylov iterations run.
Our proof will set \(r=k-b+1\).
Notably, this means that \(\mS_r\) can have more than \(k\) columns, which is not allowed by the definition of \((k,L)\)-good in \Cref{def:lgood}.
So, we first present a generalization of \Cref{def:lgood} that also suffices for convergence under \Cref{impthm:poly-eps-low-rank} and \Cref{impthm:spectral-decay-low-rank}. In particular, it suffices for \(\mS_r\) to contain a $n \times k$ size \((k,L)\)-good matrix within its span. The Krylov subspace generated by this matrix will be contained in the subspace generated by \(\mS_r\), and thus any guarantees that hold for it apply to \(\mS_r\) as well. See \Cref{app:musco_explanation} for a formal argument.

\begin{definition}[$(k,L)$-good Starting Matrix (Generalized)]
	\label{def:lgood-small-block}
		Let $\mA \in \R^{n \times d}$ be a matrix with top \(k\) left singular vectors $\mU_k \in \R^{n \times k}$.
	A matrix $\mB\in \R^{n \times \ell}$ is a $(k,L)$-good starting matrix for $\mA$ if  \(\normof{(\mU_k^\intercal\mQ)^{-1}}_2^2 \leq L\) for some orthonormal \(\mQ\in\bbR^{n \times k}\) whose columns lie in \(\colspan(\mB)\).
\end{definition}
Observe that when the starting block has exactly \(\ell=k\) columns, \Cref{def:lgood-small-block} matches \Cref{def:lgood} exactly.
To prove the \((k,L)\)-good guarantee for small block methods, we first note an equivalent formulation of the generalized \((k,L)\)-good guarantee:

\begin{lemma}
\label{lem:generalized-l-good-equiv}
\(\mB\in\bbR^{n \times \ell}\) is \((k,L)\)-good for $\mA \in \R^{n \times n}$ if and only if there exists a matrix \(\mM\in\bbR^{\ell \times k}\) with \(\mU_k^\intercal\mB\mM=\mI\) and \(\normof{\mB\mM}_2^2 \leq L\).
\end{lemma}
\begin{proof}
Suppose we are given such an \(\mM\).
Since \(\colspan(\mB\mM)\) is a \(k\)-dimensional subspace of \(\colspan(\mB)\), we can let \(\mQ\) be an orthonormal basis for \(\mB\mM\).
Then we have \(\mB\mM = \mQ\mX\) for some invertible \(\mX\in\bbR^{k \times k}\).
Since \(\mU_k^\intercal\mQ\mX= \mU_k^\intercal\mB\mM =\mI\), we have \(\mX^{-1} = \mU_k^\intercal\mQ\).
 And so, we have \(\normof{(\mU_k^\intercal\mQ)^{-1}}_2^2 = \normof{\mX}_2^2 = \normof{\mQ\mX}_2^2 = \normof{\mB\mM}_2^2 \leq L\).

In the other direction, we are given \mQ which spans a subspace of \(\colspan(\mB)\).
So, for any invertible matrix \(\mX\in\bbR^{k \times k}\), we can write \(\mB\mM=\mQ\mX\) for some \(\mM\in\bbR^{\ell \times k}\).
If we take \(\mX=(\mU_k^\intercal\mQ)^{-1}\), then we have \(\mU_k^\intercal\mB\mM = \mU_k^\intercal\mQ\mX = \mI\) and \(\normof{\mB\mM}_2^2 = \normof{\mQ\mX}_2^2 = \normof{\mX}_2^2 = \normof{(\mU_k^\intercal\mQ)^{-1}}_2^2 \leq L\).
\end{proof}

With this foundation in place, we prove the following guarantee on $\bv{S}_r$:
\begin{reptheorem}{thm:block-low-rank-approx}
Fix any PSD matrix \(\mA\in\bbR^{n \times n}\) with singular values $\sigma_1 \geq \ldots \geq \sigma_n$, \(k \in[n]\), and \(b\in[k]\).
Let \(r = k-b+1\).
Let $\mG\in\bbR^{n \times b}$ be a matrix i.i.d. \(\cN(0,1)\) entries, and let $\mS_r = \bmat{\mG & \mA^2\mG & \mA^4\mG & \ldots & \mA^{2(r-1)}\mG}$.
For any $\delta \in (0,1)$, with probability at least $(1-\delta)$, $\mS_r$ is a $(k,L)$-good starting matrix for $\mA$ for $L = \tsfrac{cb^2k^2n \log(\nicefrac 1\delta)}{\delta^2 g_{min,b}^{4(k-b)}}$.
Here $c$ is a fixed constant.
\end{reptheorem}
\begin{proof}
We prove the generalized \((k,L)\)-good bound by constructing a matrix \(\mM\in\bbR^{\ell \times k}\) with $\bv{U}_k^\intercal \bv{S}_r  \bv{M} = \bv{I}$ and then applying \Cref{lem:generalized-l-good-equiv}.
Consider first the matrix \(\mU^\intercal\mS_r\mM\). Writing the SVD \(\mA=\mU\mSigma\mU^\intercal\), we can define \(\hat\mG \defeq \mU^\intercal\mG \in \bbR^{n \times b}\), which is distributed as a iid \(\cN(0,1)\) matrix because of the rotational invariance of the Gaussian.
We can then write \(\mU^\intercal\mA^{2i}\mG = \mU^\intercal\mU\mSigma^{2i}\mU^\intercal \mG = \mSigma^{2i}\hat\mG\). Thus,
\[
	\mU^\intercal\mS_r
	= \bmat{
		\hat\mG & \mSigma^2\hat\mG & \mSigma^4\hat\mG & \ldots & \mSigma^{2(r-1)}\hat\mG
	}.
\]
If we let \(\hat\vg_i\) be the \(i^{th}\) column of \(\hat\mG\) and permute the columns of \(\mU^\intercal\mG\), we can write \(\mU^\intercal\mS_r\) as
\[
	\mU^\intercal\mS_r\mP
	= \sbmat{
		\hat\vg_1 & \mSigma^2\hat\vg_1 & \ldots & \mSigma^{2(r-1)}\hat\vg_1
		&|& \hat\vg_2 & \mSigma^2\hat\vg_2 & \ldots & \mSigma^{2(r-1)}\hat\vg_2
		&|& \ldots
		&|& \hat\vg_b & \mSigma^2\hat\vg_b & \ldots & \mSigma^{2(r-1)}\hat\vg_b
	}.
\]
where \mP is the permutation matrix that reorders the columns as such.
So, for some vector \(\vm_i\), we can decompose \(\mP^\intercal\vm_i = \sbmat{\vc_1 \\ \ldots \\ \vc_b}\) for \(\vc_j\in\bbR^{r}\), and let \(p_{_{i,j}}(t)\) be the degree \(r-1\) polynomial with coefficients \(\vc_j\).
Then, we get
\begin{align}
	\mU^\intercal\mS_r\vm_i
	= \mU^\intercal\mS_r\mP\mP^\intercal\vm_i
	= p_{_{i,1}}(\mSigma^2)\hat\vg_1
	+ p_{_{i,2}}(\mSigma^2)\hat\vg_2
	+ \ldots
	+ p_{_{i,b}}(\mSigma^2)\hat\vg_b.
	\label{eq:small-block-expansion}
\end{align}
And so, the matrix \(\mU^\intercal\mS_r\mM\) is the concatenation of \(k\) such vectors \(\mU^\intercal\mS_r\vm_1,\ldots,\mU^\intercal\mS_r\vm_k\).

Observe that \(\mU_k^\intercal\mS_r\mM\) is just the top \(k\) rows of \(\mU^\intercal\mS_r\mM\). So to apply \Cref{lem:generalized-l-good-equiv}, we need to find polynomials \(p_{1,1},\ldots,p_{k,b}\) which make the top \(k\) rows of \(\mU^\intercal\mS_r\mM\) into an identity matrix.
I.e., we need to show \(\mU_k^\intercal\mS_r\vm_i = \ve_i\) for \(i=1,\ldots,k\).
We do this by first designing the degree \(k-b\) filter polynomials \(f_1,\ldots,f_k\) by
\[
	f_i(\sigma_j^2) = \begin{cases}
		1 & j=i \\
		0 & j\notin\cN_{i} \cup \{i\}
	\end{cases}
\]
That is, \(f_i\) is 1 at \(\sigma_i^2\) and is 0 on the squares of the \(k-b+1\) singular values furthest from \(\sigma_i\).
The squares are there because the Krylov subspace uses polynomial of \(\mSigma^2\).
We then take 
\[
	p_{i,j}(t) \defeq f_i(t) z_{i,j}
\]
for some values of \(z_{i,j}\) which will be specified later.
Note these polynomials are all degree \(r-1=k-b\), which is why we constrain \(r=k-b+1\).
If we plug this into \Cref{eq:small-block-expansion}, we get
\begin{align*}
	\mU^\intercal\mS_r\vm_i
	&= \pij{i}{1}(\mSigma^2)\,\hat\vg_1 + \pij{i}{2}(\mSigma^2)\,\hat\vg_2 + \ldots + \pij{i}{b}(\mSigma^2)\,\hat\vg_b \\
	&= f_i(\mSigma^2)\,\left(\hat\vg_1z_{i,1} + \hat\vg_2z_{i,2} + \ldots + \hat\vg_b z_{i,b}\right) \\
	&= f_i(\mSigma^2)\,\hat\mG\vz_i,
\end{align*}
where \(\vz_i = \bmat{z_{i,1} & \ldots & z_{i,b}}^\intercal\).
So, we need to find a choice of \(\vz_i\in\bbR^b\) such that \(\mU_k^\intercal\mS_r\vm_i = \ve_i\).
Letting \(\hat\mG_k\in\bbR^{k \times b}\) be the top \(k\) rows of \(\hat\mG\), we write
\[
	\mU_k^\intercal\mS_r\vm_i
	= f_i(\mSigma_k^2)\,\hat\mG_k\vz_i.
\]
Recalling that \(f_i\) is a filter polynomial, we know that \([\mU_k^\intercal\mS_r\vm_i]_j = 0\) for \(j\notin\cN_{i}\cup\{i\}\).
So, we just need to pick \(\vz_i\in\bbR^b\) such that \([\mU_k^\intercal\mS_r\vm_i]_j = 0\) for \(j\in\cN_{i}\) and \([\mU_k^\intercal\mS_r\vm_i]_i = 1\).
That is, we need to pick \(\vz_i\) such that the product \(\hat\mG_k\vz_i\) is zero for the \(b-1\) rows in \(\cN_i\) and is 1 for row \(i\).
This is just a linear system, so we pick the unique \(\vz_i=[\hat\mG_k]_{\cN_{i}\cup\{i\}}^{-1}\ve_i\), where \([\hat\mG_k]_{\cN_{i}\cup\{i\}}\) just selects the \(b\) rows indexed by \(\cN_{i}\cup\{i\}\).

Once we have these \(\vz_i\) vectors, we get \(\mU_k^\intercal\mS_r\vm_i = f_i(\mSigma_k^2)\,\hat\mG_k\vz_i = \ve_i\), and so \(\mU_k^\intercal\mS_r\mM=\mI\), as required by \Cref{lem:generalized-l-good-equiv}.
Now we just have to bound \(\normof{\mS_r\mM}_2^2\).
We do this by bounding \(\normof{\mS_r\mM}_2^2 = \normof{\mU^\intercal\mS_r\mM}_2^2 \leq \normof{\mU^\intercal\mS_r\mM}_F^2 \leq \sum_{i=1}^k \normof{\mU^\intercal\mS_r\vm_i}_2^2\).

Let \(\mSigma_{-k} \in\bbR^{(n-k) \times (n-k)}\) be the bottom \(n-k\) singular values of \mA, and let \(\hat\mG_{-k}\in\bbR^{(n-k) \times b}\) bottom \(n-k\) rows of \(\hat\mG\).
Then, we can decompose
\[
	\mU^\intercal\mS_r\vm_i
	= f_i(\mSigma^2) \, \hat\mG \vz_i
	= \bmat{
		f_i(\mSigma_k^2) \hat\mG_k\vz_i \\
		f_i(\mSigma_{-k}^2) \hat\mG_{-k}\vz_i
	}
	= \bmat{
		\ve_i \\
		f_i(\mSigma_{-k}^2) \, \hat\mG_{-k}[\hat\mG_{k}]_{\cN_i\cup\{i\}}^{-1}\ve_i
	}.
\]
We now bound the  lower elements.
First, we look at \(\hat\mG_{-k}[\hat\mG_{k}]_{\cN_i\cup\{i\}}^{-1}\ve_i\), which is the product of a \((n-k) \times b\) Gaussian matrix and inverse of an independent \(b \times b\) Gaussian matrix.
Since these matrices are independent, we can directly bound
\begin{align}
	\normof{\hat\mG_{-k}[\hat\mG_{k}]_{\cN_i\cup\{i\}}^{-1}\ve_i}_2
	\leq \normof{\hat\mG_{-k}}_2 \cdot \frac1{\sigma_{min}([\hat\mG_k]_{\cN_i\cup\{i\}})} \cdot \normof{\ve}_2
	\leq O\left( \frac{b\sqrt{kn}}{\delta} \sqrt{\log\left(\frac1\delta\right)} \right),
	\label{eq:small-block-lower-norm}
\end{align}
where we use the bound \(\normof{\mG_{-k}}_2 \leq O(\sqrt{(n-k)b\ln(\frac1\delta)})\) from Equation (2.3) of \cite{RudelsonVershynin:2010} and the bound \(\sigma_{min}([\hat\mG_k]_{\cN_i\cup\{i\}}) \geq \Omega(\frac{\delta}{k\sqrt{b}})\).
This latter bound comes from a union-bound argument using the fact that any square Gaussian \(\tilde\mG\in\bbR^{b \times b}\) has \(\Pr[\sigma_{min}(\tilde\mG) \geq \Omega(\frac{\delta}{\sqrt b})] \geq 1-\delta\) \cite{huang2020remark}, and that \([\hat\mG_k]_{\cN_1\cup\{1\}}, \ldots, [\hat\mG_k]_{\cN_k\cup\{k\}}\) are all square Gaussian matrices.
Next, we bound \(f_i(\mSigma_{-k}^2)\) by writing \(f_i\) as a Lagrange interpolating polynomial:
\[
	f_i(t) = \prod_{j\in[k] \setminus \cN_i, j \neq i} \frac{t - \sigma_j^2}{\sigma_i^2-\sigma_j^2}.
\]
We can then use our gap assumption to bound \(f_i\) on all singular values below \(\sigma_k\).
That is, for \(0\leq t \leq \sigma_k\) we have
\[
	\abs{f_i(t)}
	\leq \prod_{j\in[k] \setminus \cN_i, j \neq i} \left|\frac{\sigma_j^2}{\sigma_i^2 - \sigma_j^2}\right|
	\leq \prod_{j\in[k] \setminus \cN_i, j \neq i} \left|\frac{\sigma_j}{\sigma_i - \sigma_j}\right|^2
	\leq \frac1{g_{min,b}^{2(k-b)}}.
\]
Putting this altogether, we then find that
\begin{align*}
	\normof{f_i(\mSigma_{-k}^2) \, \hat\mG_{-k}[\hat\mG_{k}]_{\cN_i\cup\{i\}}^{-1}\ve_i}_2^2
	&\leq \normof{f_i(\mSigma_{-k}^2)}_2^2 ~~ \normof{\hat\mG_{-k}[\hat\mG_{k}]_{\cN_i\cup\{i\}}^{-1}\ve_i}_2^2 \\
	&\leq O\left( \frac{b^2kn}{g_{min,b}^{4(k-b)}\delta^2} \log\left(\frac1\delta\right) \right).
\end{align*}
And therefore
\[
	\normof{\mS_r\mM}_2^2
	\leq \normof{\mU^\intercal\mS_r\mM}_F^2
	\leq O\left(1 + \frac{b^2k^2n}{g_{min,b}^{4(k-b)}\delta^2} \log\left(\frac1\delta\right) \right),
\]
which completes the proof by \Cref{lem:generalized-l-good-equiv}.
\end{proof}


\section{Convergence Results from \texorpdfstring{\cite{MuscoMusco:2015}}{[Musco Musco '15]}}
\label{app:musco_explanation}

In this section, we show how \Cref{impthm:poly-eps-low-rank} and \Cref{impthm:spectral-decay-low-rank} follow from \cite{MuscoMusco:2015}.
We prove the result using a generalization of a $(k,L)$-good starting block (\Cref{def:lgood}), which is useful in the small block analysis of \Cref{app:small-block-analysis}:

\begin{repdefinition}{def:lgood-small-block}[$(k,L)$-good Starting Matrix (Generalized)]
	Let $\mA \in \R^{n \times d}$ be a matrix with top \(k\) left singular vectors $\mU_k \in \R^{n \times k}$.
	A matrix $\mB\in \R^{n\times \ell}$ is a $(k,L)$-good starting matrix for $\mA$ if  \(\normof{(\mU_k^\intercal\mQ)^{-1}}_2^2 \leq L\) for some orthonormal \(\mQ\in\bbR^{n \times k}\) that lies in \(\colspan(\mB)\).
\end{repdefinition}

Note that if \(\mB\) has exactly \(k\) columns then this generalized definition exactly matches \Cref{def:lgood}.
This is the case for the analysis of single vector Krylov in \Cref{sec:gap-dep-analysis}, where we take \(\mB=\mS_k\) with exactly \(k\) columns and show that \(\mS_k\) is \((k,L)\)-good.
In \Cref{app:small-block-analysis}, we show an equivalent formulation which is easier to use in our analysis:

\begin{replemma}{lem:generalized-l-good-equiv}
\(\mB\in\bbR^{n \times \ell}\) is \((k,L)\)-good for \mA if and only if there exists a matrix \(\mM\in\bbR^{\ell \times k}\) with \(\mU_k^\intercal\mB\mM=\mI\) and \(\normof{\mB\mM}_2^2 \leq L\).
\end{replemma}

Given this definition, we will show that \Cref{impthm:poly-eps-low-rank} and \Cref{impthm:spectral-decay-low-rank} follow from \cite{MuscoMusco:2015}.
Before diving in, we first state a guarantee on the polynomials used in \cite{MuscoMusco:2015}.
They use polynomials of the form \(p(x) \defeq \frac{(1+\gamma)\alpha}{T_q(1+\gamma)} T_q(\frac x\alpha)\) for some \(\alpha>0\), \(\gamma\in(0,1)\), and \(q\in\bbN\), and where \(T_q(x)\) is the degree \(q\) Chebyshev polynomial of the first kind.
We first show that such polynomials are monotonic for large enough \(x\):
\begin{lemma}
\label{lem:cheby-weak-monotonic}
Let \(p(x) = \frac{(1+\gamma)\alpha}{T_q(1+\gamma)} T_q(\frac x\alpha)\) where \(\alpha > 0\) and \(\gamma\in(0,1)\).
Then, \(\max_{x\in[0,\alpha]} p(x) = p(\alpha)\) and \(p(x)\) is monotonically increasing on \((\alpha,\infty)\).
\end{lemma}
\begin{proof}
The result follows from two well-known properties of \(T_q\): that \(T_q\) is monotonically increasing on \((1,\infty)\) and that \(\max_{t\in[0,1]} T_q(t) = T_q(1) = 1\).
Since \(\gamma > 0\), we know that \(T_q(1+\gamma) > T_q(1) = 1\), so we have \(\frac{(1+\gamma)\alpha}{T_q(1+\gamma)} > 0\).
Therefore, we get that \(p(x)\) is monotonically increasing on \((\alpha,\infty)\), and that \(\max_{x\in[0,\alpha]} p(x) = p(\alpha)\).
\end{proof}

In order to relate the generalized \((k,L)\)-good definition to the convergence analysis of \cite{MuscoMusco:2015}, we rely on a slight generalization of a theorem used in the appendix of \cite{MuscoMusco:2015}:

\begin{importedlemma}[Lemma 48 of \cite{Woodruff:2014}]
\label{implem:generalized-woodruff}
Let \(\mP = \mP\mU\mU^\intercal + \mE \in \bbR^{n \times n}\) be a low-rank factorization of $\mA \in \R^{n\times n}$, with \(\mU\in\bbR^{n \times k}\) and \(\mU^\intercal\mU=\mI_k\).
Let \(\mB\in\bbR^{n \times \ell}\) (\(\ell \geq k\)) be any matrix with \(\rank(\mU^\intercal\mS) = \rank(\mU) = k\).
Let \(\mM\in\bbR^{\ell \times k}\) with \(\mU^\intercal\mB\mM=\mI_k\).
Let \(\mC=\mA\mS \in \bbR^{n \times \ell}\).
Then,
\[
	\normof{\mP-\Pi_{\mC,k}(\mP)}_F^2 \leq \normof{\mE}_F^2 + \normof{\mE\mB\mM}_F^2.
\]
Here, \(\Pi_{\mC,k}(\mP) = \mY_{opt}\mY_{opt}^\intercal\mA \in \bbR^{n\times n}\) is the best rank \(k\) approximation to \mA in the column space of \mC
So, \(\mY_{opt}\in\bbR^{n \times k}\) is an orthogonal matrix that lies in the column span of \mC.
\end{importedlemma}
Lemma 48 from \cite{Woodruff:2014} is stated with \(\mM=(\mU^\intercal\mB)^+\), but the proof only requires that \(\mU^\intercal\mB\mM=\mI\).
We now prove the imported theorems:

\begin{repimportedtheorem}{impthm:poly-eps-low-rank}[Theorem 1 of \cite{MuscoMusco:2015}]
Let \(\mB\in\bbR^{n \times \ell}\) be any \((k,L)\)-good starting matrix (\Cref{def:lgood-small-block}) matrix for \mA.
If we run Block Krylov iteration (\Cref{alg:block-krylov}) for \(q = O(\frac{1}{\sqrt{\eps}} \log(\frac{nL}{\eps}))\) iterations with starting block \mB, then the output  \(\mQ\in\bbR^{n \times k}\) satisfies
\begin{align*}
	\normof{\mA-\mQ\mQ^\intercal\mA}_{\xi} &\leq (1+\eps) \normof{\mA-\mA_k}_{\xi} & &\text{and} & \left| \vq_i^\intercal \mA\mA^\intercal \vq_i -\sigma_i(\mA)^2 \right| &\leq \eps\sigma_{k+1}(\mA)^2.
\end{align*}
\end{repimportedtheorem}
\begin{proof}
To recover this guarantee from \cite{MuscoMusco:2015}, we recover Properties 1-3 of their Lemma 9.
We start by recovering Property 1.
We let \(p_1(x) = \frac{(1+\gamma)\alpha}{T_q(1+\gamma)} T_q(\frac x\alpha)\) where \(\alpha = \sigma_{k+1}(\mA)\), \(\gamma = \frac\eps2\), and \(q = O(\frac1{\sqrt\eps}\log(\frac{nL}{\eps}))\).
Let \(\mM \in \bbR^{\ell \times k}\) be the matrix guaranteed to exist by \Cref{lem:generalized-l-good-equiv}.
We then instantiate \Cref{implem:generalized-woodruff} with \(\mP = p_1(\mA)\), \(\mB = \mB\), \(\mM=\mM\), and \(\mU = \mU_k\) being the top \(k\) singular vectors of \mA.
By \Cref{lem:cheby-weak-monotonic}, since \(\alpha=\sigma_{k+1}(\mA)\), we know that the top \(k\) singular vectors of \(\mA\) are also the top \(k\) singular vectors of \(p_1(\mA)\), and therefore that \(\mP\mU\mU^\intercal = p_1(\mA)_k\).
We then get that some orthonormal \(\mY_1\in\bbR^{n \times k}\) in the span of \(p(\mA)\mB\) has
\begin{align*}
	\normof{p_1(\mA) - \mY_1\mY_1^\intercal p_1(\mA)}_F^2
	&\leq \normof{p_1(\mA)-p_1(\mA)_k}_F^2 + \normof{(p_1(\mA)-p_1(\mA)_k)\mB\mM}_F^2 \\
	&\leq \normof{p_1(\mA)-p_1(\mA)_k}_F^2 + \normof{p_1(\mA)-p_1(\mA)_k}_F^2\normof{\mB\mM}_2^2 \\
	&\leq (L+1) \normof{p_1(\mA)-p_1(\mA)_k}_F^2 \\
	&= (L+1) \sum_{i=k+1}^n p(\sigma_i(\mA))^2 \\
	&\leq (L+1)n \cdot \frac{4\sigma_{k+1}(\mA)^2}{2^{2q\sqrt\gamma}} \tag{Lemma 5 of \cite{MuscoMusco:2015}} \\
	&\leq \frac\eps2\sigma_{k+1}(\mA)^2, \numberthis \label{eq:mm15-eq5}
\end{align*}
where the last line uses \(q \geq \frac1{2\sqrt\gamma} \log_2(\frac{8(L+1)n}{\eps}) = O(\frac{1}{\sqrt\eps} \log(\frac{Ln}{\eps}))\).
The inequality \(\normof{p_1(\mA)-\mY_1\mY_1^\intercal p_1(\mA)} \leq \frac\eps2\sigma_{k+1}(\mA)^2\) is exactly Equation (5) on page 11 of \cite{MuscoMusco:2015}.
Once Equation (5) is achieved, the rest of the proof of Property 1 then follows without any alteration.

We next move onto proving properties 2 and 3.
The proofs of properties 2 and 3 in \cite{MuscoMusco:2015} both involve defining the matrix \(\mA_{outer}\).
If \(\mA=\mU\mSigma\mU^\intercal\) is the SVD of \mA, then \(\mA_{outer} \defeq \mU\mSigma_{outer}\mU^\intercal\) where \(\mSigma_{outer}\) contains all the singular values of \mA with either \(\sigma_i(\mA) \geq \sigma_k(\mA)\) or \(\sigma_i(\mA) < \frac{1}{1+\eps/2}\sigma_k(\mA)\).
All the other singular values are set to equal zero.
Crucially, the top \(k\) singular vectors of \mA are still the top \(k\) singular vectors of \(\mA_{outer}\).

We let \(p_2(x) = \frac{(1+\gamma)\alpha}{T_q(1+\gamma)} T_q(\frac x\alpha)\) where \(\alpha = \frac1{1+\eps/2}\sigma_{k+1}(\mA)\), \(\gamma = \frac\eps2\), and \(q = O(\frac1{\sqrt\eps}\log(\frac{nL}{\eps}))\).
We next instantiate \Cref{implem:generalized-woodruff} with \(\mP = p_2(\mA_{outer})\), \(\mB = \mB\), \(\mM=\mM\), and \(\mU = \mU_k\).
By \Cref{lem:cheby-weak-monotonic}, since \(\alpha\leq\sigma_{k}(\mA_{outer})\), we know that the top \(k\) singular vectors of \(\mA_{outer}\) are also the top \(k\) singular vectors of \(p_2(\mA_{outer})\), and therefore that \(\mP\mU\mU^\intercal = p_2(\mA_{outer})_k\).
We then let \(\mY_{outer}\in\bbR^{n \times k}\) be the orthogonal basis that constructs \(\Pi_{\mC,k}(\mA)\), so that
\begin{align*}
	\normof{p_2(\mA_{outer}) - \mY_{outer}\mY_{outer}^\intercal p_2(\mA_{outer})}_F^2
	&\leq \normof{p_2(\mA_{outer})-p_2(\mA_{outer})_k}_F^2 \\&\hspace{2cm}+ \normof{(p_2(\mA_{outer})-p_2(\mA_{outer})_k)\mB\mM}_F^2 \\
	&\leq (L+1)\normof{p_2(\mA_{outer})-p_2(\mA_{outer})}_F^2 \\
	&\leq \frac\eps2 \sigma_{k+1}(\mA_{outer})^2.
\end{align*}
Where the inequalities follow from the same logic as earlier, for \Cref{eq:mm15-eq5}.
We again find ourselves at Equation (5) of \cite{MuscoMusco:2015}, and follow the rest of the proof on page 11 to get the following guarantee:
\[
	\normof{(\mA_{outer})_k}_F^2 - \normof{\mY_{outer}\mY_{outer}^\intercal(\mA_{outer})_k}_F^2 \leq \frac{\eps}{2}\sigma_{k+1}(\mA_{outer})^2.
\]
Since \((\mA_{outer})_k = \mA_k\), the above inequality then recovers Equation (10) of \cite{MuscoMusco:2015}.
Given this proof of Equation (10), the rest of the proof of Property 2 holds as written.
Equation (10) is also used on page 14 of \cite{MuscoMusco:2015} to prove Property 3.
The rest of the proof of Property 3 also holds without alteration given Equation (10).

Overall, we have recovered the proofs of Properties 1-3 of Lemma 9 of \cite{MuscoMusco:2015}.
That is, we have shown that Lemma 9 holds for block Krylov iteration starting from \((k,L)\)-good starting block \mB with \(q=O(\frac{1}{\sqrt\eps}\log(\frac{nL}{\eps}))\) iterations.
The proofs in Section 6.2 of \cite{MuscoMusco:2015} then show that Properties 1-3 suffice to achieve the spectral, Frobenius, and singular value guarantees.
\end{proof}

We now move onto \Cref{impthm:spectral-decay-low-rank} from \cite{MuscoMusco:2015}, which depends on spectral decay.
Note there are two different variables \(\ell\) and \(\ell_0\) in this context.
In order to perform rank-\(k\) approximation, we want to recover a convergence bound in terms of \(g_{k\rightarrow\ell}\) for some \(\ell \geq k\).
Our starting block \(\mB\in\bbR^{n \times \ell_0}\) has \(\ell_0 \geq \ell\) columns and is an \((\ell,L)\)-good starting block.
We need to consider this case because, when running block size \(b \leq k\) Krylov iteration, the analysis in \Cref{app:small-block-analysis} considers a simulated starting block \(\mS_r\) which uses \(\ell_0 \approx b(\ell-b) \geq \ell\) columns to simulate block size \(\ell\) Krylov iteration.

\begin{repimportedtheorem}{impthm:spectral-decay-low-rank}[Theorem 13 of \cite{MuscoMusco:2015}]
	Let \(\mB\in\bbR^{n \times \ell_0}\) be any \((\ell,L)\)-good starting matrix (\Cref{def:lgood-small-block}) matrix for \mA, for some \(\ell \geq k\).
	If we run Block Krylov iteration (\Cref{alg:block-krylov}) for \(q = O(\frac{1}{\sqrt{g_{k \rightarrow \ell}}} \log(\frac{nL}{\eps}))\) iterations with starting block \mB, where \(g_{k \rightarrow \ell} = \frac{\sigma_{k}-\sigma_{\ell+1}}{\sigma_{k}}\), then the output  \(\mQ\in\bbR^{n \times k}\) satisfies
\begin{align*}
	\normof{\mA-\mQ\mQ^\intercal\mA}_{\xi} &\leq (1+\eps) \normof{\mA-\mA_k}_{\xi} & &\text{and} & \left| \vq_i^\intercal \mA\mA^\intercal \vq_i -\sigma_i(\mA)^2 \right| &\leq \eps\sigma_{k+1}(\mA)^2.
\end{align*}
\end{repimportedtheorem}
\begin{proof}
We again show that Properties 1-3 of Lemma 9 of \cite{MuscoMusco:2015} hold, but now with \(q = \frac{1}{\sqrt{g_{k\rightarrow\ell}}} \log(\frac{nL}{\eps})\).
Following Section 7 of \cite{MuscoMusco:2015}, we will prove Property 1 for all \(l\in[k]\), which in turn implies that Properties 2 and 3 hold.
To begin, let \(p_3(x) = \frac{(1+\gamma)\alpha}{T_q(1+\gamma)} T_q(\frac x\alpha)\) where \(\alpha = \sigma_{\ell+1}(\mA)\), \(\gamma = g_{k\rightarrow\ell}\), and \(q = O(\frac1{\sqrt{g_{k\rightarrow\ell}}}\log(\frac{nL}{\eps}))\).
We let \(\mM\in\bbR^{\ell_0 \times \ell}\) be the matrix guaranteed to exist by \Cref{lem:generalized-l-good-equiv}.
Then, noticing that the top \(\ell\) singular vectors of \mA are also the top \(\ell\) eigenvectors of \(p_3(\mA)\), we again appeal to \Cref{implem:generalized-woodruff} with \(\mP=p_3(\mA)\), \(\mB=\mB\), \(\mM=\mM\), and \(\mU=\mU_k\).
We get that some orthogonal \(\mY_3\in\bbR^{n \times \ell}\) in the span of \(p_3(\mA)\mB\) has:
\begin{align*}
	\normof{p_3(\mA) - \mY_3\mY_3^\intercal p_3(\mA)}_F^2
	&\leq \normof{p_3(\mA)-p_3(\mA)_\ell}_F^2 + \normof{(p_3(\mA)-p_3(\mA)_\ell)\mB\mM}_F^2 \\
	&\leq \normof{p_3(\mA)-p_3(\mA)_\ell}_F^2 + \normof{p_3(\mA)-p_3(\mA)_\ell}_F^2\normof{\mB\mM}_2^2 \\
	&\leq (L+1) \normof{p_3(\mA)-p_3(\mA)_\ell}_F^2 \\
	&= (L+1) \sum_{i=\ell+1}^n p(\sigma_i(\mA))^2 \\
	&\leq (L+1)n \cdot \frac{4\sigma_{\ell+1}(\mA)^2}{2^{2q\sqrt\gamma}} \tag{Lemma 5 of \cite{MuscoMusco:2015}} \\
	&\leq \frac\eps2\sigma_{\ell+1}(\mA)^2,
\end{align*}
This recovers Equation (5) on page 10 of \cite{MuscoMusco:2015}, and from there the rest of the proof of Property 1 holds.
As discussed on page 16 of \cite{MuscoMusco:2015}, Property 1 holds here for all \(l \in [k]\), and therefore Properties 2 and 3 also hold.
Then, the analysis in Section 6.2 shows how Properties 1-3 imply the spectral, Frobenius, and singular value guarantees.
\end{proof}

\section{Single Vector Simultaneous Iteration}
\label{sec:single-vector-simultaneous-iteration}
We briefly present a single vector algorithm for low-rank approximation based on the standard simultaneous iteration algorithm.
Simultaneous iteration, or block power method, extracts a low-rank approximation from the span of  \(\mK = \mA^{2q}\mB\) where \(\mB\) is a starting block.
We present a prototypical pseudocode for simultaneous iteration in \Cref{alg:block-power}.
\cite{MuscoMusco:2015} show that \Cref{alg:block-power} converges from any \((k,L)\)-good starting block \mB, giving the two following theorems, which follow from the same arguments as in \Cref{app:musco_explanation} and \cite{MuscoMusco:2015}:

\begin{algorithm}[h]
	\caption{Simultaneous Iteration for Low-Rank Approximation}
	\label{alg:block-power}
	{\bfseries input}: Matrix \(\mA\in\bbR^{n \times d}\). Target rank \(k\). Starting block \(\mB\in\bbR^{n \times \ell}\). Number of iterations \(t\). \\
	{\bfseries output}: Orthogonal matrix \(\mQ\in\bbR^{n \times k}\).\\
	\vspace{-1em}
	\begin{algorithmic}[1]
		\STATE Compute an orthonormal basis \mZ for \(\mK = (\mA\mA^\intercal)^t \mB\).
		\STATE Compute \(\mU_k\), the \(k\) top eigenvectors of \(\mM = \mZ^\intercal\mA\mA^\intercal\mZ\)
		\STATE {\bfseries return} \(\mQ = \mZ\mU_k\).
	\end{algorithmic}
\end{algorithm}

\begin{algorithm}[h]
	\caption{Single Vector Simultaneous Iteration for Low-Rank Approximation}
	\label{alg:single-vec-block-power}
	{\bfseries input}: Matrix \(\mA\in\bbR^{n \times d}\). Target rank \(k\). Starting vector \(\vx\in\bbR^{n}\). Number of iterations \(t\). Memory budget \(\ell \geq k\). \\
	{\bfseries output}: Orthogonal matrix \(\mQ\in\bbR^{n \times k}\).\\
	\vspace{-1em}
	\begin{algorithmic}[1]
		\STATE Compute an orthonormal basis \mZ for \(\mK = [\, (\mA\mA^\intercal)^{t-\ell+1}\vx,~ (\mA\mA^\intercal)^{t-\ell+2}\vx, ~\ldots,~ (\mA\mA^\intercal)^t \vx \,]\).
		\STATE Compute \(\mU_k\), the \(k\) top eigenvectors of \(\mM = \mZ^\intercal\mA\mA^\intercal\mZ\)
		\STATE {\bfseries return} \(\mQ = \mZ\mU_k\).
	\end{algorithmic}
\end{algorithm}

\begin{importedtheorem}[Theorem 1 of \cite{MuscoMusco:2015}]
\label{impthm:poly-eps-low-rank-block-power}
Let \(\mB\in\bbR^{n \times \ell}\) be any \((k,L)\)-good starting matrix (\Cref{def:lgood-small-block}) matrix for \mA.
If we run Simultaneous Iteration (\Cref{alg:block-power}) for \(q = O(\frac{1}{\eps} \log(\frac{nL}{\eps}))\) iterations with starting block \mB, then the output  \(\mQ\in\bbR^{n \times k}\) satisfies
\begin{align*}
	\normof{\mA-\mQ\mQ^\intercal\mA}_{\xi} &\leq (1+\eps) \normof{\mA-\mA_k}_{\xi} & &\text{and} & \left| \vq_i^\intercal \mA\mA^\intercal \vq_i -\sigma_i(\mA)^2 \right| &\leq \eps\sigma_{k+1}(\mA)^2.
\end{align*}
\end{importedtheorem}

\begin{importedtheorem}[Theorem 13 of \cite{MuscoMusco:2015}]
\label{impthm:spectral-decay-low-rank-block-power}
Let \(\mB\in\bbR^{n \times \ell_0}\) be any \((\ell,L)\)-good starting matrix (\Cref{def:lgood-small-block}) matrix for \mA, for some \(\ell \geq k\).
If we run Simultaneous Iteration (\Cref{alg:block-power}) for \(q = O(\frac{1}{g_{k \rightarrow \ell}} \log(\frac{nL}{\eps}))\) iterations with starting block \mB, where \(g_{k \rightarrow \ell} = \frac{\sigma_{k}-\sigma_{\ell+1}}{\sigma_{k}}\), then the output  \(\mQ\in\bbR^{n \times k}\) satisfies
\begin{align*}
	\normof{\mA-\mQ\mQ^\intercal\mA}_{\xi} &\leq (1+\eps) \normof{\mA-\mA_k}_{\xi} & &\text{and} & \left| \vq_i^\intercal \mA\mA^\intercal \vq_i -\sigma_i(\mA)^2 \right| &\leq \eps\sigma_{k+1}(\mA)^2.
\end{align*}
\end{importedtheorem}

When compared to block Krylov Iteration, subspace iteration uses less memory, but converges slower. Specifically, in the theorems above we obtain a dependence on $1/\epsilon$ and $1/g_{k \rightarrow \ell}$ in comparison to $1/\sqrt{\epsilon}$ and $1/\sqrt{g_{k \rightarrow \ell}}$ in the comparable \Cref{impthm:poly-eps-low-rank} and \Cref{impthm:spectral-decay-low-rank} for block Krylov iteration. 

In \Cref{alg:single-vec-block-power}, we present a ``single vector'' variant of simultaneous iteration that similarly saves memory over the single vector Krylov method from \Cref{alg:single-vec-krylov}.
When run for $t$ iterations, instead of storing the entire length $t$ Krylov subspace as in \Cref{alg:single-vec-krylov}, the method only stores the last $\ell \geq k$ columns for a specified memory budget $\ell$. 

Taking \(\ell=k\), we can analyze \Cref{alg:single-vec-block-power} by letting \(\mS_k = \bmat{\vx & \mA^2\vx & \mA^4\vx & \ldots & \mA^{2(k-1)}\vx}\) and noticing that \Cref{alg:block-power} run with starting matrix \(\mS_k\) for \(q\) iterations produces the matrix \(\mK = \mA^{2q}\mS_k = \bmat{\mA^{2q}\vx & \mA^{2(q+1)}\vx & \ldots & \mA^{2(q+k)}\vx}\).
Since \Cref{thm:low-rank-approx} already tells us that \(\mS_k\) is \((k,L)\)-good for \mA, applying \Cref{impthm:poly-eps-low-rank-block-power} immediately gives the following convergence guarantees for \Cref{alg:single-vec-block-power}:
\begin{theorem}
\label{thm:single-vec-guarantee-block-power}
For \(\mA \in \bbR^{n \times d}\), let \(g_{min} = \min_{i \in \{1,\ldots,k-1\}} \frac{\sigma_i - \sigma_{i+1}}{\sigma_{i+1}}\).
For any \(\eps, \delta \in (0,1)\), \Cref{alg:single-vec-block-power} initialized with \(\vx\sim\cN(\vec0,\mI)\) and run for \(t = O(\tsfrac{k}{\eps} \log(\tsfrac1{g_{min}}) + \tsfrac1{\eps} \log(\tsfrac{n}{\eps\delta}))\) iterations with memory budget \(k\) returns an orthogonal $\mQ \in \R^{n \times k}$ such that, with probability at least \(1-\delta\),
\begin{align*}
	\normof{\mA-\mQ\mQ^\intercal\mA}_{\xi} &\leq (1+\eps) \normof{\mA-\mA_k}_{\xi} & &\text{and} & \left| \vq_i^\intercal \mA\mA^\intercal \vq_i -\sigma_i(\mA)^2 \right| &\leq \eps\sigma_{k+1}(\mA)^2.
\end{align*}
\end{theorem}
Additionally, if  \Cref{alg:single-vec-block-power}  is run with memory budget $\ell \geq k$, we can obtain a spectrum dependent convergence bound comparable to the result proven in \Cref{sec:spectral-decay} for single vector Krylov iteration.
Specifically, since \Cref{thm:low-rank-approx} already tells us that \(\mS_\ell\) is \((\ell,L)\)-good for \mA, applying \Cref{impthm:spectral-decay-low-rank-block-power} immediately gives the following convergence guarantees for \Cref{alg:single-vec-block-power}:
\begin{theorem}
\label{thm:decay-dependence-block-power}
For \(\mA \in\bbR^{n \times d}\) and \(\ell \geq k\), let \(g_{min} = \min_{i\in\{1,\ldots,\ell-1\}} \frac{\sigma_i - \sigma_{i+1}}{\sigma_{i+1}}\) and \(g_{k \rightarrow \ell} = \frac{\sigma_k - \sigma_{\ell+1}}{\sigma_{k}}\).
For any \(\eps,\delta\in(0,1)\), \Cref{alg:single-vec-block-power} initialized with \(\vx\sim\cN(\vec0,\mI)\) and run for \(t = O(\frac{\ell}{g_{k\rightarrow\ell}} \log(\frac1{g_{\min}}) + \frac{1}{g_{k\rightarrow\ell}}\log(\frac{n}{\delta\eps}))\) iterations with memory budget \(\ell\) returns an orthogonal \(\mQ\in\bbR^{n \times k}\) such that, with probability at least \(1-\delta\),
\begin{align*}
	\normof{\mA-\mQ\mQ^\intercal\mA}_{\xi} &\leq (1+\eps) \normof{\mA-\mA_k}_{\xi} & &\text{and} & \left| \vq_i^\intercal \mA\mA^\intercal \vq_i -\sigma_i(\mA)^2 \right| &\leq \eps\sigma_{k+1}(\mA)^2.
\end{align*}
\end{theorem}


\end{document}